\documentclass[AEJ]{AEA} \usepackage[spanish,english]{babel} \usepackage{xcolor}  \usepackage{amsthm}\newtheorem*{definition}{Definition}\usepackage{multirow,adjustbox,natbib,color,soul,relsize,dirtytalk}\usepackage{cancel,enumerate}\usepackage{subcaption}\makeatletter\AtBeginDocument{%
  \begingroup  \normalsize  \let\tmp@n@s\f@size  \let\tmp@n@b\f@baselineskip  \small  \let\tmp@s@s\f@size  \let\tmp@s@b\f@baselineskip \xdef\semismall@size{\fpeval{(\tmp@n@s+\tmp@s@s)/2}}%
  \xdef\semismall@baselineskip{\fpeval{(\tmp@n@b+\tmp@s@b)/2}}%
  \endgroup
}\newcommand{\semismall}{\fontsize{\semismall@size}{\semismall@baselineskip}\selectfont}
\usepackage{anyfontsize,resizegather,comment}
\theoremstyle{plain}
\newtheorem{theorem}{Theorem}\newtheorem{lemma}{Lemma} \allowdisplaybreaks \usepackage{accents}\usepackage{booktabs}\usepackage{lscape} \usepackage{animate} \usepackage[utf8]{inputenc} \usepackage{txfonts}  \usepackage{longtable} \usepackage{bm} \usepackage{bbm}\usepackage[figuresright]{rotating}  \makeatletter\usepackage{pseudocode}\usepackage{algorithm}\usepackage{float}\usepackage{algpseudocode} \usepackage{comment} \AtBeginDocument{\g@addto@macro{\appendix}{\renewcommand{\p@subsection}{}}} \makeatother \usepackage[para,online,flushleft]{threeparttable}\usepackage{multirow}\usepackage{multicol} \usepackage{graphicx, comment}\definecolor{ahjcolor}{rgb}{0.0, 0.13, 0.40}			\usepackage{hyperref} 

\hypersetup{colorlinks	=true,									 linkcolor	=ahjcolor,									         urlcolor	=ahjcolor,									            citecolor	=ahjcolor,            bookmarksnumbered=true}\newtheorem*{assumption*}{\assumptionnumber}
\providecommand{\assumptionnumber}{}\makeatletter\newenvironment{assumption1}[2]
 {%
  \renewcommand{\assumptionnumber}{Assumption #1-#2}%
  \begin{assumption*}%
  \protected@edef\@currentlabel{#1-#2}%
 }
 {%
  \end{assumption*} }
\newtheorem{proposition}{Proposition}\newtheorem{corollaryx}{Corollary}[proposition]\newenvironment{corollaryp}[1][]{  \begin{corollaryx}[#1]\itshape}{  \end{corollaryx}}
\draftSpacing{1.5}
\begin{document}
\title{Identification and Estimation of Intergenerational Income Mobility Measures\footnote{I am grateful to Juan Carlos Escanciano and Jan Stuhler for
their guidance and support, and to Christophe Gaillac, Antonio Raiola, Stephen Jenkins, and Nazarii Salish for their insightful discussions and comments. I also thank participants of the International Association for Applied Econometrics 2025 Annual Conference, the Eleventh Italian Congress of Econometrics and Empirical Economics, the Econometrics Brown Bag Seminar at University College London, the PhD Workshops at Universidad Carlos III de Madrid, the ENTER Jamboree 2025 Conference, and the III at 10: New Directions in Inequality Research for their valuable feedback.}} \author{Alejandro Puerta-Cuartas\thanks{Department of Economics, Universidad Carlos III, Madrid, Spain, \href{E-mail: alpuerta@eco.uc3m.es}{E-mail: alpuerta@eco.uc3m.es} }  \\\vspace{5mm}\today  
} \pubMonth{}\pubYear{}\pubVolume{}\pubIssue{}
\begin{abstract} 
Measuring the intergenerational transmission of lifetime economic status is complicated by researchers often only observing snapshots of income at specific ages. Consequently, standard practice estimates intergenerational mobility using income averages, introducing life-cycle bias that compromises reliability and comparability across studies, time, and place. I develop a missing data framework that exploits available income data and observable characteristics to eliminate life-cycle bias. This method combines nonparametric identification with Neyman-orthogonal moments to construct debiased machine learning estimators for intergenerational income mobility measures under plausible missing-at-random and testable independence assumptions. I apply this framework to estimate the intergenerational elasticity for the U.S. using the Panel Study of Income Dynamics across birth cohorts from 1954 to 1977 with rolling 10-year windows. While existing approaches estimate values between 0.41 and 0.54, the proposed method yields substantially higher estimates ranging from 0.6 to 0.7, averaging 0.64. These results align closely with recent evidence using long time averages over mid-career periods, reinforcing high U.S. intergenerational persistence.
\vspace{2mm}
\end{abstract}\maketitle \clearpage \section{Introduction}\label{sec:1}
Many economic and causal parameters depend on lifetime outcomes such as income, earnings, or consumption. However, survey and administrative data typically cover only a limited segment of individuals’ working lives, posing significant challenges across fields such as household economics, education, and labor. For example, this limitation complicates measuring the degree to which income shocks transmit to consumption \citep{jappelli2010consumption}, distorts estimates of long-run earnings returns to education \citep{heckman2006earnings}, and introduces life-cycle bias when estimating intergenerational income persistence \citep{solon1992intergenerational}. This paper develops a missing data framework to construct consistent, locally robust estimators for intergenerational income mobility measures from incomplete income data and individual characteristics. The method combines nonparametric identification with Neyman-orthogonal moments to eliminate the life-cycle bias arising from incomplete income data. I illustrate the practical advantages of this approach by estimating the intergenerational elasticity (IGE)  in the United States.\par 
Existing approaches to measuring intergenerational income persistence rely on income averages, introducing life-cycle bias. This occurs because annual income snapshots only partially capture lifetime economic status, particularly when observed at early or late career stages \citep{haider2006life}. The problem is further exacerbated by parents and children typically being observed at different life stages \citep{jenkins1987snapshots}, and by individuals exhibiting heterogeneous income growth that varies with parental characteristics \citep{halvorsen2022earnings}.\par 
While correcting individual sources of bias improves estimates, this piecemeal approach hinders comparability. Recent literature has made substantial progress by separately addressing life-cycle bias: \citet{mazumder2016estimating} use long-time averages for parents, while \citet{mello2022lifecycle} propose a life-cycle (LC) estimator that predicts children's income profiles accounting for income growth that varies with parental characteristics.  Nevertheless, questions remain about the robustness of existing estimates and the reliability of comparisons across time and place \citep{mogstad2023family}. I formalize why comparability remains elusive: bias magnitudes vary with research design, income dynamics, and sampling rules, causing estimators relying on income averages to converge to context-specific parameters even after correcting specific biases. Estimates from different studies thus target fundamentally different parameters, compromising both the assessment of mobility levels and comparisons across studies and countries. This underscores the need to move beyond proxy refinement and focus on identification, which jointly addresses all sources of bias.\par 
This paper shows that, despite lifetime income being unobserved, intergenerational income mobility measures can be nonparametrically identified from incomplete income data and observable characteristics. The key insight is that although individual income profiles are only partially observed, cross-sectional variation across ages and individual characteristics can be exploited to recover them. To illustrate this framework concretely, I establish nonparametric identification for the IGE, the most common measure of income persistence across generations \citep{nybom2017biases}. The covariance between children's and parents' lifetime incomes, the IGE numerator, is identified as the average conditional covariance of their annual incomes across all age pairs, and the denominator, the variance of parental lifetime income, is recovered from conditional autocovariances at nearby ages and predicted income profiles at distant ages.\par 
Identification relies on two sets of assumptions. First, to recover income profiles and the autocovariance of parental income, a missing-at-random (MAR) condition is imposed, requiring that, conditional on individual characteristics, whether income is observed at a given age is unrelated to the income level itself. Additionally, a common support condition requires that in the population, the probability of observing income at a given age, conditional on observable characteristics, is bounded away from zero and one.  Second, to eliminate life-cycle bias from both generations, two independence assumptions are imposed. For children, I assume that income prediction errors—the part of annual income unexplained by observable characteristics—are uncorrelated with parental lifetime income. This condition can be satisfied by modeling children's income profiles as functions of parental characteristics (such as income and education) interacted with age, capturing systematic differences in income growth across family backgrounds. For parents, I assume that prediction errors at distant periods (more than 10 years apart in the application) are uncorrelated. By conditioning on observable characteristics that capture persistent income components, these prediction errors reflect only transitory shocks that dissipate over time.\par
Building on nonparametric identification, I construct debiased machine learning estimators for intergenerational income mobility measures. For the IGE, the identification result shows that it can be recovered from conditional expectations of income profiles and parental income autocovariances. Estimating these nuisance parameters requires flexibly accommodating high-dimensional individual characteristics and heterogeneous, nonlinear income dynamics. While machine learning (ML) methods can adaptively learn these complex relationships, they introduce regularization and model selection bias that would propagate to IGE estimation. Following \cite{chernozhukov2018double} and more specifically \cite{chernozhukov2022locally}, I address this by constructing Neyman-orthogonal moments that incorporate the influence function of the first-stage ML estimates. This orthogonalization ensures local robustness: first-stage estimation errors affect the IGE estimate only at second order, thereby mitigating regularization and model selection bias. Additionally, I employ cross-fitting, which prevents overfitting and own-observation bias. This approach enables valid inference while leveraging the flexibility of modern ML methods.\par  
Consistency and asymptotic normality are established for the resulting estimators, enabling valid inference that accounts for first-stage machine learning estimation of nuisance parameters. Simulations suggest the estimators exhibit sound finite-sample performance, with negligible bias that vanishes as sample size increases and coverage rates close to nominal levels. In contrast, a naive ML plug-in estimator uncorrected for first-step estimation errors and existing approaches relying on income averages exhibit sizable bias and severe undercoverage. Additionally, I develop a locally robust test for the assumption that children’s prediction errors are uncorrelated with parental lifetime income. I show that it attains correct asymptotic size, is consistent against fixed alternatives, and exhibits nontrivial power against local alternatives. To facilitate implementation, the proposed methods will soon be available in a companion user-friendly \texttt{R} package, \texttt{LRIGE}.\par 
I apply the framework to estimate the IGE in the United States using the Panel Study of Income Dynamics (PSID). This dataset is particularly well-suited for this analysis: it has been widely used to study income mobility, provides long-term income histories, and includes rich individual characteristics relevant for income prediction. Importantly, the literature suggests that the MAR assumption for income missingness in the PSID is empirically plausible, provided relevant observables are accounted for \citep{fitzgerald2011attrition,schoeni2015implications}. The analysis focuses on birth cohorts spanning 1954 to 1977, using rolling 10-year windows and covering the lifetime period from ages 25 to 55. This sample design ensures both parents and children are observed during working life, guaranteeing both missing and non-missing observations at each age and satisfying the common support condition required for identification. Across cohort windows, the proposed test strongly supports that children's prediction errors are uncorrelated with parental permanent income: in 14 of 15 cases, the correlation is not statistically different from zero. Additionally, the results suggest that autocorrelation in parental income prediction errors becomes negligible beyond ten years in all 15 windows. Together with the empirical plausibility of the MAR assumption documented in prior work, these results validate the identifying assumptions underlying the framework. \par 
The proposed method yields IGE estimates ranging from 0.6 to 0.7 across cohorts, with an average of 0.64. These results align closely with recent PSID-based studies that address life-cycle bias and find values exceeding 0.6 \citep{gouskova2010estimating,chau2012intergenerational,mazumder2016estimating}. In contrast, a conventional approach using three-year income averages during mid-life for both generations yields an average IGE of only 0.43 (ranging from 0.41 to 0.49). The LC estimator substantially improves upon this, producing an average IGE of 0.50 (ranging from 0.46 to 0.54).  Decomposing estimates into their covariance and variance components reveals that the LC's covariance estimates closely match the locally robust benchmark, providing direct evidence that its explicit modeling of heterogeneous income profiles successfully addresses children's life-cycle bias. Finally, a plug-in ML estimator yields an average IGE of 0.60 (ranging from 0.46 to 0.71), with bias reaching 0.24 in one cohort and confidence intervals 41\% narrower on average than the debiased approach, underscoring the importance of orthogonalization for both bias correction and valid inference.\par
This paper makes two contributions. First, it develops a method for obtaining reliable and comparable intergenerational income mobility estimates across studies, time, and place. Although I focus on the IGE, the framework applies more broadly to other mobility measures. For example, the intergenerational correlation equals the IGE scaled by the ratio of parents’ to children’s income standard deviations. Extending the IGE result to the correlation thus requires only two additional steps: identifying children's income variance by the same argument used for parents, and constructing a Neyman-orthogonal moment for the correlation parameter rather than the elasticity. \par 
Second, it introduces a missing data framework that enables locally robust inference when parameters depend on partially observed lifetime outcomes. Standard debiased machine learning  assumes the parameter of interest is identified from observable data \citep{chernozhukov2022locally}. Satisfying this assumption is not straightforward when lifetime outcomes, such as permanent income, consumption, or educational returns, are only partially observed across the life cycle. I address this challenge by establishing that these parameters can be nonparametrically identified from incomplete income data and observable characteristics under MAR and orthogonality conditions, thereby enabling construction of Neyman-orthogonal moments for locally robust inference. The framework directly extends to estimating long-run educational returns and measuring intergenerational transmission of well-being, where lifetime outcomes for one or both generations are partially observed. It also applies to more complex settings, such as partial insurance models \citep{blundell2008consumption}. When rich longitudinal data are available, nonparametric identification makes it possible to exploit high-dimensional data through machine learning to recover the variance-covariance structure of consumption and income while maintaining locally robust inference.\par 
The remainder of the paper is as follows: Section \ref{sec:2} formalizes that proxy-based estimators converge to context-specific limits, compromising comparability across studies. Section \ref{sec:3} presents nonparametric identification of the IGE, a locally robust estimator, and inference results including asymptotic normality and testing. Section \ref{sec:sims} reports simulations, and Section \ref{sec:app} applies the estimator to measure the IGE in the United States. Proofs are provided in the Appendix.
\section{Biases and Comparability in Intergenerational Elasticity Estimates}\label{sec:2} 
The biases introduced by using income averages to assess intergenerational transmission of lifetime economic status affect all mobility measures. To derive explicit characterizations, however, I focus on a specific measure: the intergenerational elasticity of income, the most common measure of income persistence across generations \citep{nybom2017biases}. The IGE captures the degree to which income differences between parents are associated with income differences among their children. Formally, the IGE is defined by the regression
\begin{align}\label{eq:igemain} Y_c^P&=\alpha_0+\beta_0 Y_f^P+u, \quad \mathbb{E}\left[u \left(1, Y_f^P\right)'\right]=0,\end{align} where $Y_f^P$ and $Y_c^P$ denote the permanent component of log annual income for fathers and children \citep{solon1992intergenerational}, $u$ is an idiosyncratic error term uncorrelated with parental income, and $\beta_0$ is the intergenerational elasticity. Henceforth, I will refer to $Y^P$ as permanent income.\par 
According to equation (\ref{eq:igemain}), the closed form solution for the IGE is given by  \begin{align}\label{eq:beta}   \beta_0&=\frac{\mathbb{E}\left[\left(Y_c^P-\mathbb{E}\left[Y_c^P\right]\right)\left(Y_f^P-\mathbb{E}\left[Y_f^P\right]\right)\right]}{\mathbb{E}\left[\left(Y_f^P-\mathbb{E}\left[Y_f^P\right]\right)^2\right]}. \end{align} 
However, because permanent income data is rarely available in practice, researchers typically rely on short-term income snapshots \citep{mazumder2005fortunate}, using either a single-year observation or a multi-year average as a proxy for permanent income. This raises the question of what exactly the available estimators in the literature measure, and whether their estimates are comparable across settings.\par 
To address this, I adopt the strategy that \cite{mogstad2024instrumental} refer to as \say{reverse engineering}. In the context of instrumental variables (IV), this approach begins with a practical problem: when treatment effects exhibit unobserved heterogeneity (UHTE), the classical linear IV model is misspecified. Yet a linear IV estimate can still be computed. The reverse engineering framework then seeks to determine what, if anything, this estimator measures. Thus, it proceeds by starting with the tool, and it attempts to reverse engineer an interpretation for it under suitable assumptions.\par 
The idea of reverse engineering has already been applied in the intergenerational mobility literature to formally characterize the behavior of estimators based on proxy measures of permanent income \citep{solon1992intergenerational,nybom2016heterogeneous}. Specifically, it has been used to establish the probability limit of such estimators, thereby identifying the distinct sources of bias introduced by relying on imperfect proxies. \par 
To set the grounds for the analysis, I follow the literature by first characterizing the probability limit of the estimator based on income proxies. I begin by defining the observed data typically available to researchers. Most empirical studies utilize longitudinal datasets such as the Panel Study of Income Dynamics (PSID), which contain only partial income trajectories for parents and children, along with additional individual and family characteristics. Formally, the observed data consist of an independent and identically distributed (i.i.d.) sample of $\bm W = \big(\bm Y_c \odot \bm D_c, \bm Y_f \odot \bm D_f, \bm D_c, \bm D_f, \bm X \big),$ where \(\bm Y_c\) and \(\bm Y_f\) are \(T\)-dimensional random vectors containing information on (log) annual child and parental income, respectively, the vectors \(\bm D_c\) and \(\bm D_f\) are \(T\)-dimensional indicator vectors, with elements \(D_{gt} = 1\) if \(Y_{gt}\) is observed and \(D_{gt} = 0\) otherwise, for \(g \in \{c,f\}\), \(\odot\) denotes the element-wise product, so that \(\bm Y_g \odot \bm D_g\) contains the observed entries of \(\bm Y_g\) and zeros elsewhere, and the vector $\bm X$ contains observed characteristics for both generations.\par 
The standard approach to estimate the IGE, which I label the mid-life income (MI) estimator, consists of a two-step approach. First, it proxies permanent income as the average of $T_f$ and $T_c$ (log) annual income observations around mid-life for the fathers and the children, respectively. In the second step, it regresses the child's proxy measure on the parent's. This practice has a long history, dating back to early contributions that recognized the problem of measurement error and sought to approximate permanent income using multi-year averages \citep{de1973relation,hauser1975socioeconomic,freeman1978black,tsai1983sex}. Its limitations were also identified early on: \citet{creedy1977distribution} and \citet{jenkins1987snapshots} pointed out that this approach suffers from life-cycle bias \citep{creedy1977distribution,jenkins1987snapshots}, which arises because income \say{snapshots} fail to capture complete lifetime earnings profiles and because parents and children are often observed at different stages of their life cycles. Despite these limitations being identified decades ago, the MI estimator remains the standard practice in empirical studies of intergenerational mobility.
\par Formally, the MI estimand is defined as the slope coefficient in the projection:
\begin{align*} \tilde{Y}_c^P&=\alpha^{MI}+\beta^{MI} \tilde{Y}_f^P+u^{MI}, \quad \mathbb{E}\left[u^{MI} \left(1, \tilde{Y}_f^P\right)'\right]=0,\\\nonumber   \tilde{Y}_g^P&\coloneqq\frac{1}{T_g}\sum_{j\in \mathcal{M}_g}Y_{gj}D_{gj}, \quad g\in \{c,f\},\end{align*} 
where $D_{gj}=1$ when $Y_{gj}$ is observed and zero otherwise,  $\mathcal{M}_g$ is a set of pre-defined mid-life years for generation $g$,  and $T_{g}\coloneqq\sum_{j\in \mathcal{M}_g}D_{gj}$ is the number of years used for the average. \par 
To establish the probability limit of the MI estimator using the reverse engineering approach \citep{mogstad2024instrumental}, I now impose standard assumptions used in the literature. A comprehensive discussion of the MI estimator’s definition, theoretical underpinnings, assumptions, and sources of bias can be found in Appendix \ref{sec:mi}.\par 
\begin{assumption1}{1}{MI}(Annual Income Process)\label{as:1}
The relationship between annual and permanent income is governed by \begin{align*}  
Y_{gt}&=\lambda_tY^P_g+v_{gt}, \quad \mathbb{E}\left[v_{gt} Y^P_g\right]=0, \quad g\in \{c,f\},  \quad t=1,...,T,\\  \lambda_t&=1, \forall t\in \mathcal{M}_g, \quad g\in \{c,f\}.\end{align*}  where $\lambda_t$ captures that the persistence of permanent income may vary over the life-cycle period, and $v_{gt}$ is an age shock.
\end{assumption1} \begin{assumption1}{2}{MI}(Conditional Mean Independence) \label{as:2mi}  The following  conditional mean restrictions hold \begin{align*}     \mathbb{E}\left[v_{ct}v_{fj}\big | D_{ct},D_{fj}\right]&=0, \quad t\in \mathcal{M}_c, \quad j\in \mathcal{M}_f,\\  \mathbb{E}\left[v_{fj}Y_c^P\big | D_{fj}\right]&=0, \quad  j\in \mathcal{M}_f, \\ \mathbb{E}\left[v_{ft}Y_f^P | D_{ft}, D_{fj}\right]&=0, \quad tj \in \mathcal{M}_f, \\ \mathbb{E}\left[v_{gj} | D_{gj}\right]&=0, \quad g\in \{c,f\}, \quad j \in \mathcal{M}_j.\end{align*}
\end{assumption1}
The inconsistency of the MI estimator is well-documented in the literature. Proposition \ref{coro:1} restates this result to make explicit the four sources of bias that will be central to the discussion. While the main characterization of these biases is familiar, the proposition extends prior work by incorporating missing income data. Appendix \ref{sec:equi} shows that Proposition \ref{coro:1} reduces to the results of \cite{solon1992intergenerational} and coincides with \cite{nybom2016heterogeneous} under certain assumption variants. In addition, the proposition formalizes the empirical observation that IGE estimates are sensitive to sample inclusion criteria and missing income, by demonstrating how the observation probabilities $p_f\left({t,j}\in \mathcal{M}_f\right)$ and $p_c\left(t\in \mathcal{M}_c\right)$ shape the asymptotic bias.
\begin{proposition}
\label{coro:1}  \fontsize{10}{12}\selectfont  Under Assumptions \ref{as:1} and \ref{as:2mi}, the probability limit of the MI estimator is given by:\fontsize{10}{12}\selectfont 
\begin{gather}\label{eq:mi_inconsistent}  
\hat{\beta}_n^{MI}\overset{p}{\to}\frac{\beta_0\mathbb{E}\left[\left(Y_f^P-\mathbb{E}\left[Y_f^P\right]\right)^2\right]+\overbrace{\frac{1}{T_c}\sum_{t} \mathbb{E}\left[Y_f^Pv_{ct}\big  | D_{ct}=1,t\in \mathcal{M}_c\right]}^\text{(c)}\times \overbrace{p_c\left(t\in \mathcal{M}_c\right)}^\text{(d)}}{\mathbb{E}\left[\left(Y_f^P-\mathbb{E}\left[Y_f^P\right]\right)^2\right]+\underbrace{\frac{1}{T_f^2}\sum_{t}\sum_{j}\mathbb{E}\left[v_{ft}v_{fj}\big | D_{ft}=1,D_{fj}=1, \left\{t,j\right\}\in \mathcal{M}_f\right]}_\text{(a)}\times  \underbrace{p_f\left(\left\{t,j\right\}\in \mathcal{M}_f\right)}_\text{(b)}}, 
\end{gather} \fontsize{10}{12}\selectfont where $v_{ct}$ and $v_{ft}$ are children and parental age shocks to (log) annual income as defined in Assumption \ref{as:1}, and
$p_c\left(t\in \mathcal{M}_c\right)$ and   $p_f\left(\left\{t,j\right\}\in \mathcal{M}_f\right)$ denote the probabilities of observing child income at mid-life year $t$, and parent income at mid-life years in years $t$ and $j$, respectively.
\end{proposition} 
Proposition \ref{coro:1} presents a formal statement of the biases already familiar from prior work,  including (a) the measurement error and life-cycle bias of parental income \citep{solon1992intergenerational,mazumder2005fortunate}, (b) the sensitivity of the IGE estimates to low, zero, and missing parental income observations \citep{couch1998sample,dahl2008association,chetty2014land,nybom2016heterogeneous},  (c) the measurement error and life-cycle bias of children's income \citep{nybom2016heterogeneous}, and (d) the sensitivity to the number of years and the selected year(s) to measure children's income \citep{mello2022lifecycle}. For ease of exposition, the proposition is stated under a classical errors-in-variables formulation, so that the age-shocks are treated as capturing both transitory fluctuations and the systematic life-cycle bias. In Appendix \ref{sec:equi} (equation \ref{eq:ns}) this restriction is relaxed by allowing for a generalized errors-in-variables structure that explicitly separates the life-cycle component. A brief discussion of each component that hinders the consistent estimation of the IGE by $\hat{\beta}_n^{MI}$ can be found in Appendix \ref{sec:mi}.\par 
Identifying the distinct biases introduced by noisy measures has enabled the literature to refine estimation procedures by addressing specific sources of bias. For example, \cite{mazumder2005fortunate} addresses the measurement error and life-cycle bias of parental income (component (a) in equation (\ref{eq:mi_inconsistent})) by using long-term parental income averages centered at age 40. More recently, \cite{mello2022lifecycle} address life-cycle bias from using snapshots of children's income, captured by terms (c) and (d), by predicting children's income profiles from standard observables such as age and education, while allowing income growth to be steeper for children from more affluent families. \cite{lubotsky2006interpretation} show that including proxy variables separately in a regression and optimally weighting their coefficients yields less attenuated estimates than constructing a summary measure of the proxy variables.\par 
Despite these advances, questions remain about the robustness of existing estimates and the reliability of comparisons across time and place \citep{mogstad2023family,mello2022lifecycle}. While the literature has rightly emphasized the sensitivity of IGE estimates to the biases in Proposition \ref{coro:1}, a more fundamental issue is that these biases systematically alter the target parameter itself. Crucially, the magnitude of each bias component varies with income dynamics, study design, and sample inclusion criteria. Due to heterogeneity in institutional contexts and study design choices, these factors differ across datasets, regions, countries, and time, rendering IGE estimates non-comparable. The following corollary makes this dependence explicit by formalizing that the target parameter is inherently dependent on the study design and the underlying income dynamics.\par 
\begin{corollaryp}[Context-Dependence of the MI Estimand] \label{coro:context} Under the assumptions of Proposition \ref{coro:1}, the probability limit of the Mid-Life Income estimator is a context-dependent parameter \begin{align}\label{eq:power} \hat{\beta}_n^{MI} \overset{p}{\to} \beta^{MI}(\eta)\coloneqq\beta_0 \cdot \Delta(\eta), \end{align} where the distortion factor $\Delta(\eta) \neq 1$ captures departure from consistency. Formally,  the research-design/context vector   \[ \eta\coloneqq \big(\mathcal{M},T p,\Sigma\big),\]
collects all research-design choices and structural features of the income process: $\mathcal{M}\coloneqq\left(\mathcal{M}_f, \mathcal{M}_c\right)$ and $T\coloneqq\left(T_f, T_c\right)$ are the set of pre-defined mid-life years, and the number of years used for the average for generation $g\in \{c,f\}$, respectively; $p=(p_f,p_c)$ captures observation/availability and selection rules; and $\Sigma$ summarizes how permanent income and transitory fluctuations in parents’ and children’s earnings $\left(Y_f^P,v_{ft},v_{ct}\right)$ vary and relate to each other. Equation \eqref{eq:mi_inconsistent} provides the explicit form of $\Delta(\eta)$.\par 
\end{corollaryp}
The central insight of Corollary \ref{coro:context} was anticipated by \cite{jenkins1987snapshots}, who demonstrated using a simple two-period life-cycle model that snapshot-based IGE estimates suffer from large life-cycle biases whose direction cannot be determined a priori. Because variance and covariance factors work in opposite directions in the bias calculation, life-cycle biases may be upwards or downwards depending on the specific parameters of the income process and the choice of estimator. This analysis revealed that same-stage-of-lifecycle estimates are not necessarily superior to contemporaneous ones, and that adjusting for within-generation age variation does not necessarily reduce bias.\par
The present analysis reaches similar conclusions through a complementary approach: reverse-engineering the standard mid-life income estimator rather than building from structural primitives. This reveals that $\hat{\beta}_n^{MI}$ converges to a context-dependent parameter $\beta^{MI}(\eta)$ where the distortion factor $\Delta(\eta)$ depends on midlife definitions, averaging windows, observation patterns, and income dynamics. While the structural model in \citep{jenkins1987snapshots} provided intuition for why biases arise and demonstrated their potential magnitude through calibrated examples, equation \eqref{eq:mi_inconsistent} shows how these biases manifest in standard practice, decomposing \(\Delta(\eta)\) into explicit components.\par 
This formalization rationalizes empirical patterns that have raised concerns about the reliability of comparisons across studies, time, and place. Specifically, it shows that reliance on income proxies introduces systematic distortions, preventing identification of the true intergenerational elasticity. These distortions limit the reliability and comparability of resulting estimates across studies, as each converges to its own context-specific value $\beta^{MI}(\bm{\eta})$. A compelling example is the wide variation in recent U.S. estimates, which range from 0.35 to 0.65 \citep{mello2022lifecycle}. Corollary \ref{coro:context} makes explicit the potential drivers of this pattern. Even when components such as (a) and (c) in equation \eqref{eq:mi_inconsistent} are held constant, differences in the definition of mid-life income $\left(\mathcal{M}_g\right)$, the number of years averaged ($T_g$), or sample selection rules ($p_g$) alter the distortion factor $\Delta(\eta)$ in equation \eqref{eq:power}. As a result, each estimate converges to a different context-specific parameter $\beta^{MI}(\eta)$.\par 
Distortions in $\Delta(\eta)$ also affect trend analyses, as both design choices and cohort-specific income dynamics can vary over time. In the U.S., the PSID’s transition from annual (1968–1997) to biennial interviews illustrates how survey design changes can alter the estimand: reduced income observations for recent cohorts modify $\Delta(\eta)$ through the observation probability $p_c$, potentially distorting mobility trends. Empirical evidence from Sweden further supports the theoretical distortions highlighted in Corollary \ref{coro:context}. \cite{mello2022lifecycle} show that MI-based estimates suggest a sharp decline in mobility for the 1950s–1970s cohorts, whereas their life-cycle estimator, which corrects for life-cycle bias in children’s income, indicates stable mobility across these cohorts, and a modest increase for those born in the 1980s.\par 
Finally, Corollary \ref{coro:context} formalizes how differences in study design and income dynamics can undermine cross-country comparisons. Even under the same definitions of mid-life income, differences in transitory shock persistence (a), children’s income growth (c), and observation probabilities (b, d) alter the estimand $\beta^{MI}(\eta)$. This calls for caution in interpreting international patterns such as the Great Gatsby Curve: unlike scale-free measures like the Gini coefficient, MI-based IGE estimates reflect both underlying mobility and study-specific distortions captured by $\Delta(\eta)$.\par
Beyond its implications for common applications, Corollary \ref{coro:context} shows that eliminating individual biases alone does not guarantee comparability. Even after correcting for specific sources of bias, differences in study design, cohort composition, or the magnitude of residual distortions can still produce inconsistent estimates. This analysis highlights a fundamental shift in perspective: rather than addressing individual sources of bias, attention should be directed toward identifying the IGE, which simultaneously removes all biases and allows for reliable, comparable estimates.
\section{Identification, Estimation, and Inference for the IGE with Incomplete Data}\label{sec:3} 
\subsection{Nonparametric Identification}
To establish identification of the intergenerational elasticity, I adopt the \say{forward-engineering} strategy proposed by \citet{mogstad2024instrumental}. In their terminology, this approach begins with a model and then constructs estimators under the assumption that the model is correctly specified. In the context of the IGE, my proposal precisely follows this logic: begin by defining permanent income and then derive the conditions under which the IGE is identified, given that definition.\par 
The literature has generally characterized permanent income rather than attempting to provide a precise definition. For instance, in line with \cite{solon1992intergenerational}, \cite{mazumder2005fortunate} interprets it as the permanent component of log earnings, capturing true long-term earning capacity. In contrast, \cite{haider2006life} describes it as a long-run income variable, such as the log of the present discounted value of lifetime earnings. Other work \citep{black2011recent,corak2013income} refer more generally to log permanent earnings without elaborating further. Reflecting this theoretical heterogeneity, empirical research has proxied permanent income differently; while some studies compute it as the log of average annual income \citep{dahl2008association,mazumder2016estimating}, others take the average of log annual income \citep{zimmerman1992regression,bratberg2007trends}. \par 
Our goal is not to define a unifying measure of permanent income or to identify a uniquely correct IGE. The intergenerational elasticity has never been directly observed; it has always been estimated under specific measurement choices and assumptions. I adopt a definition of permanent income that is theoretically grounded, empirically tractable, and consistent with standard practice in the literature. By settling on a workable definition, researchers can generate estimates of the IGE that are more reliable and comparable across studies and datasets. Thus, the forward-engineering approach provides a foundation for comparability and reliability. \par 
\begin{definition}[Permanent income]\label{def1}
For an individual of generation $g$ (where $g \in \{c, f\}$ for child or father), permanent income $\left(Y^P_g\right)$ is defined as their average log annual income over a specific lifetime period from $t=1$ to T:
\begin{align}\label{eq:def_perm_inc}  Y^P_g&\coloneqq\frac{1}{T}\sum_{t=1}^TY_{gt}, \quad g\in\{c,f\},\end{align}
where $Y_{gt}$ is log annual income in year $t$, with $t=1$ indicating the start age and $T$ the number of years covered.\end{definition}
Our definition of permanent income aligns with the literature that conceptualizes it as the permanent component of log earnings. This definition provides an empirically tractable measure that facilitates the identification of the intergenerational elasticity. The linearity of the sum-of-logs specification is crucial, as it permits the use of standard missing-at-random assumptions to recover permanent income from partially observed data. An alternative definition involving the log of the average introduces nonlinearities that preclude a similar identification strategy and require stronger assumptions about the joint distribution of income over lifetime. For a detailed discussion of these considerations, see Appendix~\ref{sec:lfinc}. When applied to the life-cycle estimator of \citet{mello2022lifecycle}, this definition yields results that are virtually identical to those obtained by defining permanent income as the log of the average, as shown in Table~\ref{tab:ige_cohort}.\par 
The fundamental challenge in estimating (identifying) the IGE is its reliance on unobserved permanent income. Traditional approaches, such as the Generalized Error-in-Variables (GEIV) model \citep{haider2006life}, motivate the use of short-term averages of income—typically during mid-life—by assuming a parametric link between observed annual income and unobserved permanent income. In contrast, this definition underpins the nonparametric nature of the identification result, enabling us to recover the intergenerational elasticity without restrictive functional form assumptions.\par 
While the IGE in equation (\ref{eq:beta}) depends on unobserved permanent income for both generations, the workable definition allows us to reformulate the target parameter in terms of partially observed (log) annual incomes:
\begin{align}\label{eq:beta_new1}
 \beta_0&=\frac{\mathbb{E}\left[\left(Y_c^P-\mathbb{E}\left[Y_c^P\right]\right)\left(Y_f^P-\mathbb{E}\left[Y_f^P\right]\right)\right]}{\mathbb{E}\left[\left(Y_f^P-\mathbb{E}\left[Y_f^P\right]\right)^2\right]}= \frac{\sum_{t=1}^T \sum_{j=1}^T \mathbb{E}\left[\left(Y_{ct} - \mathbb{E}\left[Y_{c}^P\right]\right)\left(Y_{fj} - \mathbb{E}\left[Y_{f}^P\right]\right)\right]}{ \sum_{t=1}^T \sum_{j=1}^T \mathbb{E}\left[\left(Y_{ft} - \mathbb{E}\left[Y_{f}^P\right]\right)\left(Y_{fj} - \mathbb{E}\left[Y_{f}^P\right]\right)\right]},
\end{align}  where the scaling component $1/T$ cancels out.
Although this is a crucial step, it is not sufficient for identification. To bridge this gap, I exploit observable characteristics by decomposing (log) annual income as
\begin{align}\label{eq:cond_mean_main}
Y_{gt} &= \mathbb{E}\left[Y_{gt} \mid \bm{X}_{gt}\right] + \epsilon_{gt}, \quad \mathbb{E}\left[\epsilon_{gt} \mid \bm{X}_{gt}\right] = 0, \quad g\in \{c,f\}, \quad t=1,\ldots,T,
\end{align} where $\bm{X}_{gt}$ denotes the elements in the observed characteristics $\bm X$ relevant for predicting (log) annual income of generation $g$ at time $t$, and \( \epsilon_{ct} \) is the nonparametric prediction error. Although  \( \epsilon_{ct} \) can be interpreted as an age shock, it differs conceptually from the age shock \( v_{ct} \) in the GEIV model of Assumption \ref{as:1}. Specifically, \( \epsilon_{ct} \) captures the component of (log) annual income that is not explained by observed parental and own characteristics, that is, the residual from a predictive model based on observables. In contrast, \( v_{ct} \) reflects transitory deviations from an individual’s permanent income and arises within a latent factor structure that distinguishes between the permanent and transitory components of income.\par 
Identification is established in two steps. Substituting the income decomposition into equation ~(\ref{eq:beta_new1}) and imposing conditional mean independence and orthogonality assumptions involving observables and prediction errors, thereby eliminating dependence on unobserved components. Then, I impose standard missing-at-random assumptions to recover the necessary conditional moments from the available data. For the income profiles, the MAR assumption enables identification of the conditional expectation $\mathbb{E}[Y_{gt} \mid \bm{X}_{gt}] = \mathbb{E}[Y_{gt} \mid \bm{X}_{gt}, D_{gt} = 1]$,
where $D_{gt}$ indicates income observability at time $t$. In a similar way, we can identify the conditional second moments arising in the denominator of equation (\ref{eq:beta_new1}) 
$\mathbb{E}[Y_{ft} Y_{fj} \mid \bm{X}_{ftj}] = \mathbb{E}[Y_{ft} Y_{fj} \mid \bm{X}_{ftj}, D_{ft} = 1, D_{fj}=1],$
where $\bm{X}_{ftj}$ comprises the elements in the observed characteristics $\bm X$ relevant for predicting the covariance between parental incomes at ages $t$ and $j$, and $\bm{X}_{ftj}$ is defined such that $\bm{X}_{ft} \subset\bm{X}_{ftj}$ for $t,j=1,...,T.$ With these foundations in place, I now formally state the complete set of identifying assumptions.\par  
\begin{assumption1}{1}{NP}(Conditional Mean Independence and Orthogonality)\label{as:ortho_np}    
\begin{enumerate}[i.]   \item The observable characteristics satisfy:  \begin{enumerate}[1.]      \item $    \mathbb{E}\left[Y_{ct} \mid \bm{X}_{ct}, \bm{X}_{cj},\bm{X}_{fj}\right] = \mathbb{E}\left[Y_{ct} \mid \bm{X}_{ct}\right] \quad \text{ for } t,j=1,...T,$    \item $    \mathbb{E}\left[Y_{ft} \mid \bm{X}_{ft}, \bm{X}_{ftj},  \bm{X}_{cj}\right] = \mathbb{E}\left[Y_{ft} \mid \bm{X}_{ft}\right] \quad \text{ for } \bm{X}_{fj} \subset\bm{X}_{ftj},\quad  t,j=1,...T.$ \end{enumerate}
  \item The average covariance between children's prediction errors and parental permanent income across all observed years is zero   \begin{align*}        \frac{1}{T}\sum_{t=1}^T\mathbb{E}\left[\epsilon_{ct}Y_{f}^P\right]&=0,\quad        \epsilon_{ct}\coloneqq Y_{ct}-\mathbb{E}\left[Y_{ct}\mid \bm X_{ct}\right],  \quad t=1,\ldots,T.   \end{align*} \item The average covariance of parental income prediction errors for  $|t-j|>h$ is zero
\begin{align*}  \frac{1}{T^2}\sum_{(t,j)\in \mathcal{H}}\mathbb{E}\left[ \epsilon_{ft} \epsilon_{fj}\right]=0,\quad \text{where } \mathcal{H}=\{(t,j): 1\leq t,j\leq T, |t-j|>h\}\end{align*} \end{enumerate} 
\end{assumption1} 
The first condition establishes that $\bm{X}_{gt}$ contains all relevant predictors for annual income for generation $g$ at time $t$, implying the remaining information in $\bm{X}$ provides no additional explanatory power.  In Section \ref{sec:app} I illustrate that the specification of the characteristics predictive of income profiles and parental income covariance, namely, \(\bm{X}_{ct}\), \(\bm{X}_{ft}\), and \(\bm{X}_{ftj}\), can be designed to satisfy Assumption \ref{as:ortho_np}.$i$ by construction.\par 
Children’s age shocks being correlated with parental permanent income constitutes a source of bias of the MI estimator (component (c) in equation (\ref{eq:mi_inconsistent})). One of the empirical patterns driving this dependence stems from children from affluent families exhibiting faster income growth, even after controlling for observables \citep{mello2022lifecycle}. The life-cycle estimator addresses this by projecting children's annual income into the space of observables. In particular, by including in $X_{ct}$ the interaction between average parental (log) annual income observations around mid-life $\left(\tilde{Y}_f^P=\frac{1}{T_f}\sum_{j\in \mathcal{M}_f}Y_{fj}D_{fj}\right)$ and children's age at time $t$, the prediction errors of children's income $\left(\epsilon_{gt}=Y_{gt}- \mathbb{E}\left[Y_{gt} \mid \bm{X}_{gt}\right]\right)$ become uncorrelated with parental permanent income $Y_f^P$. Accordingly, Assumption \ref{as:ortho_np}.$ii$ imposes that thee average covariance between children’s prediction errors and parental permanent income across all observed years is zero, once the relevant family characteristics are controlled for. In Section \ref{sec:test}, a test for Assumption \ref{as:ortho_np}$ii$ is proposed, and in the U.S. application, the test does not reject the validity of this assumption. \par 
The requirement of Assumption \ref{as:ortho_np}.$iii$ arises from the fundamental mismatch between the complete income profiles required by equation (\ref{eq:beta_new1}) and the income snapshots typically available in practice. Specifically, joint observation of parental incomes \( (Y_{ft}, Y_{fj}) \) (i.e., \( D_{ft}=1, D_{fj}=1 \)) occurs only for relatively close time periods, such as incomes observed between ages 25 and 35 for a given individual. Consequently, income pairs for distant periods (\( |t-j| > h \)) are systematically absent in available data. Assumption \ref{as:ortho_np}.$iii$ addresses this empirical constraint by imposing that conditional on family characteristics 
$\bm{X}_{ftj}$, parental income shocks (prediction errors $\epsilon_{ft}$ and $\epsilon_{fj}$) are uncorrelated for periods separated by more than $h$ years. The availability of rich family characteristics $\bm{X}$ makes this assumption empirically plausible, as it allows us to account for the persistent components of intertemporal dependence.\par 
Income autocorrelation captures two distinct sources: a permanent component driven by family characteristics (e.g., wealth, neighborhood quality, and race), and a transitory component, driven by short-term shocks (e.g., unemployment spells, economic crises, or health events). Crucially, while the influence of transitory shocks decays as the time gap $(t - j)$  widens, the effect of family background characteristics remains over time. Assumption \ref{as:ortho_np}.$iii$ states that parental annual income from periods more than \(h\) years in the past influences current income solely through observed characteristics. This specification serves dual purposes: it realistically captures the (conditional) short-memory of transitory shocks while accommodating the limitations inherent in available longitudinal datasets. In the application, the autocorrelation in parental income prediction errors becomes negligible beyond ten years in all 15 windows, supporting the plausibility of this assumption. \par 
The following assumption formalizes some necessary conditions for identifying the intergenerational elasticity using partial income data and family characteristics. First, it requires that income realizations, for both generations and across nearby ages for fathers, are independent of their observability conditional on family characteristics. This ensures that survey attrition or non-reporting is not systematically associated with unobserved income determinants, ruling out selection bias. Second, it imposes an overlap condition guaranteeing sufficient data coverage across individuals and age windows, preventing estimates from being driven by specific reporting patterns or missing subpopulations. Together, these conditions prevent two key threats to validity: estimates being distorted either by systematic missingness (e.g., concentrated among low-income families) or by over-reliance on narrow age clusters. When satisfied, they ensure that inference is driven by income dynamics rather than data availability.\par 
According to equation (\ref{eq:beta_new1}), the IGE depends on two distinct components: the covariance between parent and child income and the covariance within parental income, which implies that identification requirements differ across generations. For children, unconfoundedness needs only to hold for single income observations since the IGE exploits contemporaneous parent-child pairs, whereas for fathers, stronger conditions on income tuples are required to capture the temporal structure of their income process.\par 
\begin{assumption1}{2}{NP}(Missing At Random)\label{as:unc_np}     \begin{enumerate}[i.]    \item The missingness of children's annual income $Y_{ct}$ is  as good as random once we control for $\bm X_{ct}$\begin{align*}       Y_{ct}&\perp D_{ct}\mid \bm X_{ct}, \quad  t=1,..., T.    \end{align*} 
    \item  Given family characteristics, there is both missing and non-missing children incomes for every age   \begin{align*}     0<&p\left(D_{ct}=1\mid \bm X_{ct} \right)<1 \quad a.s, \quad t=1,..., T.    \end{align*}     \item The missingness of parental annual income  pairs $\left(Y_{ft}, Y_{fj}\right) $ is  as good as random once we control for $\bm X_{ftj}$\begin{align*}        \left(Y_{ft}, Y_{fj}\right) \perp \left(D_{ft}, D_{fj}\right) \mid \bm{X}_{ftj},  \quad  \text{ for all } t - j > h> 0,
    \end{align*}     where $\bm{X}_{ftj}$ are the family characteristics predictive of parental income covariance between years $t$ and $j$, and $\bm{X}_{ftj}\coloneqq \bm X_{ft}$ for $j=t$.   \item Given family characteristics, there is both missing and non-missing parental incomes for every age and its neighboring ages \begin{align*}       0 < p\left(D_{ft}=1, D_{fj}=1 \mid \bm{X}_{ftj}\right) < 1 \quad \text{a.s.}, \quad  \text{ for all } t - j > h> 0.    \end{align*}\end{enumerate}\end{assumption1}
Assumption \ref{as:unc_np} imposes a missing-at-random structure for child and parental incomes and a positivity condition for identification, similar  to the conditional independence assumptions in \citet{angrist1995identification}. The assumption that income missingness in the PSID is missing at random is supported by empirical evidence. \cite{fitzgerald1998analysis} finds that attrition in the PSID is selective, primarily affecting lower socioeconomic individuals and those with unstable earnings, marriage, and migration histories, but these factors explain little of the overall attrition, and regression-to-the-mean effects mitigate selection bias. This conclusion is reinforced by \cite{lillard1998panel}, who find that ignoring attrition induces only very mild biases in household income models.\par 
\cite{fitzgerald2011attrition} examines attrition in intergenerational models of health, education, and earnings, finding that sibling correlations in outcomes are marginally higher among individuals who remain in the panel longer, though the differences are not statistically significant. Models of intergenerational links with covariates show negligible attrition bias for females. In contrast, the evidence for males is mixed but generally weak, suggesting that conditioning on observables largely mitigates selective attrition. The study finds little evidence of attrition bias, though analyses of educational and earnings outcomes for men appear to benefit from conditioning on observables.\par
\cite{schoeni2015implications} show that applying sample weights reduces differences in intergenerational income elasticity estimates between the full sample, the attriting sample, and the non-attriting sample, rendering these differences statistically insignificant. Their findings highlight that attrition, particularly higher among lower-income individuals, is influenced by the correlation between child and parental income outcomes, emphasizing the importance of incorporating both parental and child characteristics in analyses of intergenerational mobility.\par 
Taken together, the literature suggests that the MAR assumption for income missingness in the PSID is empirically plausible, provided analyses carefully account for relevant observables. To address the concerns raised by \cite{schoeni2015implications}, particularly the influence of the correlation between parental and child income outcomes on observability, the analysis incorporates both parental and child characteristics in the conditioning set, thereby strengthening the plausibility of the MAR assumption.\par 
The following Theorem establishes the nonparametric identification of the intergenerational elasticity in the presence of incomplete income data and family characteristics. This fundamental result ensures that estimates derived from the identification result are comparable across studies, providing a building block to analyze intergenerational mobility under valid inference. 
\begin{theorem}\label{thm:2}     Under assumptions  \ref{as:ortho_np} and \ref{as:unc_np} and the definition of permanent income in equation (\ref{eq:def_perm_inc}), the IGE is nonparametrically identified as
\begin{gather}\label{eq:identification}        
\beta_0=\frac{\mathbb{E}\left[\sum_{t=1}^T\left(\mu_{ct}\left(\bm{X}_{ct},1\right)-\mu_c^P\right)\sum_{j=1}^T\left(\mu_{fj}\left(\bm{X}_{fj},1\right)-\mu_f^P\right)\right]}{ \mathbb{E}\left[\sum_{|t-j| \leq h}\sigma_{tj}\left(\bm{X}_{ftj},1,1\right)+ \sum_{|t-j| > h}\left(\mu_{ft}\left(\bm{X}_{ft},1\right)-\mu_f^P\right)\left(\mu_{fj}\left(\bm{X}_{fj},1\right)-\mu_f^P\right)\right]},
\end{gather}     where  $\mu_{gt}(\bm{X}_{gt},1) \coloneqq \mathbb{E}\left[ Y_{gt} \mid \bm{X}_{gt}, D_{gt} = 1 \right]$, $\mu_g^P \coloneqq \mathbb{E}\left[  \sum_{t=1}^T \mu_{gt}(\bm{X}_{gt},1) \right]$, and  $\sigma_{tj}\left(\bm{X}_{ftj},1,1\right) \coloneqq \mathbb{E}\left[\left(Y_{ft}-\mu_f^P\right)\left(Y_{fj}-\mu_f^P\right)\mid \bm X_{ftj}, D_{ft}=1, D_{fj}=1\right]$.
\end{theorem} 
Theorem \ref{thm:2} establishes the identification of the intergenerational elasticity in the presence of incomplete income data. Specifically, it shows that, under the conditional mean independence and orthogonality conditions in Assumption \ref{as:ortho_np} and the standard missing-at-random assumptions in \ref{as:unc_np}, the IGE can be recovered from conditional expectations, including the conditional income profiles of parents  and children, and the conditional covariance matrix of parental income.\par 
To the best of my knowledge, the only existing identification result in this framework is that of \cite{an2022nonparametric}, who nonparametrically identify the mobility function relating children's to parents permanent income. While their more general framework nests the linear IGE as a special case, since they leave the relationship of parental and child incomes unspecified, my approach offers three important advantages. First, I relax their classical errors-in-variables model (Assumption \ref{as:1} with $\lambda_t=1$) for two measurement periods, by exploiting the definition of permanent income. Second, I relax the assumption that transitory shocks to children's income are uncorrelated with parental permanent income and parental transitory shocks. In contrast, I assume that the prediction error of the children's annual income is uncorrelated to parental permanent income conditional on family characteristics (Assumption \ref{as:ortho_np}). Finally, the proposed framework explicitly addresses the missing data structure inherent in real-world income observations, while incorporating all available information on both income dynamics and family characteristics.\par 
Our identification result provides two valuable contributions to the study of intergenerational mobility.  First, it resolves persistent methodological challenges by establishing sufficient conditions for identifying the intergenerational elasticity from incomplete income observations and family characteristics. Second, it provides the theoretical foundation for constructing a consistent estimator, enabling researchers to obtain valid and comparable estimates of the intergenerational elasticity. 
\subsection{Locally Robust Estimation of the IGE}
I estimate the intergenerational elasticity $\beta_0$ using a Generalized Method of Moments (GMM) approach based on Theorem \ref{thm:2}. The analysis proceeds by rearranging equation (\ref{eq:identification}) to derive the moment condition used to identify $\beta_0$: 
\fontsize{10}{12}\selectfont
 \begin{align}\label{eq:ident}  \nonumber  \mathbb{E}\left[g_1\left(W,\gamma,\beta,\mu_c^P,\mu_f^P\right)\right]&=0, \\
\nonumber
g_1\left(W,\gamma,\beta,\mu_{c}^P,\mu_{f}^P\right)&=\beta\sum_{|t-j| \leq h} \sigma_{tj}\left(\bm X_{ftj}, 1, 1\right)+\beta\sum_{|t-j| > h}\left(\mu_{ft}\left(\bm X_{ft}, 1\right)-\mu_f^P\right)\left(\mu_{fj}\left(\bm X_{fj}, 1\right)-\mu_f^P\right)\\
&-\sum_{t=1}^T\left(\mu_{ct}\left(\bm X_{ct}, 1\right)-\mu_c^P\right)\sum_{j=1}^T\left(\mu_{fj}\left(\bm X_{fj}, 1\right)-\mu_f^P\right),
\end{align} 
\normalsize where $\gamma\coloneqq (\sigma_{tj}, \mu_{ft}, \mu_{fj})$. Thus, the moment identifying the IGE depends on the income profiles and parental income covariance structure, captured by the nuisance parameter $\gamma$, as well as the population mean permanent incomes $\left(\mu_c^P, \mu_f^P\right)$. These means are themselves identified by the moment conditions (see equation (\ref{eq:cond_means})):
\begin{align*}\nonumber
\mathbb{E}\left[g_2\left(W,\gamma,\mu_c^P\right)\right]&=0,\quad 
g_2\left(W,\gamma,\mu_c^P\right)=\sum_{t=1}^T\mu_{ct}\left(\bm X_{ct}, 1\right)-\mu_c^P,\\  
\mathbb{E}\left[g_3\left(W,\gamma,\mu_f^P\right)\right]&=0,\quad
g_3\left(W,\gamma,\mu_f^P \right)=\sum_{t=1}^T\mu_{ft}\left(\bm X_{ft}, 1\right)-\mu_f^P.  
\end{align*}
Finally, I define the augmented parameter $\theta\coloneqq \left(\beta,\mu_c^P,\mu_f^P\right)$, and combine the moment conditions into a single system for GMM estimation
\begin{align*}
g\left(W,\gamma,\theta\right)=\left(
g_1\left(W,\gamma,\theta\right) \
g_2\left(W,\gamma,\theta\right) \
g_3\left(W,\gamma,\theta\right)\right)'.
\end{align*} \par 
Because the nuisance parameter $\gamma$ is unknown, a natural two-step estimation procedure is to first estimate the conditional expectations in $\gamma$, and then perform GMM estimation based on $g\left(W,\hat{\gamma},\theta\right)$. As in any two-step procedure, errors in the first step affect inference in the second. This issue is especially pronounced when machine learning (ML) is used, because regularization and model selection allow for bias to attain smaller variance. As a result, bias from the first step propagates to the second. This is formally captured by the sensitivity of the moment condition to small changes in the nuisance parameter:
\[
\frac{d}{d\tau} \mathbb{E}[g(W, \gamma_\tau, \theta)]\big |_{\tau=0} \neq 0,
\] indicating that the moment identifying $\theta$ is not locally robust to estimation error in $\gamma_0$. \par 
\cite{chernozhukov2022locally} provide a general procedure for constructing orthogonal moment functions for GMM, where the moment conditions are locally insensitive to first-step estimation (see Appendix \ref{sec:illust} for an illustration). This property ensures that the resulting estimator is locally robust,  meaning that estimation errors in the first step have no effect, locally, on the estimation of the parameter of interest. The authors show that  an orthogonal (locally robust) moment function $\psi$ can be constructed by augmenting the identifying moment function $g$ with a correction term $\phi$: 
 \begin{align*}
\psi\left(W,\gamma,\alpha,\theta\right)=g\left(W,\gamma,\theta\right)+\phi\left(W,\gamma,\alpha,\theta\right),  
 \end{align*}  where $\alpha$ encompasses additional nuisance parameters introduced by $\phi$.\par 
 \begin{proposition}\label{prop:1}    
There exists a function $\phi$ such that the augmented moment condition
\begin{align*}
\psi\left(W,\gamma,\alpha,\theta\right)=g\left(W,\gamma,\theta\right)+\phi\left(W,\gamma,\alpha,\theta\right)
\end{align*} identifies the intergenerational elasticity, as well as the population mean of permanent incomes, and satisfies local robustness:
\[
\frac{d}{d\tau} \mathbb{E}[\psi(W, \gamma_\tau,\alpha, \theta)]\big |_{\tau=0} = 0,
\] where $\psi$ is given by equation (\ref{eq:phi_ref}).  This latter property ensures that the estimator of the intergenerational elasticity is first-order insensitive to estimation error in the nuisance parameter $\gamma$.
\end{proposition} 
The locally robust closed-form solution for the IGE follows directly from Proposition \ref{prop:1}. In particular, solving for $\beta$ in the orthogonal moment condition  yields:
 \begin{gather}\label{eq:lr_cf}   \beta=\frac{\mathbb{E}\left[\sum_{t=1}^T\left(\mu_{ct}\left(\bm X_{ct}, 1\right)-\mu_c^P\right)\sum_{j=1}^T\left(\mu_{fj}\left(\bm X_{fj}, 1\right)-\mu_f^P\right)\right]+\mathbb{E}\left[\phi_4+\phi_5\right]}{\mathbb{E}\left[\sum_{|t-j| \leq h} \sigma_{tj}\left(\bm X_{ftj}, 1, 1\right)+\sum_{|t-j| > h}\left(\mu_{ft}\left(\bm X_{ft}, 1\right)-\mu_f^P\right)\left(\mu_{fj}\left(\bm X_{fj}, 1\right)-\mu_f^P\right)\right]+\mathbb{E}\left[\phi_1+\phi_2+\phi_3\right]},
  \end{gather} where 

   \begin{align*}\nonumber    \phi_1&=\beta \sum_{|t-j| \leq h}\frac{D_{ft}D_{fj}}{p\left(D_{ft}=1, D_{fj}=1 |\bm X_{ftj}\right)}\left(\left(Y_{ft}-\mu_f^P\right)\left(Y_{fj}-\mu_f^P\right) -\sigma_{tj}\left(\bm X_{ftj}, 1, 1\right)\right),\\ 
 \phi_2&=\sum_{|t-j| > h}\left(\mu_{fj}\left(\bm X_{fj}, 1\right)-\mu_f^P\right)\frac{D_{ft}}{p\left(D_{ft}=1|\bm X_{ft}\right)}\left(Y_{ft}-\mu_{ft}\left(\bm X_{ft}, 1\right)\right),\\
  \phi_3&=\sum_{|t-j| > h}\left(\mu_{ft}\left(\bm X_{ft}, 1\right)-\mu_f^P\right)\frac{D_{fj}}{p\left(D_{fj}=1|\bm X_{fj}\right)}\left(Y_{fj}-\mu_{fj}\left(\bm X_{fj}, 1\right)\right),\\
 \phi_4&=\sum_{t=1}^T\sum_{t=j}^T\left(\mu_{ct}\left(\bm X_{ct}, 1\right)-\mu_c^P\right)\frac{D_{fj}}{p\left(D_{fj}=1|\bm X_{fj}\right)}\left(Y_{fj}-\mu_{fj}\left(\bm X_{fj}, 1\right)\right),\\
 \phi_5&=\sum_{t=1}^T\sum_{t=j}^T\left(\mu_{fj}\left(\bm X_{fj}, 1\right)-\mu_f^P\right)\frac{D_{ct}}{p\left(D_{ct}=1|\bm X_{ct}\right)}\left(Y_{ct}-\mu_{ct}\left(\bm X_{ct}, 1\right)\right), \\\mu_c^P&=\mathbb{E}\left[\sum_{t=1}^T\mu_{ct}\left(\bm X_{ct}, 1\right)\right]+\mathbb{E}\left[\sum_{t=1}^T\frac{D_{ct}}{p\left(D_{ct}=1|\bm X_{ct}\right)}\left(Y_{ct}-\mu_{ct}\left(\bm X_{ct}, 1\right)\right)\right], \\
   \mu_f^P&=\mathbb{E}\left[\sum_{t=1}^T\mu_{ft}\left(\bm X_{ft}, 1\right)\right]+\mathbb{E}\left[\sum_{t=1}^T\frac{D_{ft}}{p\left(D_{ft}=1|\bm X_{ft}\right)}\left(Y_{ft}-\mu_{ft}\left(\bm X_{ft}, 1\right)\right)\right].\\
  \end{align*}
\par 
The locally robust closed-form solution for $\beta$ in equation (\ref{eq:lr_cf}) corresponds to the expression given in Theorem \ref{thm:2}, augmented with correction terms that make it first-order insensitive to estimation errors in the nuisance parameters. The term $\phi_1$ corrects for errors in estimating the conditional covariance of parental income for closely spaced periods ($|t-j|\leq h$), while $\phi_2$ and $\phi_3$ address errors in estimating parental income profiles for more distant periods ($|t-j|>h$). Similarly, $\phi_4$ and $\phi_5$ correct for errors in estimating the conditional income profiles of children and parents, respectively.  A critical feature of all correction terms ($\phi_1$ to $\phi_5$) is their inherent adjustment for non-random missingness by weighting prediction errors by the inverse propensity score. Finally,  the closed-form expressions for the population means of permanent incomes ($\mu_c^P$ and $\mu_f^P$) also incorporate the corresponding prediction errors, ensuring that the estimator remains locally robust to first-step estimation mistakes.\par
Equation (\ref{eq:lr_cf}) motivates the definition of the augmented parameter 
\(\theta \coloneqq (\beta, \mu_c^P, \mu_f^P)\) rather than including \(\mu_c^P\) and \(\mu_f^P\) in the nuisance parameter \(\gamma\). Each population mean \(\mu_g^P\) depends not only on the conditional income profiles \(\mu_{g,t}\) but also on the underlying population distribution. Consequently, small changes in the population distribution affect \(\mu_g^P\) both through the conditional profiles and through the expectation itself. Thus, including \(\mu_g^P\) in \(\gamma\) would therefore make the closed-form solution for \(\beta\) considerably more complex. By keeping \(\mu_c^P\) and \(\mu_f^P\) in \(\theta\), we separate the estimation of population permanent means from the first-step nuisance functions, which makes the locally robust solution more tractable.\par
To construct a debiased machine learning estimator for the IGE, I use the orthogonal moment condition in Proposition \ref{prop:1} (see equation (\ref{eq:phi_ref})) combined with cross-fitting to ensure robustness and mitigate overfitting. Following \citet{semenova2023inference}, cross-fitting in settings with dependence should be performed at the level of independent sampling units, in this case, families, rather than individual child--father pairs. Accordingly, let $f \in \{1, \dots, n_f\}$ index families, with $\mathcal{P}_f$ denoting the set of all child--father pairs in family $f$. The set of family indices is partitioned into $L$  mutually exclusive and exhaustive folds $\{\mathcal{F}_\ell\}_{\ell=1}^L$. For each fold $\ell = 1, \dots, L$, the nuisance parameters $\hat{\gamma}^{(\ell)}$ and $\hat{\alpha}^{(\ell)}$ are estimated using only data from families not in $\mathcal{F}_\ell$, thereby preserving independence between the samples used for first-stage estimation and those used for evaluation.

The debiased moment function is then computed as
\[
\hat{\psi}(\theta) = \frac{1}{n} \sum_{\ell=1}^L \sum_{f \in \mathcal{F}_\ell} \sum_{i \in \mathcal{P}_f} \sum_{(t,j) \in \mathcal{J}_i} \hat{\psi}_{i,tj}^{(\ell)},\quad \hat{\psi}_{i,tj}^{(\ell)} \coloneqq g\big(W_{i,tj}, \hat{\gamma}^{(\ell)}, \theta\big) + \phi\big(W_{i,tj}, \hat{\gamma}^{(\ell)}, \hat{\alpha}^{(\ell)}, \theta\big),
\]
where $\mathcal{J}_i$ denotes the set of all tuples $(t, j)$ observed for child--father pairs $i$, noting that a family may contribute multiple such pairs. Since the system is exactly identified, there is no need to compute fold-specific $\hat{\theta}^{(\ell)}$. The locally robust estimator of the IGE is thus obtained by solving
\[
\hat{\theta}^{LR}_n = \arg\min_{\theta \in \Theta \subset \mathbb{R}^3} \hat{\psi}(\theta)'\hat{\Upsilon}\hat{\psi}(\theta).
\]where $\hat{\Upsilon}$ is a positive semi-definite weighting matrix, and $\Theta$ denotes the set of parameter values. This objective function incorporates orthogonal moments and cross-fitting. While the influence function corrects for prediction errors in estimating the conditional expectations, cross-fitting eliminates overfitting in nuisance parameter estimation. Furthermore, by grouping folds at the family level, this approach aligns with the principle of leaving out dependent \say{neighbor} units in panel settings \citep{semenova2023inference}, ensuring that dependence within families does not bias the orthogonalization step.
\subsection{Asymptotic Theory and Inference}\subsubsection{Asymptotic Properties of the Locally Robust Estimator} 
To provide rigorous justification for the empirical implementation of the proposed estimator, its large-sample behavior is examined. I begin by establishing consistency, which follows from standard M-estimation theory, adapted to the locally robust framework of \citet{chernozhukov2022locally}. While their main asymptotic results assume consistency, Theorem A3 provides primitive conditions under which it holds. The following Lemma adapts these conditions to the proposed setting.
\begin{lemma}[Consistency of the Locally Robust Estimator]\label{lem:consistency}
Let \( \hat{\theta}^{LR}_n \) be the solution to the cross-fitted orthogonal moment condition:
\begin{align*}
\hat{\theta}^{LR}_n &= \arg\min_{\theta \in \Theta\subset \mathbb{R}^3} \hat{\psi}(\theta)'\hat{\Upsilon}\hat{\psi}(\theta),\quad
\hat{\psi}(\theta) = \frac{1}{n} \sum_{\ell=1}^L \sum_{f \in \mathcal{F}_\ell} \sum_{i \in \mathcal{P}_f} \sum_{(t,j) \in \mathcal{J}_i} \hat{\psi}_{i,tj}^{(\ell)},\\ \hat{\psi}_{i,tj}^{(\ell)} &\coloneqq g\big(W_{i,tj}, \hat{\gamma}^{(\ell)}, \theta\big) + \phi\big(W_{i,tj}, \hat{\gamma}^{(\ell)}, \hat{\alpha}^{(\ell)}, \theta\big), 
\end{align*}  where $\hat{\Upsilon}$ is a positive semi-definite weighting matrix.
Then \( \hat{\theta}^{LR}_n \overset{p}{\to} \theta_0 \), by Theorem A3 in \citet{chernozhukov2022locally}, provided Assumptions \ref{as:ortho_np}, \ref{as:unc_np} and \ref{ass:clr} hold.
\end{lemma}
Lemma \ref{lem:consistency} shows that under mild regularity conditions \( \hat{\theta}^{LR}_n=\left(\hat{\beta}^{LR}_n, \hat{\mu}_{c,n}^P, \hat{\mu}_{F,n}^P\right) \) converges in probability to the true parameter \( \theta_0 \). The consistency of the locally robust estimator guarantees that, under the specified conditions, the estimated intergenerational elasticity \( \hat{\beta}^{LR}_n \) converges to the true value \( \beta_0 \) as the sample size increases. This ensures that the estimator remains stable even when machine learning methods are used to estimate nuisance parameters. As a result, the estimates of the intergenerational elasticity are both reliable and comparable across different studies.\par 
Under the regularity conditions described in Appendix \ref{sec:as_p}, I establish the asymptotic normality of the proposed estimator, which explicitly accounts for uncertainty from the first-stage estimation of the nuisance parameters. This yields confidence intervals with valid coverage, a crucial requirement for drawing meaningful conclusions about intergenerational mobility patterns.\par 
The following Lemma formalizes the validity of inference for the estimator \( \hat{\theta}^{LR}_n \), even when nuisance components are estimated using high-dimensional or nonparametric methods. This robustness is achieved through the use of orthogonal moment conditions, which ensure that estimation errors in the first stage enter the moment function only at second order. As a result, standard $\sqrt{n}$ asymptotic normality can be established under relatively weak conditions. Crucially, cross-fitting plays a central role in mitigating own-observation bias and avoids the need for stringent entropy or Donsker-type conditions, which are not known to hold for many machine learning first steps. Together, these features allow us to utilize flexible first-stage methods while maintaining valid inference.
\begin{lemma}[Asymptotic Normality of the Locally Robust Estimator]\label{lemma:asymptotic_normality}
Under Assumptions \ref{as:ortho_np}-\ref{ass:lr_new} and \ref{ass:lr8}, $\hat{\theta}^{LR}_n\xrightarrow{p}\theta_0$, and non-singularity of $G' \Upsilon G$, the asymptotic normality of the estimator \( \hat{\theta}^{LR}_n \) directly follows from Theorem 9 of \cite{chernozhukov2022locally}. Specifically, we have:
\[
\sqrt{n}(\hat{\theta}^{LR}_n - \theta_0) \xrightarrow{d} \mathcal{N}(0, V),
\]
where\( V = \left(G' \Upsilon G\right)^{-1} \), \( G = \mathbb{E}[\partial_\theta g(W, \gamma, \alpha, \theta)] \), and $\hat{\Upsilon}$ is the estimated efficient weighting matrix defined as $\hat{\Upsilon} = \hat{\Psi}^{-1}$ for $\hat{\Psi} = \frac{1}{n} \sum_{\ell=1}^L \sum_{i \in \mathcal{I}_\ell} \sum_{(tj) \in \mathcal{J}_i} \hat{\psi}_{i,tj}^{(\ell)}\hat{\psi}_{i,tj}^{(\ell)'}$.
In addition, if Assumption \ref{ass:lr_other} holds, then  \( \hat{V} \xrightarrow{p} V \).
\end{lemma}
\par 
Lemma \ref{lemma:asymptotic_normality} completes the theoretical framework by integrating the three contributions: (i) the nonparametric identification of the intergenerational elasticity in the presence of incomplete income data; (ii) a consistent, locally robust estimator that corrects for first-step prediction errors; and (iii) valid inference that accounts for uncertainty from the first-stage estimation of nuisance parameters. Appendix \ref{sec:as_p} characterizes the asymptotic variance 
$V$ associated with this result. Together, these advances provide a theoretically grounded toolkit for studying income persistence through the lens of the intergenerational elasticity. \par
The asymptotic normality result in Lemma \ref{lemma:asymptotic_normality} is derived under the assumption of independently and i.i.d observations. In practice, however, datasets commonly include multiple children from the same family, introducing correlation within families. Accordingly, the asymptotic variance $V$ should  be estimated using a cluster-robust approach that accounts for this dependence structure. While the i.i.d. assumption is adopted here for ease of exposition and to align with the general theoretical framework of \cite{chernozhukov2022locally}, the core identification and estimation strategy remains sound. The cluster-robust extension is a straightforward implementation detail for the variance estimation, where the moment functions $\hat{\psi}_{i,tj}$ is aggregated at the family level before constructing the variance-covariance matrix $\hat{\Psi}$, accounting for correlation within families in the standard errors.\par 
\par 
\subsection{Testing the Identification Assumptions}\label{sec:test} This section develops formal hypothesis tests for Assumption \ref{as:ortho_np}.$ii$ and discusses how to assess in practice \ref{as:ortho_np}.$iii.$. As illustrated In Section \ref{sec:app} the specification of the characteristics predictive of income profiles and parental income covariance, namely, \(\bm{X}_{ct}\), \(\bm{X}_{ft}\), and \(\bm{X}_{ftj}\), can be designed to satisfy Assumption \ref{as:ortho_np}.$i$ by construction. The MAR conditions in Assumptions \ref{as:unc_np}.$i$ and \ref{as:unc_np}.$iii$ are not directly testable from the observed data, as they involve unobserved missingness mechanisms. Nevertheless, as discussed above, the literature suggests that the MAR assumption for income missingness in the PSID is empirically plausible, provided that analyses carefully account for relevant observables. Consistent with the findings in  \citet{schoeni2015implications}, I include both child and father characteristics in the conditioning set, thereby strengthening the plausibility of the MAR assumption in the analysis.  Finally, the boundedness condition on the propensity score in Assumptions \ref{as:unc_np}.$ii$ and \ref{as:unc_np}.$iv$ can be assessed informally through visual inspection.\par 
I start by considering a test for the orthogonality between children's prediction errors and parental permanent income   \begin{align}\label{eq:test1} H_0: \frac{1}{T}\sum_{t=1}^T\mathbb{E}\left[\epsilon_{ct}Y_{f}^P\right]=0, \quad  \quad \text{vs}\quad  H_1: \frac{1}{T}\sum_{t=1}^T\mathbb{E}\left[\epsilon_{ct}Y_{f}^P\right] \neq 0,   
\end{align}
where $\epsilon_{ct}\coloneqq Y_{ct}-\mathbb{E}\left[Y_{ct}\mid \bm X_{ct}\right]$ denotes the children's income prediction errors at time $t$ and $Y_f^P$ represents parental permanent income. The main challenge in testing this hypothesis is that both random variables are unobserved, and their machine learning estimation introduces regularization and model selection bias when testing $H_0$. To address these issues, a three-stage procedure is proposed. Establishing identification of the object of interest \(\theta_{cf} \coloneqq \frac{1}{T}\sum_{t=1}^T\mathbb{E}\left[\epsilon_{ct} Y_f^P\right]\). Next, constructing a locally robust estimator, and finally providing a $t-$test based on $\hat{\theta}_{cf}$.\par
A locally robust $t-$test for $H_0$ in (\ref{eq:test1}) is given by 
\begin{align*} t_{cf,n} &= \frac{\hat{\theta}_{cf,n}}{\sqrt{\hat{V}_{cf,n}/n}}, \end{align*} where $\hat{\theta}_{cf,n}$ is the argument solving the cross-fitted locally robust moment in equation (\ref{eq:cf}), and $\hat{V}_{cf,n}$ is a consistent estimator of the asymptotic variance of $\hat{\theta}_{cf,n}$, that accounts for dependence within families (see Appendix \ref{sec:as_test} for the step-by-step derivation). Similar to the locally robust estimator for the IGE, this cluster-robust variance estimator is constructed by aggregating moment functions at the family level to allow for arbitrary correlation between observations from the same family, while maintaining independence across different families.\par 
The following Theorem establishes the asymptotic properties of this locally robust $t-$test.
\begin{theorem}\label{th:3} (Size, Consistency, and Local Power of the Locally Robust $t-$Test I)\label{cor:wald_consistency_local} Under Assumptions \ref{as:ortho_np_n}, \ref{as:unc_np_new}, \ref{ass:4lr}-\ref{ass:lr8} and \ref{ass:clr}, the asymptotic properties of the locally robust $t$ statistic \begin{align*}
    t_{cf,n} &= \frac{\hat{\theta}_{cf,n}}{\sqrt{\hat{V}_{cf,n}/n}}
\end{align*} are given by the following statements:
\begin{enumerate}  \item (\emph{Asymptotic size}) Under $H_0:\theta_{cf0}=0$,  \[
t_{cf,n} \xrightarrow{d} \mathcal{N}(0,1) \quad\text{and}\quad \lim_{n\to\infty}\Pr\left(|t_{cf,n}| > z_{1-\alpha/2}\right) = \alpha,
\]
where $z_{1-\alpha/2}$ is the $(1-\alpha/2)$-quantile of the standard normal distribution.   \item (\emph{Consistency under fixed alternatives})  For any fixed alternative with $\theta_{cf0} \neq 0$,
\[
\lim_{n\to\infty}\Pr\!\left(|t_{cf,n}| > z_{1-\alpha/2}\right) = 1.
\]  \item (\emph{Local alternatives}) Under $H_{1n}:\theta_{cf0} = \delta/\sqrt{n}$ with fixed $\delta \in \mathbb{R}$,
\[
t_{cf,n} \xrightarrow{d} \mathcal{N}\left(\frac{\delta}{\sqrt{V_{cf}}}, 1\right),
\]
so the limiting power is
\[
\lim_{n\to\infty}\Pr\!\left(|t_{cf,n}| > z_{1-\alpha/2}\right)
= 2\left[1 - \Phi\left(z_{1-\alpha/2} - \frac{|\delta|}{\sqrt{V_{cf}}}\right)\right]
> \alpha \quad \text{whenever } \delta \neq 0,
\]
where $\Phi(\cdot)$ denotes the standard normal cumulative distribution function.\end{enumerate}
\end{theorem}\par 
While a similar locally robust test could be constructed for Assumption~\ref{as:ortho_np}.$iii$, implementing such a test faces a fundamental empirical constraint: jointly observed income pairs $(Y_{ft}, Y_{fj})$ for distant periods ($|t-j|>h$) are systematically sparse in longitudinal data, undermining the reliability and power of formal hypothesis testing. Instead, this assumption can be assessed in practice by analyzing the autocorrelation structure and variance decomposition of parental income prediction errors.
Since the variance of permanent income decomposes as $
\text{Var}(Y_f^P) = \text{Var}\big(\mathbb{E}[Y_{ft} \mid \mathbf{X}_{ft}]\big) + \text{Var}(\epsilon_{ft}),$ the contribution of distant-lag residual products $\sum_{|t-j|>h} \mathbb{E}[\epsilon_{ft}\epsilon_{fj}]$ to the IGE denominator depends on both their contribution to residual variance and the relative importance of residual variance itself. \par 
I suggest an empirical assessment in two steps. first, examining whether the autocorrelation function $\text{Corr}(\epsilon_{ft},\epsilon_{fj})$ exhibits rapid decay. Second, quantifying the weighted contribution of distant lags to the residual variance structure. Rapid decay combined with small variance contribution provides transparent evidence that the orthogonality assumption is empirically plausible and has a negligible impact on IGE estimates.
Empirically, one should examine whether the autocorrelation function $\text{Corr}(\epsilon_{ft},\epsilon_{fj})$ decays to low levels beyond lag $h$, and what fraction of the residual variance structure is attributable to distant lags, computed as the weighted contribution $\sum_{k>h} (T-k) \mathbb{E}[\epsilon_{ft}\epsilon_{fj}]$ relative to the total. If both the autocorrelation decay is rapid and the variance contribution from distant lags is negligible, this provides evidence that imposing the orthogonality assumption has negligible impact on IGE estimates. \par While the proposal offers a straightforward approach to selecting $h$ based on observed autocorrelation patterns and variance contributions, a more data-driven procedure could be developed by adapting bandwidth selection methods from the time series literature to this setting. I leave such extensions for future research and implement the empirical assessment procedure in the application.
\section{Simulations}\label{sec:sims} 
I consider the following data generating process (DGP) for generation $g$ at age $t$: \fontsize{10.5}{12.5}
\begin{align*}
Y_{gt}&=\gamma_{0,g}+\gamma_{1,g}X_{1,g}+\gamma_{2,g}X_{2,g}+\gamma_{3,g}t+\gamma_{4,g}t^2+\gamma_{5,g}X_{1,g}t+\epsilon_{gt},\quad  t=1,...,T,  \quad g\in\{c,f\},\\
\epsilon_{gt}&\sim \mathcal{N}\left(0,\sigma^2_\epsilon\right)\\
\begin{pmatrix}X_{j,c} \\ X_{j,f}
\end{pmatrix} &\sim \mathcal{N}\left(\begin{pmatrix}0  \\0
\end{pmatrix}, \begin{pmatrix}1 & \sigma_j \\\sigma_j & 1\end{pmatrix}\right), \quad j=1,2.
\end{align*}\normalsize
Using the definition of permanent income:
\begin{align}\label{eq:perm_sim}
Y_g^P&= \gamma_{0,g} + \gamma_{1,g} X_{1,g} + \gamma_{2,g} X_{2,g} + \gamma_{3,g} \bar{t} + \gamma_{4,g}\bar{t^2} + \gamma_{5,g} X_{1,g}\bar{t} + \bar{\epsilon}_{g}, \quad \quad g\in\{c,f\}, \\\nonumber
\bar{t}&=\frac{1}{T}
\sum_{t=1}^T t, \quad \bar{t^2}=\frac{1}{T}\sum_{t=1}^T t^2, \quad \bar{\epsilon}_{g}=\frac{1}{T}\sum_{t=1}^T\epsilon_{gt},
\end{align}
so the covariance between permanent incomes is given by \fontsize{10.5}{12.5}
\begin{align}\label{eq:covsim}
\text{Cov}\left(Y_c^P, Y_f^P\right) &=\text{Cov}\left(\gamma_{1,c} X_{1,c}, \gamma_{1,f} X_{1,f}\right)  + \text{Cov}\left(\gamma_{2,c} X_{2,c}, \gamma_{2,f} X_{2,f}\right)  + \text{Cov}\left(\gamma_{5,c} X_{1,c} \bar{t}, \gamma_{5,f} X_{1,f} \bar{t}\right)\\\nonumber&+ \text{Cov}\left(\gamma_{1,c} X_{1,c}, \gamma_{5,f} X_{1,f} \bar{t}\right) + \text{Cov}\left(\gamma_{5,c} X_{1,c} \bar{t}, \gamma_{1,f} X_{1,f}\right)\\\nonumber&= \gamma_{1,c} \gamma_{1,f} \sigma_1 + \gamma_{2,c} \gamma_{2,f} \sigma_2 + \gamma_{5,c} \gamma_{5,f} \bar{t^2} \sigma_1 + \gamma_{1,c} \gamma_{5,f} \bar{t} \sigma_1 + \gamma_{5,c} \gamma_{1,f} \bar{t} \sigma_1,\end{align}\normalsize where we have used that the covariates come from a bivariate normal distribution with zero mean and correlation $\sigma_j$ among generations.\par 
According to equation (\ref{eq:perm_sim}), the variance of parental income corresponds to
\begin{align}\label{eq:varsim}
\text{Var}\left(Y_f^P\right) = \gamma_{1,f}^2 + \gamma_{2,f}^2 + \gamma_{5,f}^2 \bar{t^2} + 2 \gamma_{1,f} \gamma_{5,f} \bar{t} + \sigma_\epsilon^2/T.
\end{align}
Finally, by plugging equations (\ref{eq:covsim}) and (\ref{eq:varsim}) into (\ref{eq:beta}) yields
\begin{align}\label{eq:beta_sim}
\beta_0&=\frac{\mathbb{E}\left[\left(Y_c^P-\mathbb{E}\left[Y_c^P\right]\right)\left(Y_f^P-\mathbb{E}\left[Y_f^P\right]\right)\right]}{\mathbb{E}\left[\left(Y_f^P-\mathbb{E}\left[Y_f^P\right]\right)^2\right]}\nonumber\\
&=\frac{ \sigma_1 \left(\gamma_{1,c} \gamma_{1,f}+\gamma_{5,c} \gamma_{5,f} \bar{t^2}+\left(\gamma_{1,c} \gamma_{5,f}+\gamma_{5,c} \gamma_{1,f} \right)\bar{t}\right)+ \gamma_{2,c} \gamma_{2,f} \sigma_2}{\gamma_{1,f}^2 + \gamma_{2,f}^2 + \gamma_{5,f}^2 \bar{t^2} + 2 \gamma_{1,f} \gamma_{5,f} \bar{t} + \sigma_\epsilon^2/T}.
\end{align} Setting the parameter values to 
\begin{align*}
\gamma_{0,c} &= 8.5, & \gamma_{0,f} &= 5, & 
\gamma_{1,c} &= 0.275, & \gamma_{1,f} &= 0.4, \\
\gamma_{2,c} &= 0.2, & \gamma_{2,f} &= 0.25,& 
\gamma_{3,c} &= 0.4, & \gamma_{3,f} &= 0.5, \\
\gamma_{4,c} &= -0.005, & \gamma_{4,f} &= -0.0045, & 
\gamma_{5,c} &= 0.01, & \gamma_{5,f} &= 0.015, \\
\sigma_1 &= 0.75, & \sigma_2 &= 0.75, &
\sigma_\epsilon &= 1, & t &= 20, \dots, 60,
\end{align*} yields $T=41$, and $\beta_0=0.50$ according to equation (\ref{eq:beta_sim}).
\par 
The simulation considers sample sizes $n$ = 100, 500, 1000, and 2000, with 20\%, 35\%, and 50\% of each sample randomly selected via income snapshots from parents and children. First, for each individual $i$, a contiguous observation period length is drawn from a right-censored Poisson distribution:
\begin{align*}
\ell_i = \min\left(\max(OW_i^P, 2), 41\right), \quad OW_i^P \sim \text{Pois}(\lambda),
\end{align*}
where $\lambda \approx 10$, $20$, andor $31$ years for 20\%, 35\%, and 50\% coverage respectively. Then, the observation window begins at a random age:
\begin{align*}
a_i \sim \mathcal{U}\left(20,\ 60-\ell_i+1\right)
\end{align*}
ensuring complete coverage within the 20-60 age range. Thus, only incomes satisfying $t \in [a_i, a_i+\ell_i)$ are observed, with other years being missing (completely at random), mimicking common data limitations in mobility studies. This creates contiguous observation blocks that mimic real-world data limitations where income histories may only be observed during certain life periods, mimicking realistic administrative or survey-based data constraints. The sampling is performed separately for children and parents.\par 
I assess the performance of the Locally Robust (LR) estimator by examining its bias and coverage properties relative to three alternative approaches: (1) the plug-in machine learning estimator, (2) the mid-life income estimator, and (3) the life-cycle estimator. Income profiles are estimated for both generations using XGBoost Regression, which also allows us to compute the conditional covariance of parental income. Propensity scores are estimated via logistic regression. The core difference between the locally robust (LR) and plug-in machine learning estimators lies in their moment conditions: the LR estimator uses a Neyman-orthogonal moment that incorporates the influence function of the first-stage estimates, while the plug-in estimator relies on the uncorrected identifying moment. For the mid-life income estimator, fathers’ permanent income is proxied by averaging earnings from ages 30 to 40, and children’s income is based on a single mid-life earnings draw. In contrast, the life-cycle estimator uses the same paternal income proxy but estimates children’s permanent income as the average of predicted earnings over the life cycle from a correctly specified OLS regression. For the LR and ML estimators, I proceed in two steps: hyperparameter tuning using 5-fold cross-validation, followed by cross-fitting to prevent overfitting. In all simulations, 500 Monte Carlo replications are used.   \par
Table \ref{tab:1} presents the finite-sample performance of four estimators for the intergenerational elasticity, evaluated through bias and coverage rates across 500 Monte Carlo replications. The true IGE is 0.5, with a nominal coverage rate of 0.95. The analysis spans three sample sizes (\(n = 100, 500, 1000, 2000\)) and three observation probabilities (\(\kappa = 0.20, 0.35, 0.50\)).\par 
\begin{table}[ht]
\caption{Bias and coverage of different estimators for the IGE for different sample sizes and observation probability.}
\centering\resizebox{0.95\textwidth}{!}{%
\begin{threeparttable}
\begin{tabular}{l  cc  cc  cc  cc}  
\hline 
 & \multicolumn{2}{c}{Locally Robust} & \multicolumn{2}{c}{Plug-in Machine Learning} & \multicolumn{2}{c}{Life-cycle} & \multicolumn{2}{c}{Mid-life} \\  \cmidrule(lr){2-3} \cmidrule(lr){4-5} \cmidrule(lr){6-7} \cmidrule(lr){8-9}
$n$ & Bias & Coverage & Bias & Coverage & Bias & Coverage & Bias & Coverage \\
\hline \hline 
$\kappa$ = 0.20 & & & & & & & & \\
\hline
100  & -0.00506 & 0.91 & -0.07358 & 0.56 & -0.17380 & 0.36 & -0.19890 & 0.86 \\ 
  500  & -0.00347 & 0.95 & -0.03991 & 0.43 & -0.16430 & 0.00 & -0.18762 & 0.43 \\ 
1000  & -0.00129 & 0.92 & -0.02981 & 0.41 & -0.16310 & 0.00 & -0.18462 & 0.17 \\ 

  2000  & -0.00132 & 0.93 & -0.03414 & 0.11 & -0.16420 & 0.00 & -0.18699 & 0.01 \\ 
\hline \hline
$\kappa$ = 0.35 & & & & & & & & \\
\hline
 100 &  -0.00685 & 0.91 & -0.05702 & 0.66 & -0.1540 & 0.27 & -0.18318 & 0.75 \\ 
  500 &  -0.00314 & 0.94 & -0.02800 & 0.66 & -0.14970 & 0.00 & -0.17540 & 0.17 \\ 
    1000 &  -0.00230 & 0.93 & -0.03077 & 0.35 & -0.14960 & 0.00 & -0.17099 & 0.01 \\ 
  2000 & 0.00109 & 0.92 & -0.01376 & 0.67 & -0.14900 & 0.00 & -0.16894 & 0.00 \\ 
\hline \hline
$\kappa$ = 0.50 & & & & & & & & \\
\hline
  100 & -0.00317 & 0.91 & -0.04778 & 0.72 & -0.16108 & 0.12 & -0.15682 & 0.73 \\ 
  500  & 0.00561 & 0.93 & -0.01547 & 0.84 & -0.14557 & 0.00 & -0.15302 & 0.12 \\ 
  1000 & 0.00460 & 0.93 & -0.01928 & 0.67 & -0.14741 & 0.00 & -0.15481 & 0.01 \\ 

  2000  & 0.00306 & 0.92 & -0.00949 & 0.82 & -0.13861 & 0.00 & -0.15465 & 0.00 \\ 
\hline
\end{tabular}
\begin{tablenotes}
\footnotesize
\item Results based on 500 Monte Carlo replications with true IGE equal to 0.5 and the nominal coverage is 0.95.
\end{tablenotes} 
\end{threeparttable}}
\label{tab:1}
\end{table}
The locally robust estimator exhibits superior performance, with bias decreasing as sample size increases (e.g., from $-0.0051$ at \(n=100\) to $-0.0015$ at \(n=2000\) for \(\kappa=0.20\)). This aligns with expected \(\sqrt{n}\)-consistency, reflecting its robustness to sample size variations. Coverage rates remain close to the nominal 0.95, ranging from 0.91 to 0.95 across all scenarios, with minor undercoverage at smaller sample sizes (\(n=100\)). Notably, both bias and coverage are largely insensitive to changes in \(\kappa\), indicating stability across varying observation probabilities.\par 
In contrast, the plug-in machine learning estimator has substantially higher bias in absolute terms (e.g., $-0.0736$ at \(n=100\) vs. $-0.0095$ at \(n=2000\) for \(\kappa=0.50\)). Its coverage rates are consistently below the nominal 0.95, improving from 0.56 to 0.82 as sample size increases for \(\kappa=0.50\), but remaining inadequate. This poor performance underscores the limitations of the plug-in approach, particularly in smaller samples or lower observation probabilities, justifying the preference for the locally robust estimator.\par 
The life-cycle and mid-life estimators exhibit substantial and persistent bias across all sample sizes and \(\kappa\) values, with LC bias ranging from  $-0.174$ to $-0.139$, and MI bias from $-0.199$ to $-0.155$. Their coverage rates deteriorate to zero for larger samples due to miss-centered confidence intervals: as sampling variability decreases, the intervals narrow around biased point estimates, missing the true IGE. The similar performance of LC and MI estimators in these simulations reflects two factors. First, although the LC estimator addresses the children’s life-cycle bias, both estimators rely on the same mid-life proxies for parental permanent income, so neither fully accounts for parental life-cycle bias. Second, the simple DGP generates limited heterogeneity in children's income growth by family background, attenuating the LC estimator's designed advantage. In the PSID application, where such heterogeneity is pronounced, the LC estimator substantially outperforms the MI approach (see Table \ref{tab:ige_cohort} and Figure \ref{fig:three_plots}). These simulations thus primarily validate the locally robust estimator's properties rather than a comprehensive comparison of all methods.\par 
Overall, the locally robust estimator emerges as the most reliable, offering low bias and near-nominal coverage across all settings. The plug-in machine learning estimator, while improving with larger samples, remains inferior due to higher bias and poor coverage. The life-cycle and mid-life estimators are consistently outperformed, highlighting the importance of consistent and locally robust estimation in analyzing the IGE.\par 
\section{Consistent Estimation of the Intergenerational Elasticity in the United States}\label{sec:app}
In this section, I implement the locally robust estimator to measure the intergenerational elasticity of income in the United States. The analysis employs the Panel Study of Income Dynamics, the world’s longest-running longitudinal household survey. Launched in 1968 with a nationally representative sample of 5,000 U.S. families (over 18,000 individuals), the PSID has continuously tracked these families and their descendants, collecting rich data on income, wealth, employment, education, health, and other socioeconomic outcomes.\par 
The analysis focuses on birth cohorts spanning 1954 to 1977, using rolling 10-year windows, yielding 15 overlapping cohort samples. This cohort selection ensures sufficient observations for both parents and children within the available PSID data span (1968-2023). Following \cite{lee2009trends}, I use the PSID core sample, corresponding to the Survey Research Center component, and define the income measure as family income, which allows us to include both male and female children. Individuals with only zero or missing income values are excluded. All dollar values are adjusted to 1968 dollars using the CPI. To handle nonpositive incomes, they are bottom-coded at the sample 1st percentile, which affects 0.12\% of the observations in the raw PSID data.\par
Following \cite{mazumder2016estimating}, I focus on the lifetime span from ages 25 to 55 (31 years), corresponding to the core working-life period. This age range is chosen to ensure adequate sample coverage for both generations within the PSID timeframe while capturing income during the stable working years. The 25-55 window allows us to observe sufficient years of income for parents (who are typically 25 years older than their children, see Table \ref{tab:sum}) while avoiding periods dominated by educational transitions at younger ages or retirement decisions at older ages. Sample sizes range from 1,038 to 1,103 child-father pairs across 574 to 757 families, with 11,747 to 19,694 child observations and 16,802 to 25,400 father observations (see Table \ref{tab:sample_sizes} for detailed counts by cohort window).\par 
The family characteristics in the analysis are drawn from the rich data provided by the PSID and are organized into several domains. Education is measured by years of schooling completed and whether the household head reported having additional training beyond standard school or college. Regional location follows the PSID’s classification into Northeast, North Central, South, or West. Family structure includes the birth order of the children, the father’s age at first birth, and the PSID’s intergenerational mapping is used to incorporate the number of offspring per father. Assets are captured through indicators of housing and business ownership. Demographics include race (classified as White or Non-White), sex of the children (given the focus on fathers), religion, and age at the time of interview. While the PSID offers a broader set of variables, the analysis focuses on these selected characteristics to ensure consistency and availability across survey waves.\par 
Based on this available data, I construct the characteristics predictive of income profiles and parental income covariance, namely, \(\bm{X}_{ct}\), \(\bm{X}_{ft}\), and \(\bm{X}_{ftj}\). This specification must account for three key requirements: handling missing data in observables, incorporating the dynamics of the income process, and satisfying Assumptions \ref{as:ortho_np} and \ref{as:unc_np}. To address the first, the variables are summarized over the lifetime span (ages 25–55) using averages for time-varying characteristics (excluding education), modes for religion and region, and the maximum value for education.\par 
To accurately model yearly income as a function of covariates, it is essential to incorporate the dynamics of the income process and capture relevant empirical patterns. To this end, the covariate specification in \cite{mello2022lifecycle} is closely followed. Specifically, a quartic polynomial in age is included to capture concavities and nonlinearities in income profiles. For the child generation, a noisy proxy for parental permanent income is incorporated, defined as the three-year average of log family income when the child was aged 15–17. If family income data for this period are unavailable, the closest available three-year window within ages 15–17 is used. Additionally, interactions between this noisy proxy and parental education with a quadratic polynomial in age are included to capture the greater variability in income growth at younger ages and the typically faster income growth among children from high-income families. \par 
Our covariate specification is designed to satisfy Assumption \ref{as:ortho_np}.$i$ by construction, while also making the remaining assumptions plausible in practice, although not guaranteed to hold. To satisfy Assumption \ref{as:ortho_np}.$i$, $\bm X_{ft}$ and $\bm X_{cj}$ are merged such that their non-overlapping components   $\left(\bm X_{ft}\cap \bm X_{cj}\right)^c=\{age_{ft},age_{cj}\}$,  are both deterministic. In doing so, I also include in \(\bm{X}_{ft}\) the noisy measure of parental permanent income along with its interaction with age. This time-invariant measure serves as a relevant predictor in the presence of missing income data, helping to compensate for the absence of income leads and lags. Moreover, it enhances the first-step estimation of income profiles (and parental income covariance), which is fundamentally a prediction task. Finally, to construct \(\bm{X}_{ftj}\), \(\bm{X}_{ft}\) and \(\bm{X}_{fj}\) are merged, ensuring  $\left(\bm X_{ft}\cap \bm X_{ftj}\right)^c=\{age_{fj}\}$. The current specification of the covariates ensures that Assumption \ref{as:ortho_np}.$i$ is satisfied by construction; that is, the covariates used to predict children’s income satisfy:

$\mathbb{E}\left[Y_{ct} \mid \bm{X}_{ct}, \bm{X}_{cj}, \bm{X}_{fj}\right] = \mathbb{E}\left[Y_{ct} \mid \bm{X}_{ct}\right]$ for $t,j=1,\ldots,T,$
and those used to predict fathers’ income satisfy $
\mathbb{E}\left[Y_{ft} \mid \bm{X}_{ft}, \bm{X}_{ftj},  \bm{X}_{cj}\right] = \mathbb{E}\left[Y_{ft} \mid \bm{X}_{ft}\right]$ for  $\bm{X}_{fj} \subset\bm{X}_{ftj},   t,j=1,...T.$ This follows from the complements of the intersections, $\left(\bm X_{ft}\cap \bm X_{ft}\right)^c=\{age_{ft},age_{cj}\}$ and $\left(\bm X_{ft}\cap \bm X_{ftj}\right)^c=\{age_{fj}\}$, consisting solely of age, which is deterministic and thus do not add stochastic variation beyond what is captured by $\bm X_{ct}$ and $\bm X_{ft}$.\par  
Our covariate specification further enhances the plausibility of the remaining assumptions in practice. The inclusion of an interaction between parental permanent income and age, for example, strengthens the orthogonality condition between children's income prediction errors and parental permanent income required by Assumption \ref{as:ortho_np}.$ii$. In Section \ref{sec:test}, I develop a test to empirically evaluate Assumption \ref{as:ortho_np}.$ii$. As regards, Assumption \ref{as:ortho_np}.$iii$, I outline how to construct a similar test but assess this assumption in practice by examining the estimated autocorrelation function, which is more reliable given the sparse availability of income pairs over long time spans..\par 
The dimensionality of the covariate set reflects the trade-off between the plausibility of the missing-at-random assumption and the boundedness of the propensity scores in Assumption \ref{as:unc_np}. While increasing the dimension of $\bm X_{gt}$, can make the conditional independence  $Y_{ct}\perp D_{ct}|\bm X_{ct}$ more plausible, it may reduce the likelihood that the propensity score remains bounded away from zero. Nonetheless, it is not merely the dimensionality, but rather the informativeness of the covariates that determines whether missingness is conditionally at random. In other words, we aim to control for the relevant features such that, conditional on them, the missingness of annual income occurs conditionally at random. While the MAR conditions in Assumptions \ref{as:unc_np}.$i$ and \ref{as:unc_np}.$iii$ cannot be directly tested from observed data, the boundedness conditions in Assumptions \ref{as:unc_np}.$ii$ and \ref{as:unc_np}.$iv$ can be assessed informally through visual inspection.\par
\begin{table}[ht]
\centering
\caption{Summary Statistics for Children and Fathers}\label{tab:sum}
\resizebox{0.95\textwidth}{!}{%
\begin{tabular}{lccccc|ccccc}
\toprule\hline  &
\multicolumn{5}{c|}{\textbf{Children}} & \multicolumn{5}{c}{\textbf{Fathers}} \\
\cmidrule(r){1-6} \cmidrule(l){7-11}
Variable & Mean & Median & SD & Min & Max & Mean & Median & SD & Min & Max \\
\midrule
Annual Income            & 9.22 & 9.31 & 0.90 & -1.42 & 13.8 & 9.30 & 9.36 & 0.87 & -1.53 & 12.9 \\
Proxy of Permanent Income & -    & -    & -    & -    & -    & 9.33 & 9.41 & 0.72 & -1.06 & 12.1 \\
Education Level          & 14.10 & 14.00 & 2.18 & 0.00 & 17.0 & 13.00 & 12.00 & 3.01 & 0.00 & 17.0 \\
House Ownership          & 0.64 & 0.73 & 0.34 & 0.00 & 1.00 & 0.78 & 0.93 & 0.31 & 0.00 & 1.00 \\
Business Ownership       & 0.16 & 0.00 & 0.25 & 0.00 & 1.00 & 0.18 & 0.00 & 0.28 & 0.00 & 1.00 \\
Additional Training      & 0.22 & 0.00 & 0.42 & 0.00 & 1.00 & 0.15 & 0.00 & 0.36 & 0.00 & 1.00 \\
Religion                 & 0.86 & 1.00 & 0.35 & 0.00 & 1.00 & 0.88 & 1.00 & 0.32 & 0.00 & 1.00 \\
White                    & 0.92 & 1.00 & 0.28 & 0.00 & 1.00 & 0.91 & 1.00 & 0.29 & 0.00 & 1.00 \\
Sex                      & 0.51 & 1.00 & 0.50 & 0.00 & 1.00 & -    & -    & -    & -    & -    \\
Birth Order              & 2.11 & 2.00 & 1.40 & 1.00 & 12.0 & -    & -    & -    & -    & -    \\
Northeast Region         & 0.20 & 0.00 & 0.40 & 0.00 & 1.00 & 0.20 & 0.00 & 0.40 & 0.00 & 1.00 \\
South Region             & 0.34 & 0.00 & 0.48 & 0.00 & 1.00 & 0.32 & 0.00 & 0.47 & 0.00 & 1.00 \\
West Region              & 0.18 & 0.00 & 0.38 & 0.00 & 1.00 & 0.17 & 0.00 & 0.37 & 0.00 & 1.00 \\
Age at First Child       & -    & -    & -    & -    & -    & 25.40 & 25.00 & 4.88 & 13.00 & 50.0 \\
\hline\hline
\end{tabular}%
}
\end{table} Table~\ref{tab:sum} provides summary statistics for socioeconomic characteristics of children and their fathers, revealing important intergenerational patterns. Fathers exhibit higher mean logged annual income (9.30 vs.\ 9.22) and home ownership rates (78\% vs.\ 64\%), while children show greater educational attainment (mean 14.1 vs. 13 years) and additional training participation (22\% vs.\ 15\%). The income measures exhibit slightly tighter dispersion for fathers, with smaller standard deviations (0.87 vs.\ 0.90 for annual income). Educational attainment shows greater variability among fathers (SD 3.01 vs.\ 2.18), potentially reflecting cohort differences in educational access. Both generations share nearly identical white composition (91\% vs.\ 92\%), religious affiliation (88\% vs.\ 86\%), and business ownership (18\% vs.\ 16\% for children). Half of the children are female, reflecting a balanced gender distribution. The median birth order indicates that most families in the data have two or more children, with relatively few only children. The mean and median values for the proxy of permanent income closely match those of annual income, but with less variation and a narrower range. Most fathers had their first child around age 25. Regional distributions show similar patterns across generations, with children slightly more concentrated in the South (34\% vs.\ 32\%) and both generations showing identical Northeast representation (20\%).\par
\begin{figure}
    \centering
    \includegraphics[width=0.9\linewidth]{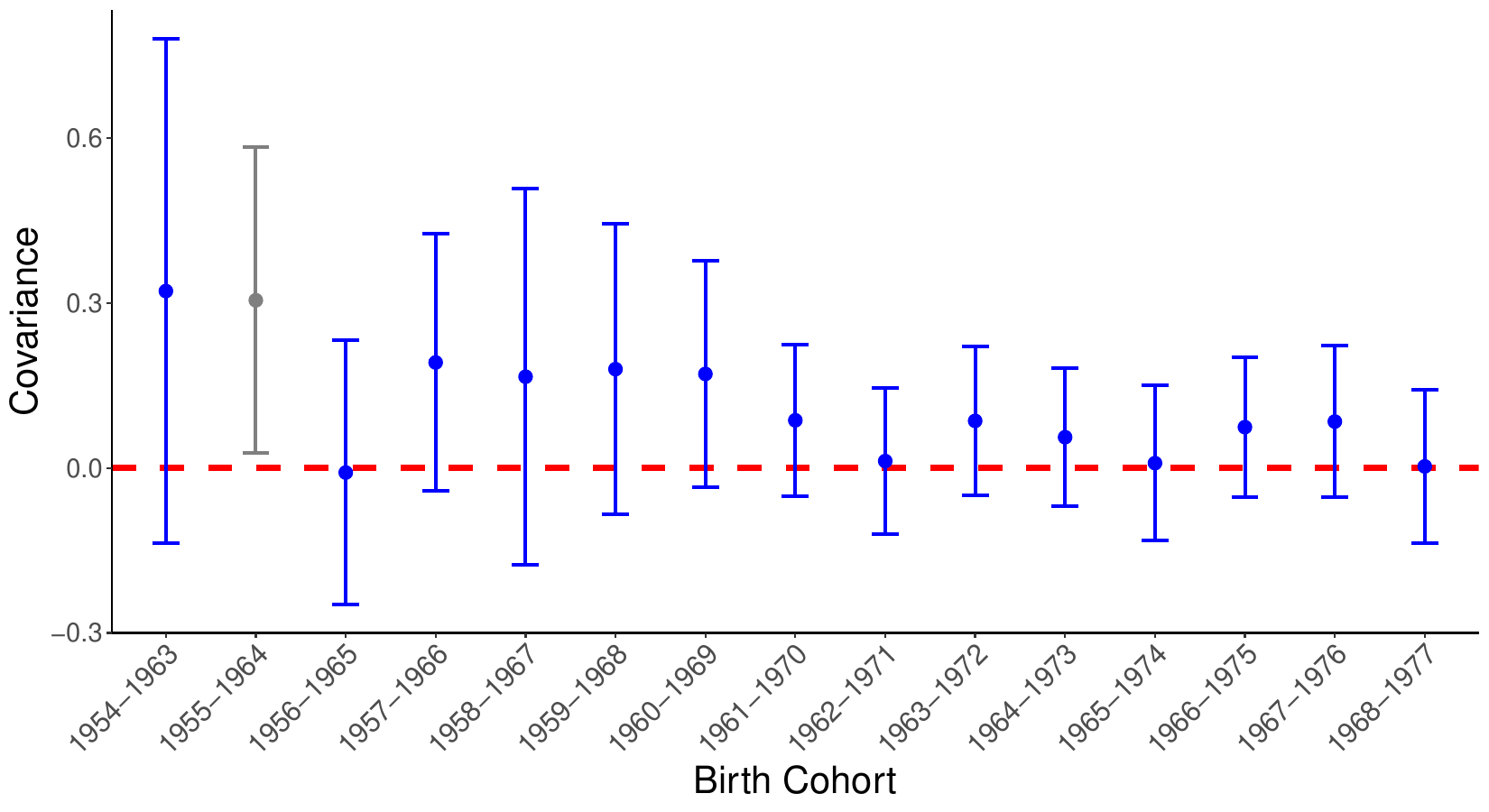}
    \caption{Covariance of Children Prediction Errors and Parental Permanent Income Across Cohorts}
    \label{fig:res_test}
\end{figure}

 
To assess the assumption that the children's prediction errors are uncorrelated with parental lifetime income, I implement the proposed locally robust test. This condition is crucial for unbiased estimation: if prediction errors systematically correlate with parental income, the IGE estimates will be biased. Results are shown in Figure \ref{fig:res_test} and Table \ref{tab:cohort_estimates}. In 14 of the 15 cohort windows, the LR test fails to reject the null hypothesis of zero covariance between children’s income prediction errors and parental permanent income, with an average $p$-value of 0.39. This finding provides empirical evidence that the life-cycle estimator of \cite{mello2022lifecycle} effectively addresses life-cycle bias from the children's side by properly accounting for how income trajectories vary with family background. Importantly, this empirical validation of the orthogonality condition strengthens confidence in both the identification strategy and the resulting IGE estimates.\par
\begin{figure}[htbp]    \centering
    \begin{subfigure}[b]{0.48\textwidth}        \centering
        \includegraphics[width=\linewidth]{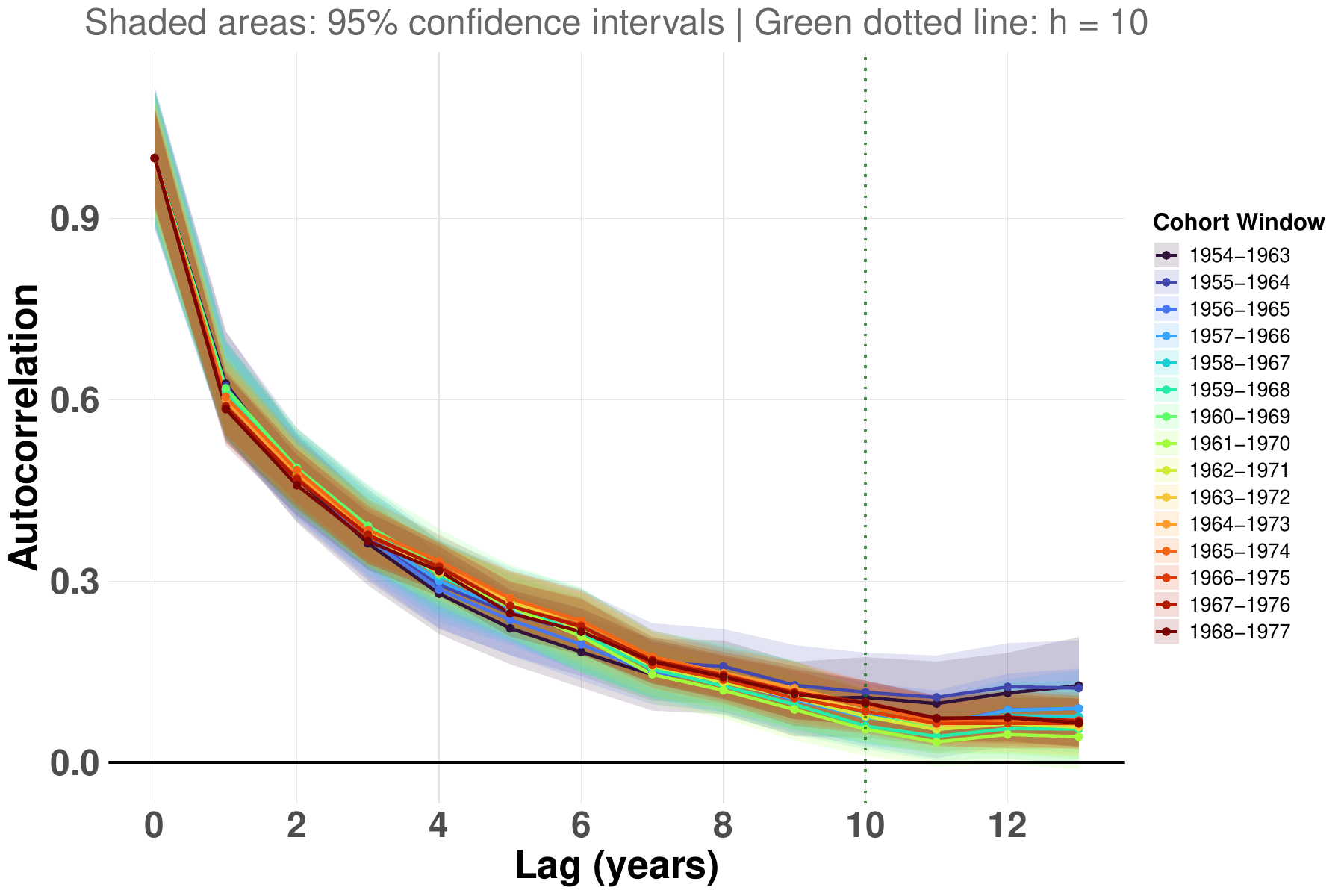}        \caption{Father Income Residual Autocorrelation by Birth Cohort Window}        \label{fig:autocor}
   \end{subfigure}   \hfill   \begin{subfigure}[b]{0.48\textwidth}       \centering       \includegraphics[width=\linewidth]{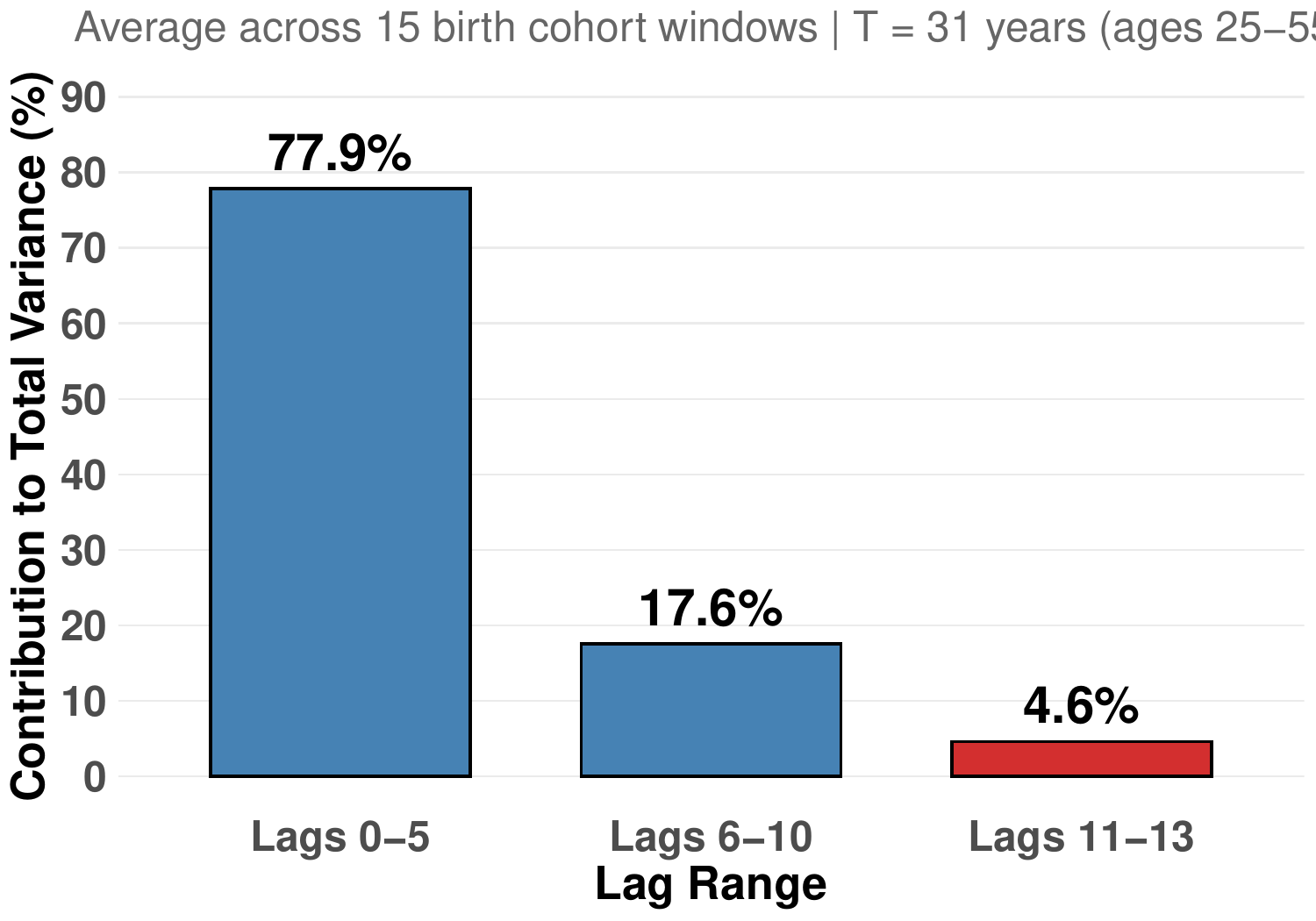}        \caption{Decomposition of Residual Income Variance by Lag Range}        \label{fig:second_figure}
    \end{subfigure}
    \caption{Dependence of Parental Income Prediction Errors}
\end{figure}

Our second orthogonality assumption requires that the average covariance of parental income prediction errors for distant periods is negligible. To assess this, I analyze the autocorrelation structure and variance decomposition of these residuals. Since the variance of permanent income (the denominator of the IGE) decomposes as $\text{Var}(Y_f^P) = \text{Var}\left(\mathbb{E}[Y_{ft} \mid \mathbf{X}_{ft}]\right) + \text{Var}(\epsilon_{ft})$,
the contribution of distant-lag residual products to the total variance depends on both their contribution to residual variance and the relative importance of residual variance itself. Figure \ref{fig:autocor}  exhibits rapid autocorrelation decay across all 15 birth cohort windows: autocorrelations fall from 0.607 at lag 1 to 0.086 at lag 10 and stabilize at 0.072 for lags 11--13. Figure~\ref{fig:second_figure} shows that lags 0--10 account for 95.4\% of the residual variance, while lags 11--13 contribute only 4.6\%. The analysis examines autocorrelations through lag 13 because observation pairs become increasingly sparse at distant lags. Unmeasured lags beyond 13 would contribute even less due to both the observed monotonic decay and their mechanically lower weights in the variance structure. These findings strongly support the orthogonality assumption for $h=10$, which I adopt for implementing the proposed estimator.\par 
Having validated the identifying assumptions empirically, I now turn to estimating the intergenerational elasticity using four alternative approaches: the locally robust (LR) estimator, the plug-in machine learning (ML) estimator, the life-cycle (LC) estimator, and the mid-life income (MI) estimator. Following \cite{mello2022lifecycle}, the LC estimator predicts children’s log income using an OLS specification that includes a quartic in age, interactions with education dummies, linear and quadratic interactions with parental income, individual fixed effects, and year fixed effects, with outliers removed. Lifetime income is then constructed as the mean of the exponentiated predicted values for each individual, multiplied by a smearing factor computed as the average of exponentiated residuals within parental income deciles, correcting for retransformation bias in the log-linear specification \citep{wooldridge2013introductory}. Fathers’ permanent income is proxied by the three-year average of log family income when the child was aged 15–17. To implement the mid-life income estimator,the same measure of parental permanent income is used, whereas children’s income is calculated as the three-year average of log income between ages 25 and 33, following \cite{solon1992intergenerational}.\par
The locally robust and plug-in machine learning estimators use the same covariates as the life-cycle estimator, augmented with the family characteristics listed in Table \ref{tab:sum}. Individual fixed effects are omitted as they cannot be accommodated under the cross-fitting procedure.
Propensity scores are estimated using a lasso-logit model, while income profiles for both generations and the conditional covariances of parental income are modeled using XGBoost regression. The key distinction between the LR and plug-in ML estimators lies in their moment conditions: the LR approach utilizes the orthogonal moment that accounts for the first-stage influence function, whereas the plug-in version relies solely on the identifying moment without such correction. For all predictive tasks in the LC, LR, and ML estimators, values are winsorized at the 1st and 99th percentiles.\par 
Table \ref{tab:ige_cohort} presents intergenerational elasticity estimates for the United States using PSID core sample data across 15 birth cohorts (1954-1977). The locally robust estimator yields an average IGE of 0.643, systematically exceeding estimates from the plug-in machine learning approach (0.600), the life-cycle estimators (0.500 and 0.502 for the log-average and average-log specifications, respectively), and the standard mid-life income estimator (0.430). These economically meaningful differences underscore the importance of combining proper identification with local robustness for reliable mobility measurement. \par
\begin{table}[ht]
\centering
\caption{Intergenerational Elasticity Estimates by Cohort Using Alternative Estimators} 
\label{tab:ige_cohort}
\begin{threeparttable}
\begin{adjustbox}{max width=0.8\textwidth,center}
\begin{tabular}{lccccc}
\toprule
Cohort & LR & ML & LC (log-avg) & LC (avg-log) & MI \\ 
  \midrule
1954-1963 & 0.598 & 0.522 & 0.509 & 0.496 & 0.413 \\ 
  (N = 1,099) & (0.414, 0.782) & (0.397, 0.647) & (0.420, 0.598) & (0.407, 0.585) & (0.317, 0.508) \\ 
  1955-1964 & 0.596 & 0.557 & 0.502 & 0.494 & 0.415 \\ 
  (N = 1,089) & (0.396, 0.796) & (0.415, 0.700) & (0.409, 0.596) & (0.401, 0.587) & (0.318, 0.513) \\ 
  1956-1965 & 0.596 & 0.545 & 0.522 & 0.519 & 0.450 \\ 
  (N = 1,083) & (0.441, 0.752) & (0.414, 0.676) & (0.428, 0.615) & (0.426, 0.612) & (0.351, 0.550) \\ 
  1957-1966 & 0.624 & 0.616 & 0.513 & 0.517 & 0.419 \\ 
  (N = 1,071) & (0.410, 0.837) & (0.482, 0.750) & (0.419, 0.608) & (0.422, 0.611) & (0.322, 0.517) \\ 
  1958-1967 & 0.634 & 0.596 & 0.536 & 0.542 & 0.486 \\ 
  (N = 1,097) & (0.453, 0.814) & (0.471, 0.721) & (0.446, 0.626) & (0.452, 0.631) & (0.394, 0.577) \\ 
  1959-1968 & 0.659 & 0.619 & 0.519 & 0.526 & 0.432 \\ 
  (N = 1,085) & (0.481, 0.836) & (0.520, 0.718) & (0.436, 0.603) & (0.444, 0.609) & (0.342, 0.521) \\ 
  1960-1969 & 0.625 & 0.590 & 0.511 & 0.514 & 0.431 \\ 
  (N = 1,062) & (0.428, 0.823) & (0.503, 0.677) & (0.429, 0.592) & (0.434, 0.595) & (0.347, 0.515) \\ 
  1961-1970 & 0.615 & 0.579 & 0.496 & 0.495 & 0.421 \\ 
  (N = 1,042) & (0.444, 0.785) & (0.499, 0.659) & (0.418, 0.573) & (0.419, 0.571) & (0.341, 0.501) \\ 
  1962-1971 & 0.630 & 0.610 & 0.496 & 0.499 & 0.415 \\ 
  (N = 1,045) & (0.468, 0.792) & (0.532, 0.688) & (0.418, 0.574) & (0.422, 0.576) & (0.333, 0.498) \\ 
  1963-1972 & 0.672 & 0.676 & 0.504 & 0.507 & 0.428 \\ 
  (N = 1,038) & (0.512, 0.831) & (0.589, 0.762) & (0.424, 0.585) & (0.427, 0.586) & (0.345, 0.512) \\ 
  1964-1973 & 0.663 & 0.667 & 0.475 & 0.481 & 0.413 \\ 
  (N = 1,060) & (0.508, 0.817) & (0.580, 0.754) & (0.396, 0.554) & (0.403, 0.559) & (0.330, 0.497) \\ 
  1965-1974 & 0.702 & 0.711 & 0.504 & 0.511 & 0.451 \\ 
  (N = 1,067) & (0.551, 0.854) & (0.623, 0.800) & (0.428, 0.580) & (0.436, 0.586) & (0.368, 0.535) \\ 
  1966-1975 & 0.700 & 0.460 & 0.487 & 0.495 & 0.439 \\ 
  (N = 1,074) & (0.587, 0.813) & (0.386, 0.535) & (0.407, 0.567) & (0.415, 0.574) & (0.361, 0.517) \\ 
  1967-1976 & 0.673 & 0.612 & 0.474 & 0.477 & 0.428 \\ 
  (N = 1,088) & (0.526, 0.819) & (0.541, 0.683) & (0.395, 0.552) & (0.399, 0.555) & (0.353, 0.504) \\ 
  1968-1977 & 0.671 & 0.641 & 0.461 & 0.467 & 0.410 \\ 
  (N = 1,103) & (0.504, 0.838) & (0.563, 0.720) & (0.381, 0.542) & (0.386, 0.547) & (0.336, 0.485) \\ 
\bottomrule
\end{tabular}
\end{adjustbox}
\begin{tablenotes}
\footnotesize 
\item \hspace{12mm} 95\% confidence intervals clustered at the family level are reported in parentheses.
\end{tablenotes} 
\end{threeparttable}
\end{table}
Our estimates of the IGE range from 0.6 to 0.7 across cohorts, with an average of 0.64. These results closely align with recent PSID-based studies reviewed by \cite{mazumder2018intergenerational} that attempt to mitigate life-cycle bias. Building on the theoretical framework developed by \cite{haider2006life}, \cite{gouskova2010estimating} produces a bias-corrected estimate of 0.63. Taking a different approach, \cite{chau2012intergenerational} embeds an earnings dynamics model with long time spans of income data and estimates the IGE to exceed 0.6. Most recently, \cite{mazumder2016estimating} leverages the full length of the PSID panel by using up to 15-year averages of fathers' income centered at age 40, estimating the IGE in family income to be greater than 0.6. Our locally robust estimates thus reinforce this convergent evidence from multiple PSID-based methodologies, all pointing to substantially higher intergenerational persistence than the conventional estimates of approximately 0.4. Moreover, we find that the IGE increases slightly across cohorts. The earliest three cohort windows average 0.597, the next six average 0.631, and the latest average 0.680. Despite wide confidence intervals, this 14\% progression provides suggestive evidence of a modest increase in persistence across the 1954-1977 birth cohorts.\par 
The naive ML estimator yields an IGE of 0.60 (ranging from 0.46 to 0.71), systematically underestimating the true elasticity.  On average, the plug-in ML estimator underestimates the IGE by 0.043 (7\%), though the bias varies considerably across cohorts. In the best case, the bias is only -0.008 (1.2\%). However, in the worst case (1966-1975 cohort),  the plug-in approach underestimates the IGE by 34\%, yielding an estimate of 0.46 versus the locally robust estimate of 0.70. This represents a meaningful difference in the assessment of intergenerational mobility. In line with the simulation results (Table \ref{tab:1}), the plug-in ML exhibits considerable undercoverage: its confidence intervals are 41\% narrower on average than the locally robust estimator's, reflecting false precision from neglecting uncertainty in nuisance parameter estimation. \par The differences between the locally robust and plug-in ML estimates and confidence intervals underscore that local robustness corrections have real consequences for empirical conclusions. While conventional machine learning approaches excel at prediction, directly applying them to estimate causal or structural parameters without orthogonalization can lead to non-negligible bias that materially affects our understanding of mobility patterns. The 34\% underestimation observed in the 1966-1975 cohort illustrates that the effects of failing to debias can be substantial, even when the average bias across cohorts appears modest.\par 
The standard mid-life income estimator produces an average intergenerational elasticity (IGE) of 0.430. Across the 15 cohort windows, the estimated IGE ranges from 0.410 to 0.486, reflecting a substantial downward bias. On average, the estimator underestimates the true IGE by 0.213, with the bias ranging from 23\% to 39\% across cohorts. This substantial underestimation stems from life-cycle bias in both generations: parental income measured at mid-life imperfectly captures permanent economic status, while children's three-year averages fail to account for differential income growth patterns across family backgrounds.\par 
The life-cycle estimator, implemented using the log of average income as originally proposed by \cite{mello2022lifecycle}, represents a substantial improvement over the mid-life income approach, producing an average IGE of 0.500 with estimates ranging from 0.46 to 0.53. By predicting children's income profiles and accounting for heterogeneous income growth across family backgrounds, the LC estimator successfully addresses the bias children's life-cycle bias inherent in the MI approach. The improvement is considerable: while the MI estimator exhibits an average bias of 33\% (ranging from 23\% to 39\%), the LC estimator reduces this to 22\% (ranging from 12\% to 31\%). Nonetheless, the remaining underestimation suggests that the LC estimator, while correcting for children's bias, still relies on the same mid-life income proxies for parents as the MI estimator.\par 
To assess the sensitivity of IGE estimates to the workable definition of permanent income, I implement the life-cycle estimator using the average of log income (our workable definition underlying the identification strategy in Theorem \ref{thm:2}). The estimates are nearly identical, with the average-log specification yielding an average IGE of 0.500 (range: 0.467-0.542) compared to 0.503 (range: 0.461-0.535) for the log-average specification. The close correspondence between these estimates—differing by less than 0.4\% on average validates the use of the average-log specification that enables nonparametric identification under the proposed missing data framework.\par
\begin{figure}[htbp]
    \centering
    \captionsetup[subfigure]{justification=centering}
    
    \begin{subfigure}[b]{0.8\textwidth}
        \centering
        \includegraphics[width=\textwidth]{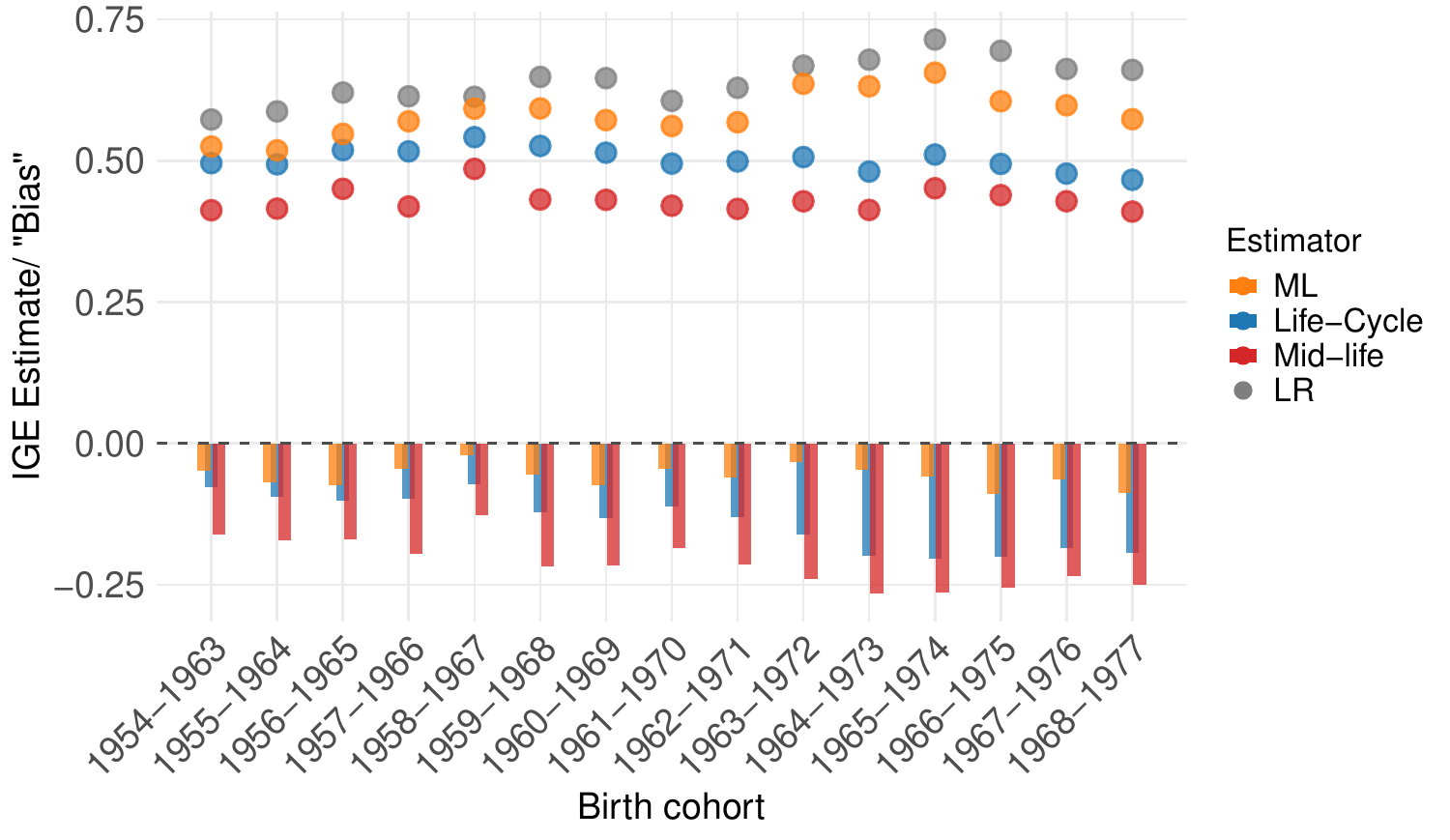}
            \caption{IGE Estimates by Birth Cohort and Estimator}
        \label{fig:betas}
    \end{subfigure}
    \begin{subfigure}[b]{0.8\textwidth}
        \centering
        \includegraphics[width=\textwidth]{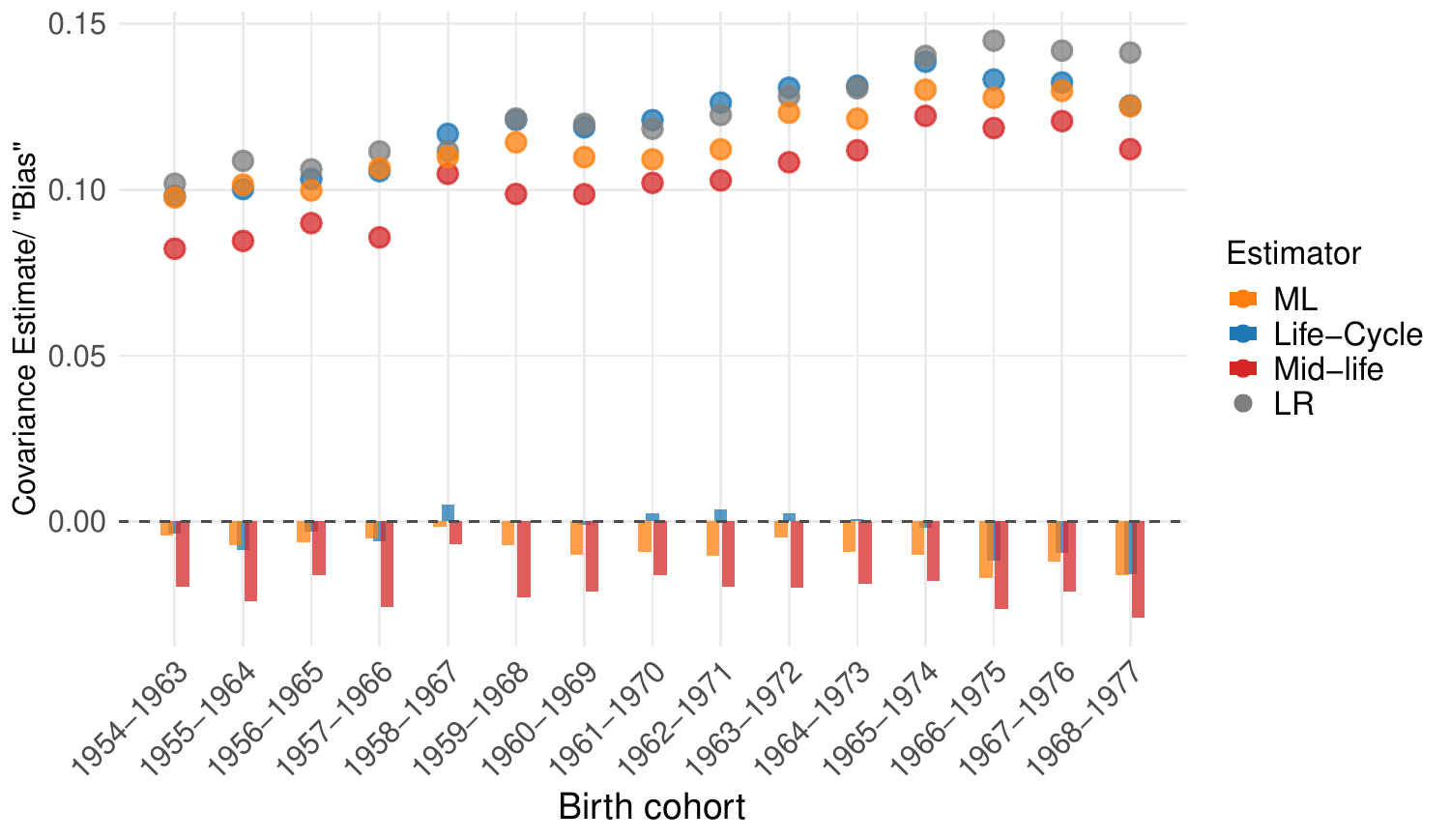}
        \caption{Covariance Estimates by Birth Cohort and Estimator}
        \label{fig:covariances}
    \end{subfigure} 
    \begin{subfigure}[b]{0.8\textwidth}
        \centering
        \includegraphics[width=\textwidth]{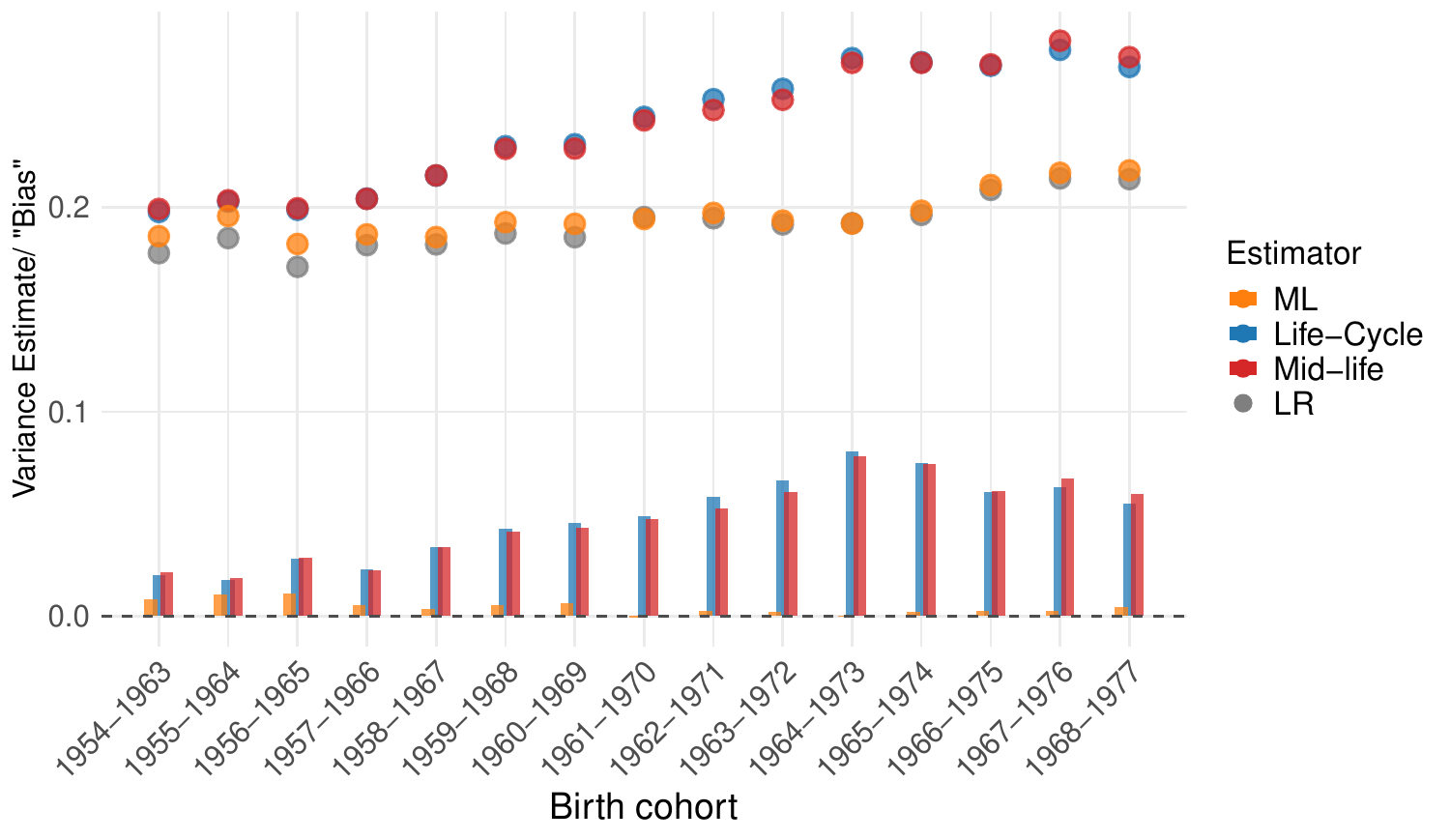}
        \caption{Variance Estimates by Birth Cohort and Estimator}
        \label{fig:variances}
    \end{subfigure}
    
    \caption{Intergenerational Elasticity, Covariances, and Variance Estimates and Bias Across Birth Cohorts and Estimators}
    \label{fig:three_plots}
\end{figure}

 
To shed light on the specific sources of bias for each estimator, I break down the IGE into its core components: the covariance of child and parent permanent income (numerator) and the variance of parental permanent income (denominator). Figure \ref{fig:three_plots} presents this decomposition across all estimators and birth cohorts, displaying both the point estimates and their deviations from the locally robust benchmark, which serve as the empirical measure of bias in this sample. Importantly, the nature of this bias differs fundamentally across estimators: the MI and LC estimators are affected by life-cycle bias and measurement error, due to their reliance on mid-life income averages as proxies for parental permanent income, whereas the plug-in ML estimator is susceptible to bias from regularization and model selection in the first-stage machine learning predictions.\par 
The decomposition in Figure \ref{fig:three_plots} reveals a striking pattern that validates the proposed theoretical framework. Panel (a) reproduces the IGE estimates from Table \ref{tab:ige_cohort}, showing the systematic ordering of estimators across cohorts. Panels (b) and (c) expose the underlying mechanisms: the LC estimator performs remarkably well at estimating the covariance between child and parent permanent income, frequently outperforming even the plug-in ML estimator despite the latter's use of flexible machine learning methods. However, the LC estimator systematically overestimates the variance of parental permanent income, with deviations from the locally robust benchmark that are consistently positive across all cohorts.\par 
Figure \ref{fig:covariances}  provides strong empirical evidence that the LC estimator successfully addresses children's life-cycle bias. The covariance estimates closely track the locally robust benchmark across all cohorts, illustrating that the LC approach accurately captures the child-parent income relationship.  This success stems from the estimator's explicit modeling of children's income profiles: by predicting how income evolves over the life cycle and accounting for heterogeneous income growth across family backgrounds, the LC estimator addresses the sensitivity to the age at which child income is measured. Remarkably, the LC estimator frequently outperforms even the plug-in ML estimator on this component, despite the latter's use of flexible machine learning methods. This superior performance is particularly noteworthy: it demonstrates that a well-specified parametric approach, informed by economic theory about income dynamics and family background effects, can outperform flexible machine learning methods in the absence of debiasing.\par 
Panel (c) highlights the critical importance of addressing all sources of bias simultaneously. Both the LC and MI estimators systematically overestimate the variance of parental permanent income across cohorts. This arises from both relying on mid-life income averages that suffer from life-cycle bias and measurement error. Thus, the inflated denominator mechanically reduces the IGE estimates, leading to substantial underestimation even when the numerator is correctly measured. These results underscore the practical relevance of addressing all sources of bias: the LC estimator’s average IGE of 0.50 falls well short of the locally robust estimate of 0.64.\par
Overall, the empirical application provides compelling evidence that properly addressing both missing income data and estimation uncertainty is essential for credible inference on intergenerational mobility. The locally robust estimator consistently produces higher intergenerational elasticity estimates across all PSID birth cohorts, indicating lower mobility in the United States than suggested by conventional methods and aligning closely with recent evidence based on long-term income data. At the same time, our results confirm that the life-cycle estimator performs precisely as intended on the children’s side: it effectively corrects for  children's life-cycle bias by modeling heterogeneous income growth across family background. However, its reliance on mid-life parental income proxies leaves residual bias from the parental side, leading to systematic underestimation of the IGE. The plug-in machine learning estimator, while flexible and powerful for prediction, fails to correct for first-stage estimation error, resulting in underestimation and undercoverage. Together, these findings illustrate the benefits of local robustness for combining the strengths of structured life-cycle modeling and flexible machine learning to produce reliable and measures of intergenerational mobility.
\section{Conclusions}\label{sec:conc}
This paper addresses the fundamental challenge of measuring the intergenerational transmission of lifetime economic status when researchers observe only snapshots of income at specific ages alongside individual characteristics. Constrained by these data limitations, standard practice estimates the intergenerational elasticity using income averages during mid-life, a procedure that introduces well-documented life-cycle bias. In response, the literature has made substantial progress in refining IGE estimates by separately addressing life-cycle bias from either the parent or child generation. I show that this piecemeal approach hinders comparability of estimates. This insight motivates our core contribution: moving beyond proxy refinement to establish identification conditions that jointly address all sources of bias, enabling consistent and comparable mobility estimates across studies, time, and place.\par 
Building on these insights, I formalize that the magnitude of each bias source depends on study design, income dynamics, and sampling rules, causing proxy-based estimators to converge to context-specific parameters rather than the population IGE, which compromises their comparability across studies, time, and place.\par 
I establish nonparametric identification of the IGE from incomplete income data and family characteristics under standard missing-at-random assumptions and testable orthogonality conditions. Moving beyond the conventional generalized errors-in-variables model, I adopt a workable definition of permanent income as the average of log annual earnings during working life—a specification that proves empirically robust, as IGE estimates remain virtually unchanged when permanent income is instead defined as the log of average lifetime income. \par 
Building on this identification result, I develop a consistent and locally robust estimator by constructing an orthogonal moment function that ensures machine learning estimation of nuisance parameters—such as conditional income expectations and propensity scores—has no first-order effect on the IGE estimate. This debiasing procedure combines Neyman-orthogonal moments with cross-fitting at the family level to prevent overfitting while accommodating the dependent structure inherent in intergenerational data. I establish the estimator's asymptotic normality and construct valid confidence intervals that account for uncertainty introduced by first-stage estimation. Additionally, I develop locally robust tests for the orthogonality assumptions underlying identification, providing researchers with tools to assess the validity of our framework in different empirical contexts.\par 
Our framework enables comparable IGE estimates across time and place in the presence of incomplete income data. By addressing the key methodological challenge of not observing lifetime income, our approach establishes a robust foundation for studying income persistence, enhancing the reliability and interpretability of mobility research across diverse economic settings.\par 
Our simulation analysis illustrates the sound finite-sample performance of the locally robust estimator, which exhibits negligible bias that vanishes as sample size increases and coverage rates close to nominal levels across different scenarios. The estimator substantially outperforms alternative approaches: the plug-in machine learning estimator exhibits both higher bias and severe undercoverage due to failing to account for first-stage uncertainty, while traditional proxy-based methods show large and persistent biases that do not diminish with sample size.\par 
Our locally robust estimates of the IGE range from 0.60 to 0.70 across cohorts, averaging 0.64. These substantially exceed conventional estimates relying on mid-life income averages, but align closely with recent evidence using long-time averages over mid-career periods, revealing considerably lower U.S. mobility than conventional estimates suggest. I also find a naive plug-in machine learning estimator exhibits considerable bias and severe undercoverage, underscoring that empirical relevance on constructing Neyman-orthogonal moments for both eliminating regularization bias and achieving valid inference.\par 
Our study highlights three important directions for future research. First, our identification strategy requires longitudinal data that are often unavailable in developing countries, where understanding income persistence is most relevant. Future work should establish alternative identification results for data-scarce environments.  Second, the shift in the literature toward rank-based measures has been partly motivated by concerns about the nonlinear relationship between log child income and log parent income. Accordingly, future research should study identification and locally robust estimation for a nonlinear version of the intergenerational elasticity. One promising direction involves estimating the regression of child permanent income on parent permanent income in levels. A quantile-specific elasticity can then be constructed by multiplying the marginal effect at each point in the parental income distribution by the ratio of average child to parent income at that quantile. This would provide a richer, distributional perspective on income persistence and allow researchers to quantify how mobility varies across the income ladder patterns while avoiding the limitations of log-linear specifications.\par 
Finally, a central empirical challenge in economics is that many key parameters, from models of life-cycle income, savings, and consumption to measures of individual well-being, depend on lifetime outcomes, while only partial observations at certain ages are typically available. The methods proposed in this paper open the door for applications in other contexts with partially observed outcomes, providing a flexible framework for addressing similar empirical challenges.
\bibliography{cites}

@techreport{mello2022lifecycle,  title={A lifecycle estimator of intergenerational income mobility},  author={Mello, Ursula and Nybom, Martin and Stuhler, Jan},  year={2025},  institution={Working Paper}}

@article{stuhler2018review,  title={A review of intergenerational mobility and its drivers},  author={Stuhler, Jan and others},  journal={Publications Office of the European Union, Luxembourg},  year={2018}}

@article{jappelli2010consumption,  title={The consumption response to income changes},  author={Jappelli, Tullio and Pistaferri, Luigi},  journal={Annual Review of Economics},  volume={2},  number={1},  pages={479--506},  year={2010},  publisher={Annual Reviews}}

@article{heckman2006earnings,  title={Earnings functions, rates of return and treatment effects: The Mincer equation and beyond},  author={Heckman, James J and Lochner, Lance J and Todd, Petra E},  journal={Handbook of the Economics of Education},  volume={1},  pages={307--458},  year={2006},  publisher={Elsevier}}

@book{dahl2008association,  title={The association between children's earnings and fathers' lifetime earnings: estimates using administrative data},  author={Dahl, Molly W and DeLeire, Thomas},  year={2008},  publisher={University of Wisconsin-Madison, Institute for Research on Poverty Madison}}

@article{nybom2016heterogeneous,  title={Heterogeneous income profiles and lifecycle bias in intergenerational mobility estimation},  author={Nybom, Martin and Stuhler, Jan},  journal={Journal of Human Resources},  volume={51},  number={1},  pages={239--268},  year={2016},  publisher={University of Wisconsin Press}}

@article{couch1998sample,  title={Sample selection rules and the intergenerational correlation of earnings},  author={Couch, Kenneth A and Lillard, Dean R},  journal={Labour Economics},  volume={5},  number={3},  pages={313--329},  year={1998},  publisher={Elsevier}}

@article{chetty2014land,  title={Where is the land of opportunity? The geography of intergenerational mobility in the {United States}},  author={Chetty, Raj and Hendren, Nathaniel and Kline, Patrick and Saez, Emmanuel},  journal={The Quarterly Journal of Economics},  volume={129},  number={4},  pages={1553--1623},  year={2014},  publisher={MIT Press}}

@article{angrist1995identification,  title={Identification and estimation of local average treatment effects},  author={Angrist, Joshua D and Imbens, Guido W},  journal={Econometrica},  volume={64},  number={2},  pages={467--475},  year={1994},  publisher={JSTOR}}

@article{halvorsen2022earnings,  title={Earnings dynamics and its intergenerational transmission: Evidence from {Norway}},  author={Halvorsen, Elin and Ozkan, Serdar and Salgado, Sergio},  journal={Quantitative Economics},  volume={13},  number={4},  pages={1707--1746},  year={2022},  publisher={Wiley Online Library}}

@article{zimmerman1992regression,  title={Regression toward mediocrity in economic stature},  author={Zimmerman, David J},  journal={The American Economic Review},  pages={409--429},  year={1992},  publisher={JSTOR}}

@article{bratberg2007trends,  title={Trends in intergenerational mobility across offspring's earnings distribution in {Norway}},  author={Bratberg, Espen and Nilsen, {\O}ivind Anti and Vaage, Kjell},  journal={Industrial Relations: A Journal of Economy and Society},  volume={46},  number={1},  pages={112--129},  year={2007},  publisher={Wiley Online Library}}

@article{chernozhukov2022locally, title={Locally robust semiparametric estimation},  author={Chernozhukov, Victor and Escanciano, Juan Carlos and Ichimura, Hidehiko and Newey, Whitney K and Robins, James M},  journal={Econometrica},  volume={90},  number={4},  pages={1501--1535},  year={2022},  publisher={Wiley Online Library}}

@article{jenkins1987snapshots,  title={Snapshots versus movies:‘Lifecycle biases’ and the estimation of intergenerational earnings inheritance},  author={Jenkins, Stephen},  journal={European Economic Review},  volume={31},  number={5},  pages={1149--1158},  year={1987},  publisher={Elsevier}}

@article{blanden2014intergenerational,  title={Intergenerational Mobility in the {United States} and {Great Britain}: A Comparative Study of Parent--Child Pathways},  author={Blanden, Jo and Haveman, Robert and Smeeding, Timothy and Wilson, Kathryn},  journal={Review of Income and Wealth},  volume={60},number={3},  pages={425--449},  year={2014},  publisher={Wiley Online Library}}

@article{semenova2023inference,  title={Inference on heterogeneous treatment effects in high-dimensional dynamic panels under weak dependence},  author={Semenova, Vira and Goldman, Matt and Chernozhukov, Victor and Taddy, Matt},  journal={Quantitative Economics},  volume={14},  number={2},  pages={471--510},  year={2023},  publisher={Wiley Online Library}}

@article{chernozhukov2018double,  title={Double/debiased machine learning for treatment and structural parameters},  author={Chernozhukov, Victor and Chetverikov, Denis and Demirer, Mert and Duflo, Esther and Hansen, Christian and Newey, Whitney and Robins, James},  journal={The Econometrics Journal},  volume={21},  number={1},  pages={C1--C68},  year={2018},  publisher={Oxford University Press}}

@article{blundell2008consumption,  title={Consumption inequality and partial insurance},  author={Blundell, Richard and Pistaferri, Luigi and Preston, Ian},  journal={American Economic Review},  volume={98},  number={5},  pages={1887--1921},  year={2008},  publisher={American Economic Association}}

@article{wooldridge2013introductory,  title={Introductory econometrics: A modern approach 5th edition},  author={Wooldridge, Jeffrey M},  journal={Mason, OH: South-Western},  year={2013}}

@incollection{mogstad2024instrumental,  title={Instrumental variables with unobserved heterogeneity in treatment effects},   author={Mogstad, Magne and Torgovitsky, Alexander},  booktitle={Handbook of Labor Economics},  volume={5},  pages={1--114},  year={2024},  publisher={Elsevier}}

@article{schoeni2015implications,  title={The implications of selective attrition for estimates of intergenerational elasticity of family income},  author={Schoeni, Robert F and Wiemers, Emily E},  journal={The Journal of Economic Inequality},  volume={13},  number={3},  pages={351--372},  year={2015},  publisher={Springer}}

@article{mazumder2018intergenerational,  title={Intergenerational mobility in the United States: What we have learned from the PSID},  author={Mazumder, Bhashkar},  journal={The Annals of the American Academy of Political and Social Science},  volume={680},  number={1},  pages={213--234},  year={2018},  publisher={SAGE Publications Sage CA: Los Angeles, CA}}

@article{chau2012intergenerational, title={Intergenerational income mobility revisited: Estimation with an income dynamic model with heterogeneous age profile}, author={Chau, Tak Wai}, journal={Economics Letters}, volume={117}, number={3}, pages={770--773}, year={2012}, publisher={Elsevier}}

@article{lillard1998panel,  title={Panel attrition from the {Panel Study of Income Dynamics}: Household income, marital status, and mortality},  author={Lillard, Lee A and Panis, Constantijn WA},  journal={Journal of Human Resources},  pages={437--457},  year={1998},  publisher={JSTOR}}

@techreport{fitzgerald1998analysis, title={An analysis of sample attrition in panel data: {The Michigan Panel Study of Income Dynamics}},  author={Fitzgerald, John and Gottschalk, Peter and Moffitt, Robert A},  year={1998},  institution={National Bureau of Economic Research}}

@article{gouskova2010estimating,
  title={Estimating the intergenerational persistence of lifetime earnings with life course matching: Evidence from the PSID},
  author={Gouskova, Elena and Chiteji, Ngina and Stafford, Frank},
  journal={Labour Economics},
  volume={17},
  number={3},
  pages={592--597},
  year={2010},
  publisher={Elsevier}
}

@article{fitzgerald2011attrition,  title={Attrition in models of intergenerational links in health and economic status in the PSID},  author={Fitzgerald, John},  journal={The BE Journal of Economic Analysis \& Policy},  volume={11},  number={3},  pages={1--61},  year={2011}}

@article{corak2013income,
  title={Income inequality, equality of opportunity, and intergenerational mobility},
  author={Corak, Miles},
  journal={Journal of Economic Perspectives},
  volume={27},
  number={3},
  pages={79--102},
  year={2013},
  publisher={American Economic Association}
}

@article{black2011recent,
  title={Recent developments in intergenerational mobility},
  author={Black, Sandra E and Devereux, Paul J},
  journal={Handbook of labor economics},
  volume={4},
  pages={1487--1541},
  year={2011},
  publisher={Elsevier}
}

@techreport{heidrich2016study,  author       = {Heidrich, Stefanie},  title        = {A Study of the Missing Data Problem for Intergenerational Mobility Using Simulations},  institution  = {Umeå University, Department of Economics},  series       = {Umeå Economic Studies},  number       = {930},  year         = {2016}}

@article{francesconi2006intergenerational,  title={Intergenerational mobility and sample selection in short panels},  author={Francesconi, Marco and Nicoletti, Cheti},  journal={Journal of Applied Econometrics},  volume={21},  number={8},  pages={1265--1293},  year={2006},  publisher={Wiley Online Library}}

@article{mazumder2005fortunate,  title={Fortunate sons: New estimates of intergenerational mobility in the {United States} using social security earnings data},  author={Mazumder, Bhashkar},  journal={Review of Economics and Statistics},  volume={87},  number={2},  pages={235--255},  year={2005},  publisher={MIT Press 238 Main St., Suite 500, Cambridge, MA 02142-1046, USA journals}}

@article{mogstad2023family,  title={Family background, neighborhoods, and intergenerational mobility},  author={Mogstad, Magne and Torsvik, Gaute},  journal={Handbook of the Economics of the Family},  volume={1},  number={1},  pages={327--387},  year={2023},  publisher={Elsevier}}

@article{an2022nonparametric,  title={A nonparametric nonclassical measurement error approach to estimating intergenerational mobility elasticities},  author={An, Yonghong and Wang, Le and Xiao, Ruli},  journal={Journal of Business \& Economic Statistics},   volume={40},  number={1},  pages={169--185},  year={2022},  publisher={Taylor \& Francis}}

@article{lee2009trends,   title={Trends in intergenerational income mobility},  author={Lee, Chul-In and Solon, Gary},  journal={The review of economics and statistics},  volume={91},  number={4},  pages={766--772},  year={2009},  publisher={The MIT Press}}

@incollection{mazumder2016estimating,  title={Estimating the intergenerational elasticity and rank association in the {United States}: Overcoming the current limitations of tax data},  author={Mazumder, Bhashkar},  booktitle={Inequality: Causes and consequences},  pages={83--129},  year={2016},  publisher={Emerald group publishing limited}}

@article{lubotsky2006interpretation,  title={Interpretation of regressions with multiple proxies},  author={Lubotsky, Darren and Wittenberg, Martin},  journal={The Review of Economics and Statistics},  volume={88},  number={3},  pages={549--562},  year={2006},  publisher={The MIT Press}}

@article{hausman2001mismeasured,  title={Mismeasured variables in econometric analysis: problems from the right and problems from the left},  author={Hausman, Jerry},  journal={Journal of Economic perspectives},  volume={15},  number={4},  pages={57--67},  year={2001},  publisher={American Economic Association}}

@article{bjorklund1997intergenerational,  title={Intergenerational income mobility in {Sweden} compared to the {United States}},  author={Bj{\"o}rklund, Anders and J{\"a}ntti, Markus},  journal={The American Economic Review},  volume={87},  number={5},  pages={1009--1018},  year={1997},  publisher={JSTOR}}

@article{nybom2017biases,  title={Biases in standard measures of intergenerational income dependence},  author={Nybom, Martin and Stuhler, Jan},  journal={Journal of Human Resources},  volume={52},  number={3},  pages={800--825},  year={2017},  publisher={University of Wisconsin Press}}

@article{bohlmark2006life,  title={Life-cycle variations in the association between current and lifetime income: replication and extension for {Sweden}},  author={B{\"o}hlmark, Anders and Lindquist, Matthew J},  journal={Journal of Labor Economics},  volume={24},  number={4},  pages={879--896},  year={2006},  publisher={The University of Chicago Press}}

@techreport{freeman1978black,  title={Black economic progress after 1964: who has gained and why?},  author={Freeman, Richard B},  year={1978},  institution={National Bureau of Economic Research}}

@article{de1973relation,  title={The relation between income, intelligence, education and social background},  author={de Wolff, Peter and van Slijpe, Arnd RD}, journal={European Economic Review}, volume={4}, number={3},  pages={235--264},  year={1973},  publisher={Elsevier}}

@article{becker1986human,  title={Human capital and the rise and fall of families},  author={Becker, Gary S and Tomes, Nigel},  journal={Journal of labor economics},  volume={4},  number={3, Part 2},  pages={S1--S39},  year={1986},  publisher={University of Chicago Press}}

@article{solon1992intergenerational,  title={Intergenerational income mobility in the {United States}},  author={Solon, Gary},  journal={American Economic Review}, pages={393--408}, year={1992}, publisher={JSTOR}}

@incollection{friedman1957permanent,  title={The permanent income hypothesis},  author={Friedman, Milton},  booktitle={A theory of the consumption function},  pages={20--37},  year={1957},  publisher={Princeton University Press}}

@article{creedy1977distribution,
  title={The distribution of lifetime earnings},
  author={Creedy, John},
  journal={Oxford Economic Papers},
  volume={29},
  number={3},
  pages={412--429},
  year={1977},
  publisher={Oxford University Press}
}

@techreport{hauser1975socioeconomic,
  title={Socioeconomic background and returns to education},
  author={Hauser, Robert M},
  year={1972},
  type={Working Paper},
  number={72-31},
  institution={Center for Demography and Ecology, University of Wisconsin--Madison},
  month={November}
}

@book{tsai1983sex,  title={Sex differences in the process of stratification},  author={Tsai, Shu-Ling},  year={1983},  publisher={The University of Wisconsin-Madison}}

@article{haider2006life,  title={Life-cycle variation in the association between current and lifetime earnings},  author={Haider, Steven and Solon, Gary},  journal={American Economic Review},  volume={96},  number={4},  pages={1308--1320},  year={2006},  publisher={American Economic Association}}
\appendix\section{Appendix A}\label{sec:proofs}  \addcontentsline{toc}{section}{Appendices} \renewcommand{\thesection}{\Alph{section}}\numberwithin{equation}{section}
\renewcommand{\theequation}{\thesection\arabic{equation}}
\subsection{The Mid-life Income Estimator}\label{sec:mi}
The standard approach to estimating intergenerational elasticity, which we label the mid-life income (MI) estimator, proxies permanent income by averaging (log) annual income snapshots around mid-life, primarily for fathers, though also applicable to children. This strategy is motivated by the errors-in-variables (EIV) framework, building on the permanent income hypothesis of \cite{friedman1957permanent}, which posits that observed income depends on a permanent and a transitory component. By averaging multiple years of income, the MI estimator isolates the permanent component, reducing the impact of transitory fluctuations on intergenerational elasticity 
estimates.\par
Following the seminal work of \cite{solon1992intergenerational}, this approach has become standard for measuring fathers’ permanent income, with researchers using a simple average of (log) yearly income. Solon’s key contribution was showing that averaging multiple income snapshots reduces attenuation bias, with the bias decreasing as the number of periods averaged increases. However, as noted by \cite{becker1986human}, earlier studies \citep{de1973relation,hauser1975socioeconomic,freeman1978black,tsai1983sex} had already employed income averaging to mitigate response errors and transitory components, laying the groundwork for this practice.\par 
Using mid-life observations to proxy permanent income is rationalized by the generalized error-in-variables (GEIV) model \citep{haider2006life}
\begin{align}\label{eq:geiv}
Y_{gt} &= \lambda_t Y_g^P + v_{gt}, \quad \mathbb{E}\left[v_{gt} Y_g^P\right] = 0, \quad g \in {c, f}, \quad t = 1, \ldots, T,
\end{align}
where $Y_g^P$ is permanent income, $\lambda_t$ captures varying persistence over the life cycle, and $v_{gt}$ is an age-specific shock. As suggested by equation (\ref{eq:geiv}), annual income at younger and older ages is a noisier measure of permanent income compared to mid-life income, as permanent income is less persistent during these periods (indicated by a smaller $\lambda_t$). This phenomenon, known as life-cycle bias, can be mitigated by measuring income during mid-life, when persistence approaches one \citep{haider2006life}. Extensive evidence supports this approach, demonstrating that mid-life income yields more accurate estimates of permanent income for fathers (e.g., \cite{bohlmark2006life,nybom2017biases}). \par  For children’s permanent income, standard practice uses a single mid-life observation, as measurement error in the dependent variable (usually) only affects efficiency, while error in the independent variable causes attenuation bias \citep{hausman2001mismeasured}.\par 
Formally, the MI estimand $\left(\beta^{MI}\right)$ corresponds to the slope coefficient of the projection of the average child's (log) income during mid-life  $\left(\tilde{Y}_c^P\right)$ on the average parental (log) income during mid-life $\left(\tilde{Y}_f^P\right)$:
\begin{align}\label{eq:mi0}
\tilde{Y}_c^P&=\alpha^{MI}+\beta^{MI} \tilde{Y}_f^P+u^{MI}, \quad \mathbb{E}\left[u^{MI} \left(1, \tilde{Y}_f^P\right)'\right]=0,\\\nonumber
   \tilde{Y}_g^P&\coloneqq\frac{1}{T_g}\sum_{j\in \mathcal{M}_g}Y_{gj}D_{gj}, \quad g\in \{c,f\},
\end{align} 
where $D_{gj}=1$ when $Y_{gj}$ is observed and zero otherwise,\footnote{
While $D_{gj} = 1$ is formally defined as indicating when $Y_{gj}$ is observed, in practice, it also implicitly requires that $Y_{gj}$ is used for estimation. This distinction arises because empirical studies often sample parental income selectively—for example, by focusing on log annual income during midlife (e.g., ages 30--50) to reduce lifecycle bias or measurement error. Thus, even if income is observed in other years, it may be excluded from estimation due to sampling design. This refinement clarifies that $D_{gj}$ reflects both data availability and inclusion criteria, ensuring consistency with standard empirical approaches.
} $\mathcal{M}_g$ is a set of pre-defined mid-life years for generation $g$,\footnote{While some papers define parental mid-life according to their offspring's age \citep{chetty2014land,blanden2014intergenerational}, others use parental age \citep{bjorklund1997intergenerational,mazumder2005fortunate}, so
$\mathcal{M}_f$ can differ from $\mathcal{M}_c$.}  and $T_{g}\coloneqq\sum_{j\in \mathcal{M}_g}D_{gj}$ is the number of years used for the average. Accordingly, the closed-form expression for the MI estimand

\begin{align}\label{eq:phi}   
\beta^{MI}&=\frac{\mathbb{E}\left[\left(\tilde{Y}_c^P-\mathbb{E}\left[\tilde{Y}_c^P\right]\right)\left(\tilde{Y}_f^P-\mathbb{E}\left[\tilde{Y}_f^P\right]\right)\right]}{\mathbb{E}\left[\left(\tilde{Y}_f^P-\mathbb{E}\left[\tilde{Y}_f^P\right]\right)^2\right]}\coloneqq\frac{\mathbb{E}\left[\tilde{y}_c^P\tilde{y}_f^P\right]}{\mathbb{E}\left[\left(\tilde{y}_f^P\right)^2\right]},
\end{align} 
where low-case letters denote the random variables in deviations from their population mean. Thus, the corresponding MI estimator is given by
\begin{align}
\label{eq:mi_estimator}
   \hat{\beta}_n^{MI}&=\frac{\mathbb{E}_n\left[\left(\tilde{Y}_c^P-\mathbb{E}_n\left[\tilde{Y}_c^P\right]\right)\left(\tilde{Y}_f^P-\mathbb{E}_n\left[\tilde{Y}_f^P\right]\right)\right]}{\mathbb{E}_n\left[\left(\tilde{Y}_f^P-\mathbb{E}_n\left[\tilde{Y}_f^P\right]\right)^2\right]},
    \end{align}
where $\mathbb{E}_n\left[X\right]\coloneqq\frac{1}{n}\sum_{i=1}^nX_i$ is the empirical expectation operator.
\par 
To study identification of the MI estimand, we now state its underlying assumptions. The first one begins by imposing that the GEIV model of equation (\ref{eq:geiv}), which motivates the MI estimator, is correctly specified and assumes that persistency of permanent income equals 1 during mid-life. This assumption serves the crucial function of mapping unobserved lifetime income to (partially) observed (log) annual income.  
\begin{assumption1}{1}{MI}(Annual Income Process)\label{as:1_app}
The relationship between annual and permanent income is governed by \begin{align*}  
Y_{gt}&=\lambda_tY^P_g+v_{gt}, \quad \mathbb{E}\left[v_{gt} Y^P_g\right]=0, \quad g\in \{c,f\},  \quad t=1,...,T,\\  \lambda_t&=1, \forall t\in \mathcal{M}_g, \quad g\in \{c,f\}.
\end{align*} 
where $\lambda_t$ captures that the persistence of permanent income may vary over the life-cycle period, and $v_{gt}$ is an age shock.
\end{assumption1}
While empirical evidence suggests that $\lambda_t$ approaches one during mid-life \citep{haider2006life,nybom2016heterogeneous}, the assumption that $\lambda_t$ equals one in this period is unlikely to hold. This motivates the use of optimally weighted income measures \citep{lubotsky2006interpretation}, which provide more accurate estimates than simple averages. Crucially, even if $\lambda_t=1$ during mid-life, the MI estimator remains inconsistent for the IGE\citep{nybom2016heterogeneous}. \par
The model in equation (\ref{eq:mi0}), together with Assumption \ref{as:1_app}, involves a random i.i.d sample of $\tilde{W} = \left(\bm Y_{c} \odot D_{c}, \bm Y_{f} \odot D_{f}, \bm D_{c}, \bm D_{f}\right)$,\footnote{Because the MI estimator does not incorporate family characteristics in its estimation procedure, we abstract from their observation in our analysis.}  unobserved components including: (i) the time-varying shocks $v_{gt}$ for both children and fathers, indexed by $g \in {c,f}$ and $t = 1, \dots, T$; (ii) the permanent income $Y_g^P$ of both generations $g \in {c,f}$; and (iii) the error term $u^{MI}$ associated with the MI estimator. The model parameters consist of the parameter of interest $\beta^{MI}$, and the nuisance parameter $\alpha^{MI}$.\footnote{While $\left\{\lambda_{t}\right\}_{t=1}^T$ would typically be nuisance parameters in an unrestricted model, our framework does not classify them as such, as we impose the restriction \( \lambda_t = 1 \) during mid-life.} To evaluate whether $\beta^{MI}$ identifies $\beta_0$, we now introduce zero conditional mean restrictions involving the observed and unobserved components in this setting. 
\begin{assumption1}{2}{MI}(Conditional Mean Independence)
\label{as:2mi_app} 
The following  conditional mean restrictions hold
\begin{align*}    
\mathbb{E}\left[v_{ct}v_{fj}\big | D_{ct},D_{fj}\right]&=0, \quad t\in \mathcal{M}_c, \quad j\in \mathcal{M}_f,\\ 
\mathbb{E}\left[v_{fj}Y_c^P\big | D_{fj}\right]&=0, \quad  j\in \mathcal{M}_f, \\
\mathbb{E}\left[v_{ft}Y_f^P | D_{ft}, D_{fj}\right]&=0, \quad tj \in \mathcal{M}_f, \\
\mathbb{E}\left[v_{gj} | D_{gj}\right]&=0, \quad g\in \{c,f\}, \quad j \in \mathcal{M}_j.
\end{align*}
\end{assumption1}
 Assumption \ref{as:2mi_app} imposes a set of orthogonality conditions involving age shocks, missingness indicators, and unobserved permanent income. The first condition states that, conditional on the missingness indicators for child and parent income, mid-life age shocks to children and parents are mean independent. The second condition requires that parental age shocks during mid-life are mean independent of the child’s permanent income, conditional on the missingness status of parental income. The third condition assumes that, given the missingness indicators for a tuple of years of parental income, age shocks are mean independent of the parent’s permanent income.\footnote{By the law of iterated expectations, this condition implies that age shocks are uncorrelated with permanent income for the parental generation, as already assumed in Assumption \ref{as:1_app}, but the reverse does not necessarily hold.} Lastly, the fourth condition states that observed age shocks have zero mean, conditional on the missingness status of the corresponding annual income observation.\par
Previous work (e.g., \cite{couch1998sample,mazumder2005fortunate,heidrich2016study}) has shown that the MI estimator does not perform well. Furthermore, there is an extensive literature discussing its sources of bias (e.g., \cite{solon1992intergenerational, mazumder2005fortunate,nybom2016heterogeneous}). We now briefly examine the sources of bias that prevent the MI estimator from being consistent. As shown in Corollary \ref{coro:1}, the probability limit of $\hat{\beta}_n^{MI}$ takes the form:
\begin{gather}\label{eq:mi_inconsistenti}   
\hat{\beta}_n^{MI}\overset{p}{\to}\frac{\beta_0\mathbb{E}\left[\left(Y_f^P-\mathbb{E}\left[Y_f^P\right]\right)^2\right]+\overbrace{\frac{1}{T_c}\sum_{t} \mathbb{E}\left[Y_f^Pv_{ct}\big  | D_{ct}=1,t\in \mathcal{M}_c\right]}^\text{(c)}\times \overbrace{p_c\left(t\in \mathcal{M}_c\right)}^\text{(d)}}{\mathbb{E}\left[\left(Y_f^P-\mathbb{E}\left[Y_f^P\right]\right)^2\right]+\underbrace{\frac{1}{T_f^2}\sum_{t}\sum_{j}\mathbb{E}\left[v_{ft}v_{fj}\big | D_{ft}=1,D_{fj}=1, \left\{t,j\right\}\in \mathcal{M}_f\right]}_\text{(a)}\times  \underbrace{p_f\left(\left\{t,j\right\}\in \mathcal{M}_f\right)}_\text{(b)}}.
\end{gather} 
The downward bias by measurement error highlighted by \cite{solon1992intergenerational} and \cite{mazumder2005fortunate} corresponds to component (a) in equation (\ref{eq:mi_inconsistenti}). Consider rewriting (a) as 
\begin{gather}
  \frac{1}{T_f^2}\sum_{t}\mathbb{E}\left[v_{ft}^2\big | D_{ft}=1,D_{fj}=1, \left\{t,j\right\}\in \mathcal{M}_f\right]+  \frac{2}{T_f^2}\sum_{t}\sum_{j\neq t}\mathbb{E}\left[v_{ft}v_{fj}\big | D_{ft}=1,D_{fj}=1, \left\{t,j\right\}\in \mathcal{M}_f\right],
\end{gather}
 where the first component is the variance of the transitory income component, causing the attenuation bias shown is \cite{solon1992intergenerational}, while the second term comprises the autoregressive nature of the transitory component illustrated in \cite{mazumder2005fortunate}. This term rationalizes why even the 10-year average is not enough for the attenuation bias to vanish due to the transitory component of income being highly serially correlated \citep{mazumder2005fortunate}. As $T_f$ grows (more years are used for the average), the first component in the last display might vanish. However, the second one does not, because the number of covariances in the second term is $T_f^2-T_f$. Consequently, the second term in the last display will not disappear when the transitory income component is highly serially correlated.
\par

Component (b) captures the sensitivity of the IGE estimates to low, zero, and missing income documented in the literature. There is extensive evidence that IGE estimates are not robust to how extreme and missing incomes are treated \citep{couch1998sample,dahl2008association,chetty2014land,nybom2016heterogeneous}. When observations with zero and low income are dropped, the probability of observing a given tuple of years for parents changes. Moreover, not observing permanent income also affects the probability of observing a given tuple of years. If we were to observe permanent income, components (d) and (b) in equation (\ref{eq:mi_inconsistenti}) would be equal to 1, and would not induce bias.\par 
The sensitivity of the IGE to sample inclusion rules is also comprised in component (d). \cite{couch1998sample} show with empirical evidence that the MI estimator is sensitive to different sample inclusion rules. As noted by \cite{francesconi2006intergenerational}, studies usually restrict their analysis to children from specific birth cohorts. The upper bound for the birth cohort is required to ensure that children’s socioeconomic status is observed as long as possible so their observed status is a reliable measure of long-run permanent status. Imposing such a restriction mechanically affects the probability of observing income in a given year, which corresponds to component (b) in equation (\ref{eq:mi_inconsistenti}).\par
Even when using the same data, changes in the definition of mid-life alter the estimand in equation (\ref{eq:phi}), further limiting comparability. The transitory component of children's income depending on parental income is encompassed by (c) in equation (\ref{eq:mi_inconsistenti}). \cite{halvorsen2022earnings} highlight that children from affluent families might experience faster income growth, which would cause the steepness of the income trajectory to depend on parental permanent income. Thus, the correlation between age shocks to children's (log) annual income and parental permanent income induces bias in estimating the IGE.\par 
Component (d) in equation (\ref{eq:mi_inconsistenti}) captures that the estimate of the IGE depends both on the number of years used to measure children's income and the selected year(s). \cite{mello2022lifecycle} provide evidence that using $\hat{\beta}_n^{MI}$ to estimate the IGE is sensitive to the span of ages where the child generation is observed and the number of income observations available for each individual. The first finding is captured by the pre-defined mid-life years $(\mathcal{M}_c)$ used to proxy children's permanent income. The second one, by the cardinality of $\mathcal{M}_c$ affecting the magnitude of component (d). \par 
As shown in equation (\ref{eq:ns}), if we were to drop the assumption that $\lambda_t=1$ during mid-life, our inconsistency result would also capture the life-cycle bias in estimating the IGE. As previously mentioned, imposing $\lambda_t=1$ allows us to obtain the closed-form solution in equation (\ref{eq:mi_inconsistenti}). However, relaxing this assumption allows our inconsistency result to capture another source of bias discussed in the literature.
\subsection{Equivalence with Previous Inconsistency Results}\label{sec:equi}
Corollary \ref{coro:1} encompasses previous formalizations of bias in estimating the IGE. In particular, we first show that the inconsistency result in \cite{solon1992intergenerational} is a particular case of Corollary \ref{coro:1} under an additional assumption. We then show that the two inconsistency results in \cite{nybom2016heterogeneous} are particular cases of Corollary \ref{coro:1} when variants of Assumption \ref{as:1_app} are considered, and Assumption \ref{as:2mi_app} is relaxed.\par 
We now show that equation (\ref{eq:mi_inconsistent}) in Corollary  \ref{coro:1} simplifies to the inconsistency result in \cite[p. 400]{solon1992intergenerational}, if we further assume that parental permanent income is uncorrelated to child age shocks, for the observed years during mid-life. In particular, consider assuming 
\begin{assumption1}{3}{S}\label{as:23}  
\begin{align*}     
\mathbb{E}\left[Y^P_f v_{ct} | D_{ct}=1, t \in \mathcal{M}_c\right]&=0\\     D_{ft}&=1 \quad \forall t \in \mathcal{M}_f,
\end{align*}  
\end{assumption1}  so that 
\begin{align*}   
\mathbb{E}\left[v_{ft}v_{fj}\big | D_{ft}=1,D_{fj}=1, \left\{t,j\right\}\in \mathcal{M}_f\right]\times  p_f\left(\left\{t,j\right\}\in \mathcal{M}_f\right)&= \mathbb{E}\left[v_{ft}v_{fj}\right]\times 1. \end{align*}
Then, under Assumptions \ref{as:1_app}, \ref{as:2mi_app},and \ref{as:23} equation  (\ref{eq:a21}) (which corresponds to equation (\ref{eq:mi_inconsistent}) in Corollary \ref{coro:1}) boils down to
\begin{align}\label{eq:equiV_{cf}}
\hat{\beta}_n^{MI}\overset{p}{\to}\frac{\beta_0\mathbb{E}\left[\left(Y_f^P-\mathbb{E}\left[Y_f^P\right]\right)^2\right]}{\mathbb{E}\left[\left(Y_f^P-\mathbb{E}\left[Y_f^P\right]\right)^2\right]+\frac{1}{T_f^2}\sum_{t}\sum_{j}\mathbb{E}\left[v_{ft}v_{fj} \right]},
\end{align}  
which is the result in \cite{solon1992intergenerational}. The shape of the second term in the denominator of the last display depends on the assumptions of the transitory shock to parental income. For instance, if we assume that is white noise, it boils down to $V^2_v/T_f$. Conversely, if we assume it follows a MA(1) it becomes $\left(V^2_v/T_f\right)\times \left[1+2\theta\left(T_f-1\right)/T_f\right]$, where $\theta$ denotes the first-order autocorrelation. Finally, if we assume a stationary AR(1) process the second term in the denominator of the last display becomes $\left(V^2_v/T_f\right)\times \left[1+2\theta\left\{T_f-(1-\theta^T_f)/(1-\theta)\right\}/\left(T_f[1-\theta]\right)\right]$ (see footnote 17 in \cite{solon1992intergenerational}). We now turn to contrasting Corollary \ref{coro:1} with more recent results. \par  
There are two inconsistency results in \cite{nybom2016heterogeneous}. Both of them impose the additional assumption that income is measured in a single year during mid-life. However, while the first result assumes an error-in-variables model,  the second one assumes a generalized error-in-variables model. By formalizing these assumptions, we show that Corollary \ref{coro:1} encompasses both of these inconsistency results.\par 
We first consider the inconsistency result in equation (2) in \cite{nybom2016heterogeneous}. For this purpose, consider the following alternative to Assumption \ref{as:1_app} \begin{assumption1}{1}{NS}(Annual Income Process)\label{as:22} The relationship between annual and permanent income is governed by 
\begin{align*}  
Y_{gt}&=\lambda_tY^P_g+v_{gt}, \quad \mathbb{E}\left[v_{gt}\right]=0, \quad g\in \{c,f\},  \quad t=1,...,T,\\  \lambda_t&=1, \forall t\in \mathcal{M}_g, \quad g\in \{c,f\}. 
\end{align*}  
We assume 
\begin{align*}
\mathcal{M}_g&=T_g, \quad g\in \{c,f\}\\
D_{gt}&=1 \quad \forall t=T_g, \quad g\in \{c,f\}. 
\end{align*}  
\end{assumption1} 
That is, we  assume that parental and child's income are measured in a given year so that every child and parent is observed in that year. Moreover, we assume that the age shock to annual income has zero mean, and relax the assumption that transitory income shocks are uncorrelated to parental permanent income. Furthermore, in their result, the authors relax the conditional mean restrictions in Assumption \ref{as:2mi_app}.\par 
Since we are only using one observation for both parents and children, we have $\tilde{Y}_c^P=Y_{ct}$ and $\tilde{Y}_f^P=Y_{fj}$, so that 
\begin{align}\label{eq:phi_p_new} 
\mathbb{E}\left[\tilde{y}_c^P\tilde{y}_f^P\right]&=\mathbb{E}\left[Y_{ct}Y_{fj}\right]-\mathbb{E}\left[Y_{ct}\right]\mathbb{E}\left[Y_{fj}\right]\nonumber\\ 
&=\mathbb{E}\left[\left(Y_c^P+v_{ct}\right)\left(Y_f^P+v_{fj}\right)\right]-\mathbb{E}\left[\left(Y_c^P+v_{ct}\right)\right]\mathbb{E}\left[\left(Y_f^P+v_{fj}\right)\right]\nonumber\\    
&=\mathbb{E}\left[Y_c^PY_f^P\right]+    \mathbb{E}\left[Y_f^Pv_{ct}\right]+\mathbb{E}\left[Y_c^Pv_{fj}\right]+\mathbb{E}\left[v_{ct}v_{fj}\right]-\mathbb{E}\left[Y_c^P\right]\mathbb{E}\left[Y_f^P\right]\nonumber\\ 
&=\mathbb{E}\left[\left(Y_c^P-\mathbb{E}\left[Y_c^P\right]\right)\left(Y_f^P-\mathbb{E}\left[Y_f^P\right]\right)\right]+\mathbb{E}\left[Y_f^Pv_{ct}\right]+\mathbb{E}\left[Y_c^Pv_{fj}\right]+\mathbb{E}\left[v_{ct}v_{fj}\right]\nonumber\\    
&=\beta_0\mathbb{E}\left[\left(Y_f^P-\mathbb{E}\left[Y_f^P\right]\right)^2\right]+    \mathbb{E}\left[Y_f^Pv_{ct}\right]+\mathbb{E}\left[Y_c^Pv_{fj}\right]+\mathbb{E}\left[v_{ct}v_{fj}\right],
\end{align}  
and 
\begin{align}\label{eq:new1} 
\mathbb{E}\left[\left(\tilde{y}_f^P\right)^2\right]&=\mathbb{E}\left[\left(Y_{ft}\right)^2\right]-\mathbb{E}\left[Y_{ft}\right]^2\nonumber\\ 
&=\mathbb{E}\left[\left(Y_f^P+v_{ft}\right)^2\right]-\mathbb{E}\left[\left(Y_f^P+v_{ft}\right)\right]^2\nonumber\\ 
&=\mathbb{E}\left[\left(Y_f^P\right)^2\right]+\mathbb{E}\left[\left(v_{ft}\right)^2\right]+2\mathbb{E}\left[Y_f^Pv_{ft}\right]-\mathbb{E}\left[Y_f^P\right]^2\nonumber\\  
&=\mathbb{E}\left[\left(Y_f^P-\mathbb{E}\left[Y_f^P\right]\right)^2\right]+\mathbb{E}\left[v_{ft}^2\right]+2\mathbb{E}\left[Y_f^Pv_{ft}\right]
\end{align} 
Then, by equation (\ref{eq:phi_p_new})

\begin{align*}       \mathbb{E}_n\left[\tilde{y}_c^P\tilde{y}_f^P\right]&\overset{p}{\to}\mathbb{E}\left[\tilde{y}_c^P\tilde{y}_f^P\right]\nonumber \\
&=\beta_0\mathbb{E}\left[\left(Y_f^P-\mathbb{E}\left[Y_f^P\right]\right)^2\right]+    \mathbb{E}\left[Y_f^Pv_{ct}\right]+\mathbb{E}\left[Y_c^Pv_{fj}\right]+\mathbb{E}\left[v_{ct}v_{fj}\right],
\end{align*} 
and by equation (\ref{eq:new1})
\begin{align*}
\mathbb{E}_n\left[\left(\tilde{y}_f^P\right)^2\right]&\overset{p}{\to}\mathbb{E}\left[\left(\tilde{y}_f^P\right)^2\right] \nonumber\\
&=\mathbb{E}\left[\left(Y_f^P-\mathbb{E}\left[Y_f^P\right]\right)^2\right]+\mathbb{E}\left[v_{ft}^2\right]+2\mathbb{E}\left[Y_f^Pv_{ft}\right],
\end{align*} 
so that, under Assumption \ref{as:22}, equation  (\ref{eq:a21})  boils down to  
\begin{gather}\label{eq:ns1}
\hat{\beta}_n^{MI}\overset{p}{\to}\frac{\beta_0\mathbb{E}\left[\left(Y_f^P-\mathbb{E}\left[Y_f^P\right]\right)^2\right]+    \mathbb{E}\left[Y_f^Pv_{ct}\right]+\mathbb{E}\left[Y_c^Pv_{fj}\right]+\mathbb{E}\left[v_{ct}v_{fj}\right]}{\mathbb{E}\left[\left(Y_f^P-\mathbb{E}\left[Y_f^P\right]\right)^2\right]+\mathbb{E}\left[v_{ft}^2\right]+2\mathbb{E}\left[Y_f^Pv_{ft}\right]},
\end{gather} 
which is equation (2) in \cite{nybom2016heterogeneous}.\par
We now turn to the second inconsistency result, which in contrast to Assumption \ref{as:1_app}, assumes 
\begin{assumption1}{1'}{NS}(Annual Income Process)\label{as:253} 
The relationship between annual and permanent income is governed by \begin{align*}  
Y_{gt}&=\lambda_tY^P_g+v_{gt}, \quad \mathbb{E}\left[v_{gt}\left(1, Y^P_g\right)'\right]=0, \quad g\in \{c,f\},  \quad t=1,...,T.
\end{align*} 
where $\lambda_t$ captures that the persistence of permanent income may vary over the life-cycle period, and $v_{gt}$ is an age shock, uncorrelated by construction with $Y^P_g$. We assume $\mathcal{M}_c=t_c$, $\mathcal{M}_f=t_f$,  $D_{ct}=1$ for $t=t_c$, and $D_{ft}=1$ for $t=t_f$. 
\end{assumption1}
That is, similar to Assumption \ref{as:1_app}, we consider the linear projection of $Y^P_g$ on $Y_{gt}$ so that $v_{gt}$ is uncorrelated to $Y^P_g$ by construction. Moreover, we relax the assumption of $ \lambda_t=1, \forall t\in \mathcal{M}_g,  g\in \{c,f\}$ in Assumption \ref{as:1_app}. Thus, we have that   
\begin{align*}
\mathbb{E}\left[\tilde{y}_c^P\tilde{y}_f^P\right]&=\mathbb{E}\left[Y_{ct}Y_{fj}\right]-\mathbb{E}\left[Y_{ct}\right]\mathbb{E}\left[Y_{fj}\right]\nonumber\\
&=\mathbb{E}\left[\left(\lambda_{t}Y_c^P+v_{ct}\right)\left(\lambda_{j}Y_f^P+v_{fj}\right)\right]-\mathbb{E}\left[\left(\lambda_{t}Y_c^P+v_{ct}\right)\right]\mathbb{E}\left[\left(\lambda_{j}Y_f^P+v_{fj}\right)\right]\nonumber\\   
&=\lambda_t\lambda_j\mathbb{E}\left[Y_c^PY_f^P \right]+\lambda_j\mathbb{E}\left[Y_f^Pv_{ct}\right]+\lambda_t\mathbb{E}\left[Y_c^Pv_{fj}\right]+\mathbb{E}\left[v_{ct}v_{fj}\right]-\lambda_t\mathbb{E}\left[Y_c^P\right]\lambda_j\mathbb{E}\left[Y_f^P\right]\nonumber\\   
&=\lambda_t\lambda_j\mathbb{E}\left[\left(Y_c^P-\mathbb{E}\left[Y_c^P\right]\right)\left(Y_f^P-\mathbb{E}\left[Y_f^P\right]\right)\right]+\lambda_j\mathbb{E}\left[Y_f^Pv_{ct}\right]+\lambda_t\mathbb{E}\left[Y_c^Pv_{fj}\right]+\mathbb{E}\left[v_{ct}v_{fj}\right]   \nonumber\\   
&=\lambda_t\lambda_j\mathbb{E}\left[\left(\beta_0\left(Y_f^P-\mathbb{E}\left[Y_f^P\right]\right)+u\right)\left(Y_f^P-\mathbb{E}\left[Y_f^P\right]\right)\right]\nonumber\\
&+\lambda_j\mathbb{E}\left[Y_f^Pv_{ct}\right]+\lambda_t\mathbb{E}\left[Y_c^Pv_{fj}\right]+\mathbb{E}\left[v_{ct}v_{fj}\right]\nonumber\\  
&=\beta_0\lambda_t\lambda_j\mathbb{E}\left[\left(Y_f^P-\mathbb{E}\left[Y_f^P\right]\right)^2\right]+\lambda_j\mathbb{E}\left[Y_f^Pv_{ct}\right]+\lambda_t\mathbb{E}\left[Y_c^Pv_{fj}\right]+\mathbb{E}\left[v_{ct}v_{fj}\right],
\end{align*} 
and 
\begin{align*}     \mathbb{E}\left[\left(\tilde{y}_f^P\right)^2\right] 
&=\mathbb{E}\left[\left(\lambda_{j}Y_f^P+v_{fj}\right)^2\right]-\mathbb{E}\left[\left(\lambda_{j}Y_f^P+v_{fj}\right)\right]^2\nonumber\\   
&=\lambda_{j}^2\mathbb{E}\left[\left(Y_f^P\right)^2\right]+2\lambda_j\mathbb{E}\left[Y_f^Pv_{fj}\right]+\mathbb{E}\left[v_{fj}^2\right]-\lambda_{j}^2\mathbb{E}\left[Y_f^P\right]^2\\   
&=\lambda_{j}^2\mathbb{E}\left[\left(Y_f^P-\mathbb{E}\left[Y_f^P\right]\right)^2\right]+\mathbb{E}\left[v_{fj}^2\right]   \end{align*} 
so that
 
\begin{align*} 
\mathbb{E}_n\left[\tilde{Y}_c^P\tilde{Y}_f^P\right]&\overset{p}{\to}\mathbb{E}\left[\tilde{Y}_c^P\tilde{Y}_f^P\right]\nonumber \\
&=\beta_0\lambda_t\lambda_j\mathbb{E}\left[\left(Y_f^P-\mathbb{E}\left[Y_f^P\right]\right)^2\right]+\lambda_j\mathbb{E}\left[Y_f^Pv_{ct}\right]+\lambda_t\mathbb{E}\left[Y_c^Pv_{fj}\right]+\mathbb{E}\left[v_{ct}v_{fj}\right],    
\end{align*} 
and
  \begin{align*}
\mathbb{E}_n\left[\left(\tilde{Y}_f^P\right)^2\right]&\overset{p}{\to}\mathbb{E}\left[\left(\tilde{Y}_f^P\right)^2\right] \nonumber\\
&=\lambda_{j}^2\mathbb{E}\left[\left(Y_f^P-\mathbb{E}\left[Y_f^P\right]\right)^2\right]+\mathbb{E}\left[v_{fj}^2\right].
\end{align*}  
Thus, under Assumption  \ref{as:253} we have that equation  (\ref{eq:a21})  becomes
\begin{gather}\label{eq:ns}
\hat{\beta}_n^{MI}\overset{p}{\to}\frac{\beta_0\lambda_t\lambda_j\mathbb{E}\left[\left(Y_f^P-\mathbb{E}\left[Y_f^P\right]\right)^2\right]+\lambda_j\mathbb{E}\left[Y_f^Pv_{ct}\right]+\lambda_t\mathbb{E}\left[Y_c^Pv_{fj}\right]+\mathbb{E}\left[v_{ct}v_{fj}\right]}{\lambda_{j}^2\mathbb{E}\left[\left(Y_f^P-\mathbb{E}\left[Y_f^P\right]\right)^2\right]+\mathbb{E}\left[v_{fj}^2\right]},
\end{gather} 
which is equation (6) in \cite{nybom2016heterogeneous}.
\subsection{Definition of Permanent Income}\label{sec:lfinc}
Permanent income is defined as the average log annual income over a specific lifetime period from $t=1$ to T:
\begin{align}\label{eq:ap_def}
  Y^P_g&\coloneqq\frac{1}{T}\sum_{t=1}^TY_{gt}, \quad g\in\{c,f\},
\end{align}
where $Y_{gt}$ is log annual income in year $t$, with $t=1$ indicating the start age and $T$ the number of years covered.
A key advantage of this formulation is that it facilitates the identification of the Intergenerational Elasticity (IGE). Specifically, under this definition, permanent income depends only on the marginal distributions of log annual income, which can be partially observed and consistently estimated from the data. This property is particularly valuable in the presence of missing data, where income is not available for all individuals or years. The linearity of $Y_g^P$ allows for the interchange of the summation and expectation operators, which is crucial for identification.\par
To illustrate this point, notice that under standard missing at random (MAR) and conditional mean independence assumptions, we can recover $\mathbb{E}\left[Y_g^P\right]$ using observed conditional means:
\begin{align*}
\mathbb{E}\left[Y_g^P\right]&=\mathbb{E}\left[\sum_{t=1}^TY_{gt}\right]\nonumber\\
 &=\sum_{t=1}^T\mathbb{E}\left[Y_{gt}\right]\nonumber\\
 &=\sum_{t=1}^T\mathbb{E}\left[\mathbb{E}\left[Y_{gt}\mid \bm X_{gt}\right]\right]\nonumber\\
 &=\sum_{t=1}^T\mathbb{E}\left[\mathbb{E}\left[Y_{gt}\mid \bm X_{gt}, D_{gt}=1\right]\right]\nonumber\\
 &=\mathbb{E}\left[\sum_{t=1}^T\mathbb{E}\left[Y_{gt}\mid \bm X_{gt}, D_{gt}=1\right]\right],
\end{align*} 
where $\bm{X}_{gt}$ represents family characteristics predictive of annual income for generation $g$ at time $t$, and $D_{gt}$ is an indicator equal to 1 if $Y_{gt}$ is observed and 0 otherwise. The MAR assumption ensures that $\mathbb{E}\left[Y_{gt} \mid \bm{X}{gt}\right] = \mathbb{E}\left[Y{gt} \mid \bm{X}{gt}, D{gt}=1\right]$, allowing us to impute missing values using observed data.\par 
In contrast, consider an alternative definition of permanent income based on the log of average absolute income:  
\begin{align*}
    Y_g^{PL} \coloneqq \log\left(\frac{1}{T}\sum_{t=1}^T e^{Y_{gt}}\right),
\end{align*}  
where $e^{Y_{gt}}$ denotes absolute annual income. This formulation complicates identification of the IGE because the nonlinearity introduced by the logarithm prevents expectations from decomposing into period-by-period components. By Jensen's inequality, we have  
\begin{align*}
\mathbb{E}\left[Y_g^{PL}\right] 
= \mathbb{E}\!\left[\log\!\left(\tfrac{1}{T}\sum_{t=1}^T e^{Y_{gt}}\right)\right] 
\;\neq\; \tfrac{1}{T}\sum_{t=1}^T \mathbb{E}\left[Y_{gt}\right] 
= \mathbb{E}\left[Y_g^{P}\right].
\end{align*}  
Thus, the left-hand side requires taking the expectation over the full (unobserved) joint distribution of the vector $\left(Y_{g1}, \dots, Y_{gT}\right)$, which captures all dependencies across time periods. Accordingly, the expectation $\mathbb{E}\left[\log\left(\sum_t e^{Y_{gt}}\right)\right]$ cannot be reduced to the mean of conditional expectations of individual $Y_{gt}$. Instead, it requires knowledge (or estimation) of the entire joint distribution of $\left(Y_{g1}, \dots, Y_{gT}\right)$ to account for correlations and higher-order moments across periods. Under missing data, this would necessitate stronger assumptions about the joint distribution and potentially complex imputation methods for the full vector of incomes, rather than period-by-period conditional means. This makes identification infeasible with our strategy, which exploits only marginal conditional expectations and covariances from partially observed data. \par 
When applied to the life-cycle estimator of \citet{mello2022lifecycle}, our definition yields results that are virtually identical to those obtained using the log-sum specification. In their framework, predicted log annual incomes for children are first exponentiated to obtain absolute incomes, which are then averaged across years and logged to form permanent income. The IGE is subsequently estimated by regressing this measure on parental average log income. Table \ref{tab:ige_cohort} depicts that  the resulting IGE estimates under their log-average definition ($Y_c^{PL}$) are nearly indistinguishable from those obtained using our average-log definition ($Y_c^P$).\par
This close correspondence can be understood by rewriting the log-mean as
\begin{align*} \log\!\left( \frac{1}{T} \sum_{t=1}^T e^{Y_{gt}} \right)  &= \log\!\left( \sum_{t=1}^T e^{Y_{gt}} \right) - \log(T) \\ &= \log\!\left( e^{Y_g^P} \sum_{t=1}^T e^{Y_{gt} - Y_g^P} \right) - \log(T) \\ &= Y_g^P + \log\!\left( \sum_{t=1}^T e^{Y_{gt} - Y_g^P} \right) - \log(T) \\ &= Y_g^P + \log\!\left( \frac{1}{T} \sum_{t=1}^T e^{Y_{gt} - Y_g^P} \right).
\end{align*} 
When annual log income deviates minimally from its average ($Y_{gt} - Y^P_g \approx 0$), then $e^{Y{gt} - Y^P_g} \approx 1$ for all $t$. Consequently, the average inside the log term is approximately 1, so that
\begin{align*}
Y_g^{PL} &=\log\left( \frac{1}{T} \sum_{t=1}^T e^{Y_{gt}} \right)\approx Y_g^P+ \log(1) \\
&= Y_g^P,
\end{align*}
explaining the similarity in IGE estimates of the life-cycle estimator. 
\subsection{Proof of Proposition 1}\label{sec:theo1}
We aim to characterize the population quantity identified by the MI estimand, defined as
\begin{align*}  
\beta^{MI}&=\frac{\mathbb{E}\left[\left(\tilde{Y}_c^P-\mathbb{E}\left[\tilde{Y}_c^P\right]\right)\left(\tilde{Y}_f^P-\mathbb{E}\left[\tilde{Y}_f^P\right]\right)\right]}{\mathbb{E}\left[\left(\tilde{Y}_f^P-\mathbb{E}\left[\tilde{Y}_f^P\right]\right)^2\right]}\coloneqq\frac{\mathbb{E}\left[\tilde{y}_c^P\tilde{y}_f^P\right]}{\mathbb{E}\left[\left(\tilde{y}_f^P\right)^2\right]},
\end{align*} where low-case letters denote the random variables in deviations from their population mean. To this end, we derive closed-form expressions for the numerator and denominator under Assumptions \ref{as:1_app} and \ref{as:2mi_app}. We start by analyzing the denominator:   \begin{align}\label{eq:phi_p} \nonumber\mathbb{E}\left[\tilde{y}_c^P\tilde{y}_f^P\right]&=\mathbb{E}\left[\tilde{Y}_c^P\tilde{Y}_f^P\right]-\mathbb{E}\left[\tilde{Y}_c^P\right]\mathbb{E}\left[\tilde{Y}_f^P\right]\\    \nonumber
&=\mathbb{E}\left[\left(\frac{1}{T_c}\sum_{t\in \mathcal{M}_c}Y_{ct}D_{ct}\right)\left(\frac{1}{T_f}\sum_{j\in \mathcal{M}_f}Y_{fj}D_{fj}\right)\right]-\mathbb{E}\left[\tilde{Y}_c^P\right]\mathbb{E}\left[\tilde{Y}_f^P\right]\\\nonumber    
&=\frac{1}{T_c T_f}\sum_{t\in \mathcal{M}_c}\sum_{j\in \mathcal{M}_f}\mathbb{E}\left[Y_{ct}Y_{fj}D_{ct}D_{fj}\right]-\mathbb{E}\left[\tilde{Y}_c^P\right]\mathbb{E}\left[\tilde{Y}_f^P\right]\\\nonumber    
&=\frac{1}{T_c T_f}\sum_{t\in \mathcal{M}_c}\sum_{j\in \mathcal{M}_f}\mathbb{E}\left[\left(\lambda_tY_c^P+v_{ct}\right)\left(\lambda_jY_f^P+v_{fj}\right)D_{ct}D_{fj}\right]\nonumber-\mathbb{E}\left[\tilde{Y}_c^P\right]\mathbb{E}\left[\tilde{Y}_f^P\right]\\\nonumber      
&=\frac{1}{T_c T_f}\mathbb{E}\left[Y_c^PY_f^P\sum_{t\in \mathcal{M}_c}\lambda_t D_{ct}\sum_{j\in \mathcal{M}_f}\lambda_jD_{fj}\right]+    \frac{1}{T_c T_f}\mathbb{E}\left[Y_f^P\sum_{t\in \mathcal{M}_c} v_{ct}D_{ct}\sum_{j\in \mathcal{M}_f}\lambda_jD_{fj}\right]\\    
&+\frac{1}{T_c T_f}\sum_{t\in \mathcal{M}_c}\lambda_t\sum_{j\in \mathcal{M}_f}\mathbb{E}\left[Y_c^Pv_{fj}D_{ct}D_{fj}\right]+\frac{1}{T_c T_f}\sum_{t\in \mathcal{M}_c}\sum_{j\in \mathcal{M}_f}\mathbb{E}\left[v_{ct}v_{fj}D_{ct}D_{fj}\right]\nonumber\\  &-\mathbb{E}\left[\tilde{Y}_c^P\right]\mathbb{E}\left[\tilde{Y}_f^P\right],    \end{align}  where we have used the definition of average (log) income during mid-life as given in equation (\ref{eq:mi0}) in the second equality, and the fourth equality follows by Assumption \ref{as:1_app}.  \par The first term in equation (\ref{eq:phi_p}) can be expressed as 
\begin{align}\label{eq:aux3}\nonumber 
\frac{1}{T_c T_f}\mathbb{E}\left[Y_c^PY_f^P\sum_{t\in \mathcal{M}_c}\lambda_t D_{ct}\sum_{j\in \mathcal{M}_f}\lambda_jD_{fj}\right]&=\frac{1}{T_c T_f}\mathbb{E}\left[Y_c^PY_f^P\sum_{t\in \mathcal{M}_c} D_{ct}\sum_{j\in \mathcal{M}_f}D_{fj}\right]\nonumber\\\nonumber 
&=\frac{1}{T_c T_f}\mathbb{E}\left[Y_c^PY_f^P\right]\times T_c T_f\\ &=\mathbb{E}\left[Y_c^PY_f^P\right],    \end{align} where the first equality follows from assuming $\lambda_t=1, \forall t\in \mathcal{M}_g$ for $g \in \{c,f\}$ in Assumption \ref{as:1_app}, and the second one by the definition of $T_{g}\coloneqq\sum_{j\in \mathcal{M}_g}D_{gj}$.  As regards the second term in equation (\ref{eq:phi_p}),  it can be simplified as follows:
\begin{align}\label{eq:aux4}\nonumber 
\frac{1}{T_c T_f}\mathbb{E}\left[Y_f^P\sum_{t\in \mathcal{M}_c} v_{ct}D_{ct}\sum_{j\in \mathcal{M}_f}\lambda_jD_{fj}\right] 
&=\frac{1}{T_c T_f}\mathbb{E}\left[Y_f^P\sum_{t\in \mathcal{M}_c} v_{ct}D_{ct}\sum_{j\in \mathcal{M}_f}D_{fj}\right]\nonumber\\\nonumber 
&=\frac{1}{T_c T_f}\mathbb{E}\left[Y_f^P\sum_{t\in \mathcal{M}_c} v_{ct}D_{ct}\right]T_f\\\nonumber      
&=\frac{1}{T_c}\sum_{t\in \mathcal{M}_c} \mathbb{E}\left[Y_f^Pv_{ct}D_{ct}\right]\\\nonumber       
&=\frac{1}{T_c}\sum_{t} \mathbb{E}\left[Y_f^Pv_{ct}\big  | D_{ct}=1,t\in \mathcal{M}_c\right]\times p\left(D_{ct}=1\big | t\in \mathcal{M}_c\right)\\ 
&\coloneqq\frac{1}{T_c}\sum_{t} \mathbb{E}\left[Y_f^Pv_{ct}\big  | D_{ct}=1,t\in \mathcal{M}_c\right]\times p_c\left(t\in \mathcal{M}_c\right),
\end{align} 
where the fourth equality follows by the law of total probability.\par        The third term in equation (\ref{eq:phi_p}) equals zero by the law of iterated expectations  (LIE) and Assumption \ref{as:2mi_app}  
\begin{align}\label{eq:aux1} 
\mathbb{E}\left[Y_c^Pv_{fj}D_{ct}D_{fj}\right]=\mathbb{E}\left[\mathbb{E}\left[Y_c^Pv_{fj}\big | D_{ct},D_{fj}\right]D_{ct}D_{fj}\right]=\mathbb{E}\left[\mathbb{E}\left[Y_c^Pv_{fj}\big | D_{fj}\right]D_{ct}D_{fj}\right]=0.   
\end{align} 
Similarly, the fourth term in equation (\ref{eq:phi_p}) also equals zero by Assumption \ref{as:2mi_app} and LIE  
\begin{align} \label{eq:aux2}
\mathbb{E}\left[v_{ct}v_{fj}D_{ct}D_{fj}\right]=\mathbb{E}\left[\mathbb{E}\left[v_{ct}v_{fj}\big | D_{ct},D_{fj}\right]D_{ct}D_{fj}\right]=\mathbb{E}\left[\mathbb{E}\left[v_{ct}v_{fj}\big | D_{ct}, D_{fj}\right]D_{ct}D_{fj}\right]=0.    
\end{align} 
\par  Finally, for the last term in equation (\ref{eq:phi_p}) we have 
\begin{align}\label{eq:last}    \mathbb{E}\left[\tilde{Y}_c^P\right]\mathbb{E}\left[\tilde{Y}_f^P\right]   &=\mathbb{E}\left[\frac{1}{T_c}\sum_{t\in \mathcal{M}_c}Y_{ct}D_{ct}\right]\mathbb{E}\left[\frac{1}{T_f}\sum_{j\in \mathcal{M}_f}Y_{fj}D_{fj}\right]\nonumber\\
&=\mathbb{E}\left[\frac{1}{T_c}\sum_{t\in \mathcal{M}_c}\left(\lambda_tY_c^P+v_{ct}\right)D_{ct}\right]\mathbb{E}\left[\frac{1}{T_f}\sum_{j\in \mathcal{M}_f}\left(\lambda_jY_f^P+v_{fj}\right)D_{fj}\right]\nonumber\\
&=\mathbb{E}\left[Y_c^P\frac{1}{T_c}\sum_{t\in \mathcal{M}_c}D_{ct}+\frac{1}{T_c}\sum_{t\in \mathcal{M}_c}v_{ct}D_{ct}\right]\mathbb{E}\left[Y_f^P\frac{1}{T_f}\sum_{j\in \mathcal{M}_j}D_{fj}+\frac{1}{T_f}\sum_{j\in \mathcal{M}_f}v_{fj}D_{fj}\right]\nonumber\\     
&=\left(\mathbb{E}\left[Y_c^P\right]+\frac{1}{T_c}\sum_{t\in \mathcal{M}_c}\mathbb{E}\left[v_{ct}D_{ct}\right]\right)\left(\mathbb{E}\left[Y_f^P\right]+\frac{1}{T_f}\sum_{j\in \mathcal{M}_f}\mathbb{E}\left[v_{fj}D_{fj}\right]\right)\nonumber\\
&=\left(\mathbb{E}\left[Y_c^P\right]+\frac{1}{T_c}\sum_{t\in \mathcal{M}_c}\mathbb{E}\left[\mathbb{E}\left[v_{ct}|D_{ct}\right]D_{ct}\right]\right)\nonumber\\     
&\times \left(\mathbb{E}\left[Y_f^P\right]+\frac{1}{T_f}\sum_{j\in \mathcal{M}_f}\mathbb{E}\left[\mathbb{E}\left[v_{fj}|D_{fj}\right]D_{fj}\right]\right)\nonumber\\  
&=\mathbb{E}\left[Y_c^P\right]\mathbb{E}\left[Y_f^P\right],    
\end{align} 
where the last equality follows by Assumption \ref{as:1_app}.\par 
By plugging equations (\ref{eq:aux4}), (\ref{eq:aux3}),  (\ref{eq:aux1}), (\ref{eq:aux2}), and (\ref{eq:last}) into equation (\ref{eq:phi_p}), we have 
\begin{align}\label{eq:phi_p1} 
\mathbb{E}\left[\tilde{y}_c^P\tilde{y}_f^P\right]&=\mathbb{E}\left[\left(Y_c^P-\mathbb{E}\left[Y_c^P\right]\right)\left(Y_f^P-\mathbb{E}\left[Y_f^P\right]\right)\right]\nonumber\\
&+\frac{1}{T_c}\sum_{t} \mathbb{E}\left[Y_f^Pv_{ct}\big  | D_{ct}=1,t\in \mathcal{M}_c\right]\times p_c\left(t\in \mathcal{M}_c\right).    
\end{align} 
We now analyze the denominator in equation (\ref{eq:phi})    \begin{align}\label{eq:phi_d}\nonumber
\mathbb{E}\left[\left(\tilde{y}_f^P\right)^2\right] &=\mathbb{E}\left[\left(\frac{1}{T_f}\sum_{t\in \mathcal{M}_f}Y_{ft}D_{ft}\right)\left(\frac{1}{T_f}\sum_{j\in \mathcal{M}_f}Y_{fj}D_{fj}\right)\right]-\mathbb{E}\left[\left(\tilde{Y}_f^P\right)^2\right]\\\nonumber    
&=\frac{1}{T_f^2}\sum_{t\in \mathcal{M}_f}\sum_{j\in \mathcal{M}}\mathbb{E}\left[Y_{ft}Y_{fj}D_{ct}D_{fj}\right]-\mathbb{E}\left[\left(\tilde{Y}_f^P\right)^2\right]\\\nonumber    
&=\frac{1}{T_f^2}\sum_{t\in \mathcal{M}_f}\sum_{j\in \mathcal{M}_f}\mathbb{E}\left[\left(\lambda_tY_f^P+v_{ft}\right)\left(\lambda_jY_f^P+v_{fj}\right)D_{ft}D_{fj}\right]-\mathbb{E}\left[\left(\tilde{Y}_f^P\right)^2\right]\\\nonumber    
&=\frac{1}{T_f^2}\mathbb{E}\left[\left(Y_f^P\right)^2\sum_{t\in \mathcal{M}_f}\lambda_tD_{ft}\sum_{j\in \mathcal{M}_f}\lambda_jD_{fj}\right]+    \frac{1}{T_f^2}\sum_{t\in \mathcal{M}_f}\sum_{j\in \mathcal{M}_f}\lambda_j\mathbb{E}\left[v_{ft}Y_f^PD_{ft}D_{fj}\right]\\    
&+\frac{1}{T_f^2}\sum_{t\in \mathcal{M}_f}\lambda_t\sum_{j\in \mathcal{M}_f}\mathbb{E}\left[Y_f^Pv_{fj}D_{ft}D_{fj}\right]+\frac{1}{T_f^2}\sum_{t\in \mathcal{M}_f}\sum_{j\in \mathcal{M}_f}\mathbb{E}\left[v_{ft}v_{fj}D_{ft}D_{fj}\right]\nonumber\\    &-\mathbb{E}\left[\left(\tilde{Y}_f^P\right)^2\right].    
\end{align} 
The first term of equation (\ref{eq:phi_d}) can be expressed as    \begin{align}\label{eq:aux6}
\nonumber \frac{1}{T_f^2}\mathbb{E}\left[\left(Y_f^P\right)^2\sum_{t\in \mathcal{M}_f}\lambda_tD_{ft}\sum_{j\in \mathcal{M}_f}\lambda_jD_{fj}\right]&=\frac{1}{T_f^2}\mathbb{E}\left[\left(Y_f^P\right)^2\sum_{t\in \mathcal{M}_f}D_{ft}\sum_{j\in \mathcal{M}_f}D_{fj}\right]\nonumber\\
&=\frac{1}{T_f^2}\mathbb{E}\left[\left(Y_f^P\right)^2\right]T_f^2\nonumber\\&=\mathbb{E}\left[\left(Y_f^P\right)^2\right]    
\end{align}
The second and third terms in the last display equal zero by Assumption \ref{as:1_app}, since     
\begin{align}\label{eq:aux5}      
\mathbb{E}\left[v_{ft}Y_f^PD_{ft}D_{fj}\right]&= \mathbb{E}\left[\mathbb{E}\left[v_{ft}Y_f^P | D_{ft}, D_{fj}\right]D_{ft}  D_{fj}\right]=0,    
\end{align}    
and the fourth term
\begin{align}\label{eq:aux7}
\nonumber \frac{1}{T_f^2}\sum_{t\in \mathcal{M}_f}\sum_{j\in \mathcal{M}_f}\mathbb{E}\left[v_{ft}v_{fj}D_{ft}D_{fj}\right]&=\frac{1}{T_f^2}\sum_{t}\sum_{j}\mathbb{E}\left[v_{ft}v_{fj}\big | D_{ft}=1,D_{fj}=1, \left\{t,j\right\}\in \mathcal{M}_f\right]\nonumber\\ 
&\times p\left(D_{ft}=1,D_{fj}=1\big |\left\{t,j\right\}\in \mathcal{M}_f\right)\nonumber\\ 
&\coloneqq\frac{1}{T_f^2}\sum_{t}\sum_{j}\mathbb{E}\left[v_{ft}v_{fj}\big | D_{ft}=1,D_{fj}=1, \left\{t,j\right\}\in \mathcal{M}_f\right]\nonumber\\
&\times p_f\left(\left\{t,j\right\}\in \mathcal{M}_f\right).    
\end{align} 
By the same arguments as those of equation (\ref{eq:last}), the last term in equation (\ref{eq:phi_d}) boils down to     
\begin{align}\label{eq:new_aux_2} 
\mathbb{E}\left[\left(\tilde{Y}_f^P\right)^2\right]=\mathbb{E}\left[\left(Y_f^P\right)^2\right]     
\end{align}
By plugging equations (\ref{eq:aux6}), (\ref{eq:aux5}),  (\ref{eq:aux7}), and (\ref{eq:new_aux_2}) into equation (\ref{eq:phi_d}), we have    
\begin{align}\label{eq:phi_d_new} 
\mathbb{E}\left[\left(\tilde{y}_f^P\right)^2\right] &=\mathbb{E}\left[\left(Y_f^P-\mathbb{E}\left[Y_f^P\right]\right)^2\right]\nonumber\\
&+\frac{1}{T_f^2}\sum_{t}\sum_{j}\mathbb{E}\left[v_{ft}v_{fj}\big | D_{ft}=1,D_{fj}=1, \left\{t,j\right\}\in \mathcal{M}_f\right]\times  p_f\left(\left\{t,j\right\}\in \mathcal{M}_f\right).    
\end{align}  
Thus, by plugging equations (\ref{eq:phi_p1}) and (\ref{eq:phi_d_new}) into (\ref{eq:phi}), we get         
\begin{gather}         \beta^{MI}=\frac{\mathbb{E}\left[\left(Y_c^P-\mathbb{E}\left[Y_c^P\right]\right)\left(Y_f^P-\mathbb{E}\left[Y_f^P\right]\right)\right]\nonumber+\frac{1}{T_c}\sum_{t} \mathbb{E}\left[Y_f^Pv_{ct}\big  | D_{ct}=1,t\in \mathcal{M}_c\right]\times p_c\left(t\in \mathcal{M}_c\right)}{\mathbb{E}\left[\left(Y_f^P-\mathbb{E}\left[Y_f^P\right]\right)^2\right]+\frac{1}{T_f^2}\sum_{t}\sum_{j}\mathbb{E}\left[v_{ft}v_{fj}\big | D_{ft}=1,D_{fj}=1, \left\{t,j\right\}\in \mathcal{M}_f\right]\times  p_f\left(\left\{t,j\right\}\in \mathcal{M}_f\right)}.   
\end{gather}
To characterize the limit in probability of the MI estimator, defined as
\begin{align*}
   \hat{\beta}_n^{MI}&=\frac{\mathbb{E}_n\left[\left(\tilde{Y}_c^P-\mathbb{E}_n\left[\tilde{Y}_c^P\right]\right)\left(\tilde{Y}_f^P-\mathbb{E}_n\left[\tilde{Y}_f^P\right]\right)\right]}{\mathbb{E}_n\left[\left(\tilde{Y}_f^P-\mathbb{E}_n\left[\tilde{Y}_f^P\right]\right)^2\right]},
    \end{align*} we study the convergence in probability of each of its components. The numerator converges in probability to 
\begin{align}\label{eq:first}     \mathbb{E}_n\left[\tilde{y}_c^P\tilde{y}_f^P\right]&\overset{p}{\to}\mathbb{E}\left[\left(\tilde{Y}_c^P-\mathbb{E}\left[\tilde{Y}_c^P\right]\right)\left(\tilde{Y}_f^P-\mathbb{E}\left[\tilde{Y}_f^P\right]\right)\right]\nonumber \\ 
&=\mathbb{E}\left[\left(Y_c^P-\mathbb{E}\left[Y_c^P\right]\right)\left(Y_f^P-\mathbb{E}\left[Y_f^P\right]\right)\right]\nonumber\\&+\frac{1}{T_c}\sum_{t} \mathbb{E}\left[Y_f^Pv_{ct}\big  | D_{ct}=1,t\in \mathcal{M}_c\right]\times p_c\left(t\in \mathcal{M}_c\right)\nonumber\\
&=\mathbb{E}\left[\left(\beta_0\left(Y_f^P-\mathbb{E}\left[Y_f^P\right]\right)+u\right)\left((Y_f^P-\mathbb{E}\left[Y_f^P\right]\right)\right]\nonumber\\
&+\frac{1}{T_c}\sum_{t} \mathbb{E}\left[Y_f^Pv_{ct}\big  | D_{ct}=1,t\in \mathcal{M}_c\right]\times p_c\left(t\in \mathcal{M}_c\right)\nonumber \\     
&=\beta_0\mathbb{E}\left[\left(Y_f^P-\mathbb{E}\left[Y_f^P\right]\right)^2\right]+\frac{1}{T_c}\sum_{t} \mathbb{E}\left[Y_f^Pv_{ct}\big  | D_{ct}=1,t\in \mathcal{M}_c\right]\times p_c\left(t\in \mathcal{M}_c\right) 
\end{align} 
where the convergence in probability follows by the Law of Large Numbers (LLN), the first equality from equation (\ref{eq:phi_p1}), and the second and third  by equation (\ref{eq:igemain}). As regards the denominator in equation (\ref{eq:mi_estimator}), it converges in probability to  
\begin{align}\label{eq:second}
\mathbb{E}_n\left[\left(\tilde{y}_f^P\right)^2\right]&\overset{p}{\to}\mathbb{E}\left[\left(\tilde{Y}_f^P-\mathbb{E}\left[\tilde{Y}_f^P\right]\right)^2\right]\nonumber\\
&=\mathbb{E}\left[\left(Y_f^P-\mathbb{E}\left[Y_f^P\right]\right)^2\right]\nonumber\\
&+\frac{1}{T_f^2}\sum_{t}\sum_{j}\mathbb{E}\left[v_{ft}v_{fj}\big | D_{ft}=1,D_{fj}=1, \left\{t,j\right\}\in \mathcal{M}_f\right]\times  p_f\left(\left\{t,j\right\}\in \mathcal{M}_f\right)
\end{align}  
where the convergence in probability follows by the LLN, and the equality from equation (\ref{eq:phi_d_new}). Thus, by plugging equations  (\ref{eq:first}) and (\ref{eq:second}) into (\ref{eq:phi}) and applying
the Continuous Mapping Theorem (CMT) yields
\begin{gather}\label{eq:a21}     \hat{\beta}_n^{MI}\overset{p}{\to}\frac{\beta_0\mathbb{E}\left[\left(Y_f^P-\mathbb{E}\left[Y_f^P\right]\right)^2\right]+\frac{1}{T_c}\sum_{t} \mathbb{E}\left[Y_f^Pv_{ct}\big  | D_{ct}=1,t\in \mathcal{M}_c\right]\times p_c\left(t\in \mathcal{M}_c\right)}{\mathbb{E}\left[\left(Y_f^P-\mathbb{E}\left[Y_f^P\right]\right)^2\right]+\frac{1}{T_f^2}\sum_{t}\sum_{j}\mathbb{E}\left[v_{ft}v_{fj}\big | D_{ft}=1,D_{fj}=1, \left\{t,j\right\}\in \mathcal{M}_f\right]\times  p_f\left(\left\{t,j\right\}\in \mathcal{M}_f\right)}. Q.E.D. \end{gather}
\section{Appendix B: nonparametric Identification of the IGE and Local Robustness}
\subsection{Proof of Theorem 1}\label{sec:theo2} 
To establish identification of the intergenerational elasticity, defined as \begin{align}\label{eq:ige_app}
    \beta_0&=\frac{\mathbb{E}\left[\left(Y_c^P-\mathbb{E}\left[Y_c^P\right]\right)\left(Y_f^P-\mathbb{E}\left[Y_f^P\right]\right)\right]}{\mathbb{E}\left[\left(Y_f^P-\mathbb{E}\left[Y_f^P\right]\right)^2\right]},
\end{align} we will express both the numerator and denominator into observable components. We start by showing identification of the conditional means of permanent income for both generations $g \in \{c,p\}$:
\begin{align}\label{eq:cond_means}
\mathbb{E}\left[Y_g^P\right]&=\mathbb{E}\left[\sum_{t=1}^TY_{gt}\right]\nonumber\\
 &=\sum_{t=1}^T\mathbb{E}\left[\frac{1}{T}Y_{gt}\right]\nonumber\\
 &=\frac{1}{T}\sum_{t=1}^T\mathbb{E}\left[\mathbb{E}\left[Y_{gt}\mid \bm X_{gt}\right]\right]\nonumber\\
 &=\frac{1}{T}\sum_{t=1}^T\mathbb{E}\left[\mathbb{E}\left[Y_{gt}\mid \bm X_{gt}, D_{gt}=1\right]\right]\nonumber\\
 &=\frac{1}{T}\mathbb{E}\left[\sum_{t=1}^T\mathbb{E}\left[Y_{gt}\mid \bm X_{gt}, D_{gt}=1\right]\right]\nonumber\\
&\coloneqq\mu_g^P. 
\end{align}
Turning attention to the numerator in equation (\ref{eq:ige_app}), we exploit Assumptions \ref{as:ortho_np} and \ref{as:unc_np} to express this term as:\fontsize{9}{11}\selectfont
\begin{align}\label{eq:base_2}\nonumber 
\mathbb{E}\left[\left(Y_c^P-\mathbb{E}\left[Y_c^P\right]\right)\left(Y_f^P-\mathbb{E}\left[Y_f^P\right]\right)\right]&\coloneqq\mathbb{E}\left[\left(Y_c^P-\mu_c^P\right)\left(Y_f^P-\mu_f^P\right)\right]\\\nonumber&=\frac{1}{T^2}\sum_{t=1}^T\sum_{j=1}^T\mathbb{E}\left[\left(Y_{ct}-\mu_c^P\right)\left(Y_{fj}-\mu_f^P\right)\right]\\  
&=\frac{1}{T^2}\sum_{t=1}^T\sum_{j=1}^T\mathbb{E}\left[\left(\mathbb{E}\left[Y_{ct}\mid \bm X_{ct}\right]+\epsilon_{ct}-\mu_c^P\right)\left(Y_{fj}-\mu_f^P\right)\right]\nonumber\\\nonumber        
&=\frac{1}{T^2}\sum_{t=1}^T\sum_{j=1}^T\mathbb{E}\left[\left(\mathbb{E}\left[Y_{ct}\mid \bm X_{ct}\right]-\mu_c^P\right)\left(Y_{fj}-\mu_f^P\right)\right]\\\nonumber  
&=\frac{1}{T^2}\sum_{t=1}^T\sum_{j=1}^T\mathbb{E}\left[\left(\mathbb{E}\left[Y_{ct}\mid \bm X_{ct},\bm X_{fj}\right]-\mu_c^P\right)\left(Y_{fj}-\mu_f^P\right)\right]\\\nonumber     
&=\frac{1}{T^2}\sum_{t=1}^T\sum_{j=1}^T\mathbb{E}\left[\left(\mathbb{E}\left[Y_{ct}\mid \bm X_{ct},\bm X_{fj}\right]-\mu_c^P\right)\left(\mathbb{E}\left[Y_{fj}\mid \bm X_{ct},\bm X_{fj}\right]-\mu_f^P\right)\right]      \\ 
&=\frac{1}{T^2}\sum_{t=1}^T\sum_{j=1}^T\mathbb{E}\left[\left(\mathbb{E}\left[Y_{ct}\mid \bm X_{ct}\right]-\mu_c^P\right)\left(\mathbb{E}\left[Y_{fj}\mid \bm X_{fj}\right]-\mu_f^P\right)\right]  \nonumber        \\ 
&=\frac{1}{T^2}\sum_{t=1}^T\sum_{j=1}^T\mathbb{E}\left[\left(\mathbb{E}\left[Y_{ct}\mid \bm X_{ct}, D_{ct}=1\right]-\mu_c^P\right)\left(\mathbb{E}\left[Y_{fj}\mid \bm X_{fj}, D_{fj}=1\right]-\mu_f^P\right)\right]\nonumber 
 \\ 
&=\frac{1}{T^2}\mathbb{E}\left[\sum_{t=1}^T\sum_{j=1}^T\left(\mathbb{E}\left[Y_{ct}\mid \bm X_{ct}, D_{ct}=1\right]-\mu_c^P\right)\left(\mathbb{E}\left[Y_{fj}\mid \bm X_{fj}, D_{fj}=1\right]-\mu_f^P\right)\right]\nonumber\\
&\coloneqq\frac{1}{T^2}\mathbb{E}\left[\sum_{t=1}^T\left(\mu_{ct}\left(\bm{X}_{ct},1\right)-\mu_c^P\right)\sum_{j=1}^T\left(\mu_{fj}\left(\bm{X}_{fj},1\right)-\mu_f^P\right)\right],
\end{align} where $\mu_{gt}(\bm{X}_{gt},1)\coloneqq\mathbb{E}\left[Y_{gt}\mid \bm X_{gt}, D_{gt}=1\right]$ for  $g \in \{c,p\}$.\par \normalsize
The first equality follows from the definition of the conditional means of permanent income in equation~\eqref{eq:cond_means}, while the second equality follows from the definition of permanent income together with the linearity of expectation. The prediction error of the children's annual income in the third equality is defined according to Assumption~\ref{as:ortho_np}$.ii$. The fourth equality utilizes the same assumption, which ensures that this error is uncorrelated with parental permanent income and has zero mean. The fifth equality exploits Assumption \ref{as:ortho_np}$.i$, ensuring  $\mathbb{E}\left[Y_{ct} \mid \bm{X}_{ct}, \bm{X}_{fj}\right] = \mathbb{E}\left[Y_{ct} \mid \bm{X}_t\right]$, while the sixth applies the same argument along with the law of iterated expectations. The seventh equality also uses Assumption \ref{as:ortho_np}$.i$, while the eigth equality follows from Assumption~\ref{as:unc_np}$.i$ and Assumption~\ref{as:unc_np}$.iii$, with the eight one resulting from another application of the linearity of expectations.\par 
Turning to the denominator of equation (\ref{eq:ige_app}), we have \begin{align}\label{eq:deno}\nonumber 
\mathbb{E}\left[\left(Y_f^P-\mathbb{E}\left[Y_f^P\right]\right)^2\right]&\coloneqq \mathbb{E}\left[\left(Y_f^P-\mu_f^P\right)^2\right] \\\nonumber&=\frac{1}{T^2}\sum_{t=1}^T\sum_{j=1}^T\mathbb{E}\left[\left(Y_{ft}-\mu_f^P\right)\left(Y_{fj}-\mu_f^P\right) \right]\\
    \nonumber
    &=\frac{1}{T^2} \sum_{t=1}^T\sum_{j=1}^T\mathbb{E}\left[\mathbb{E}\left[\left(Y_{ft}-\mu_f^P\right)\left(Y_{fj}-\mu_f^P\right)\mid \bm X_{ftj}\right]\right]\nonumber\\\nonumber
    &=\frac{1}{T^2} \sum_{|t-j| \leq h}\mathbb{E}\left[\mathbb{E}\left[\left(Y_{ft}-\mu_f^P\right)\left(Y_{fj}-\mu_f^P\right)\mid \bm X_{ftj}\right]\right]\\&+ \frac{1}{T^2}\sum_{|t-j| > h}\mathbb{E}\left[\mathbb{E}\left[\left(Y_{ft}-\mu_f^P\right)\left(Y_{fj}-\mu_f^P\right)\mid \bm X_{ftj}\right]\right],
\end{align}   where the third equality follows from LIE. Focusing on the second term of the last equality, we have
\begin{align}\label{eq:second_term}\nonumber
  &\sum_{|t-j| > h}\mathbb{E}\left[\mathbb{E}\left[\left(Y_{ft}-\mu_f^P\right)\left(Y_{fj}-\mu_f^P\right)\mid \bm X_{ftj}\right]\right]\\=&\sum_{|t-j| > h}\mathbb{E}\left[\mathbb{E}\left[\left(\mathbb{E}\left[Y_{ft}| \bm X_{ft}\right]+\epsilon_{ft}-\mu_f^P\right)\left(\mathbb{E}\left[Y_{fj}| \bm X_{fj}\right]+\epsilon_{fj}-\mu_f^P\right)\mid \bm X_{ftj}\right]\right]\nonumber\\
  =&\sum_{|t-j| > h}\mathbb{E}\left[\left(\mathbb{E}\left[Y_{ft}| \bm X_{ft}\right]-\mu_f^P\right)\left(\mathbb{E}\left[Y_{fj}| \bm X_{fj}\right]-\mu_f^P\right)\right]+\mathbb{E}\left[\mathbb{E}\left[\epsilon_{ft}\epsilon_{fj}\mid \bm{X}_{ftj}\right]\right]\nonumber\\
  =&\sum_{|t-j| > h}\mathbb{E}\left[\left(\mathbb{E}\left[Y_{ft}| \bm X_{ft}\right]-\mu_f^P\right)\left(\mathbb{E}\left[Y_{fj}| \bm X_{fj}\right]-\mu_f^P\right)\right]+\mathbb{E}\left[\epsilon_{ft}\epsilon_{fj}\right]\nonumber\\
  =&\sum_{|t-j| > h}\left(\mathbb{E}\left[Y_{ft}| \bm X_{ft}\right]-\mu_f^P\right)\left(\mathbb{E}\left[Y_{fj}| \bm X_{fj}\right]-\mu_f^P\right).
\end{align} In the second equality we have used Assumption \ref{as:ortho_np}$.i$, and the fact that $\mu_f^P$ is constant w.r.t $\bm X_{ftj}$. The cross-terms involving $\epsilon_{ft}$ and $\epsilon_{fj}$  therefore vanish, since $\mathbb{E}\left[\epsilon_{ft}\mid \bm X_{ftj}\right]=\mathbb{E}\left[\epsilon_{ft}\mid \bm X_{ft}\right]=0$. In the last equality, we have used Assumption \ref{as:ortho_np}$.iii$.\par 
Plugging equation (\ref{eq:second_term}) into (\ref{eq:deno}) identifies the covariance of parental permanent income: \fontsize{10}{12}\selectfont
\begin{align}\label{eq:deno_res}\nonumber 
\mathbb{E}\left[\left(Y_f^P-\mathbb{E}\left[Y_f^P\right]\right)^2\right]&=\frac{1}{T^2} \sum_{|t-j| \leq h}\mathbb{E}\left[\mathbb{E}\left[\left(Y_{ft}-\mu_f^P\right)\left(Y_{fj}-\mu_f^P\right)\mid \bm X_{ftj}\right]\right]\\&+\frac{1}{T^2} \sum_{|t-j| > h}\mathbb{E}\left[\left(\mathbb{E}\left[Y_{ft}| \bm X_{ft}\right]-\mu_f^P\right)\left(\mathbb{E}\left[Y_{fj}| \bm X_{fj}\right]-\mu_f^P\right)\right]\nonumber \\\nonumber
&=\frac{1}{T^2} \sum_{|t-j| \leq h}\mathbb{E}\left[\mathbb{E}\left[\left(Y_{ft}-\mu_f^P\right)\left(Y_{fj}-\mu_f^P\right)\mid \bm X_{ftj}, D_{ft}=1, D_{fj}=1\right]\right]\\&\nonumber+\frac{1}{T^2} \sum_{|t-j| > h}\mathbb{E}\left[\left(\mathbb{E}\left[Y_{ft}| \bm X_{ft}, D_{ft}=1\right]-\mu_f^P\right)\left(\mathbb{E}\left[Y_{fj}| \bm X_{fj}, D_{fj}=1\right]-\mu_f^P\right)\right]\\
&\coloneqq\frac{1}{T^2} \mathbb{E}\left[\sum_{|t-j| \leq h}\sigma_{tj}\left(\bm{X}_{ftj},1,1\right)+ \sum_{|t-j| > h}\left(\mu_{ft}\left(\bm{X}_{ft},1\right)-\mu_f^P\right)\left(\mu_{fj}\left(\bm{X}_{fj},1\right)-\mu_f^P\right)\right],
\end{align} \normalsize where we have defined $\sigma_{tj}\left(\bm{X}_{ftj},1,1\right) \coloneqq \mathbb{E}\left[\left(Y_{ft}-\mu_f^P\right)\left(Y_{fj}-\mu_f^P\right)\mid \bm X_{ftj}, D_{ft}=1, D_{fj}=1\right]$, and used Assumptions \ref{as:unc_np}$.i$ and \ref{as:unc_np}$.iii$.
\par   Finally, plugging equations (\ref{eq:base_2}) and (\ref{eq:deno_res}) into \ref{eq:ige_app}) yields
\begin{gather}
\nonumber\beta_0=\frac{\mathbb{E}\left[\sum_{t=1}^T\left(\mu_{ct}(\bm{X}_{ct},1)-\mu_c^P\right)\sum_{j=1}^T\left(\mu_{fj}\left(\bm{X}_{fj},1\right)-\mu_f^P\right)\right]}{ \mathbb{E}\left[\sum_{|t-j| \leq h}\sigma_{tj}\left(\bm{X}_{ftj},1,1\right)+ \sum_{|t-j| > h}\left(\mu_{ft}\left(\bm{X}_{ft},1\right)-\mu_f^P\right)\left(\mu_{fj}\left(\bm{X}_{fj},1\right)-\mu_f^P\right)\right]}.  Q.E.D.
\end{gather}
\subsection{Locally Robust Moments}\label{sec:illust} 
Before proposing a locally robust estimator for the IGE, we first illustrate the construction of locally robust moments, as proposed by \citet{chernozhukov2022locally}. The point of departure is GMM estimation of a parameter of interest $\theta$, which depends on a nuisance parameter $\bm \gamma$, and $\bm W$, a data observation with unknown cumulative distribution function (CDF) $F_0$. We assume that there is a known function $g\left(W,\gamma,\theta\right)$ of a possible realization $\bm W$ of $\bm W$, $\bm \gamma$ and $\theta$ such that
\begin{align}\label{eq:riesz_exp}   \mathbb{E}\left[g\left(W,\gamma_0,\theta_0\right)\right]=0,
\end{align} 
where $ \mathbb{E}\left[\cdot\right]$ is the expectation under $F_0$ and $\gamma_0$ is the probability limit (plim) under $F_0$ of a first step estimator $\hat{\bm \gamma}$. We also assume that $\theta_0$ is identified by this moment, meaning that $\theta_0$ is the unique solution to (\ref{eq:riesz_exp}) over $\theta$ in some set $\Theta$.\par 
\cite{chernozhukov2022locally} provide a general procedure to construct 
orthogonal moment functions for GMM, where first steps have no effect, locally, on average moment functions. In particular, the authors show that  an orthogonal (locally robust) moment function $\left(\psi\right)$ can be constructed by adding the first step influence function $\left(\phi\right)$ to the identifying moment function $\left(g\right)$ 
 \begin{align}\label{eq:lr_moment} 
\psi\left(W,\gamma,\alpha,\theta\right)=g\left(W,\gamma,\theta\right)+\phi\left(W,\gamma,\alpha,\theta\right),
 \end{align} 
 where $\alpha$ is a function called the Riesz representer\footnote{We have assumed that $\theta_0$ is identified by equation (\ref{eq:riesz_exp}). Thus, our object of interest can be expressed as   $\theta_0=\mathbb{E}\left[m\left(W,\gamma_0\right)\right]$. Under a continuity condition, we can express $\theta_0$ as
 \begin{align*}    
 \theta_0=\mathbb{E}\left[\gamma_0\alpha_0\right], \quad \text{for all possible } \gamma_0,
 \end{align*}  
 where $\alpha_0$ is called the Riesz representer of the functional $\gamma_0$.} of the functional $\gamma$, on which only the first step influence function\footnote{The first step influence function gives the effect of $\bm \gamma$ on average identifying moment functions under general misspecification. Therefore, adding the FSIF $\left(\phi\left(W,\gamma,\alpha,\theta\right)\right)$ to the identifying moment $g\left(W,\gamma,\theta\right)$, provides an orthogonal moment, where first step estimation of $\bm \gamma$ has no effect, locally, on $\mathbb{E}\left[g\left(W,\gamma,\theta\right)\right]$.}  depends.\par 
The vector of moment functions $\psi\left(W,\gamma,\alpha,\theta\right)$ is considered to be locally robust when (i) varying $\gamma$ away from $\gamma_0=\gamma(F_0)$ has no local effect on $\mathbb{E}\left[\psi\left(W,\gamma,\alpha_0,\theta\right)\right]$, and (ii) varying  $\alpha$ away from $\alpha_0$  has no local effect on $\mathbb{E}\left[\psi\left(W,\gamma_0,\alpha,\theta\right)\right]$, where $\gamma(F)$ is the limit in probability of $\hat{\gamma}$ for a possible CDF of the data $\bm W$, denoted by $F$. The first condition is met when the set $\Gamma$ of possible directions of departure of $\gamma(F)$ of $\gamma_0$ satisfy  
 \begin{align*}
     \frac{d}{dt}\mathbb{E}\left[\psi\left(W,\gamma_0+t\delta ,\alpha_0,\theta\right)\right]=0 \text{ for all } \delta \in \Gamma, \text{ and } \theta \in \Theta,
 \end{align*} where $t$ is a scalar, $\delta$ is a direction of deviation of  $\gamma(F)$ of $\gamma_0$, and the derivative is evaluated at $t=0$. The second condition is met when 
 \begin{align*}     \mathbb{E}\left[\phi\left(W,\gamma_0,\alpha,\theta\right)\right]=0 \text{ for all } \theta \in \Theta, \text{ and } \alpha \in \mathcal{A},  \end{align*} where the set $\mathcal{A}$ is given by the $\alpha_0$'s satisfying
 \begin{align*}     \frac{d}{d\tau}\mathbb{E}\left[g\left(W,\gamma\left(F_\tau\right),\theta\right)\right]&=\int \phi\left( \omega,\gamma_0,\alpha_0,\theta\right)H(d\omega),\\
     \mathbb{E}\left[\phi\left(W,\gamma_0,\alpha_0, \theta\right)\right]
&=0,\quad \mathbb{E}\left[\phi\left(W,\gamma_0,\alpha_0, \theta\right)^2\right]<\infty,
 \end{align*} 
 for all $H$ and all $\theta \in \Theta$, where $H$ is an alternative distribution of $\bm W$ different from its true distribution $F_0$, and  $F_\tau=(1-\tau)F_0+\tau H$ for $\tau \in [0,1]$, where $H$ is such that  $\gamma(F_\tau)$ exists for $\tau$ small enough and regularity conditions are met.
\subsection{A Locally Robust Moment for the IGE in the Presence of Incomplete Income Data}\label{sec:lrm} 
Theorem~\ref{thm:2} establishes an identification result for the intergenerational elasticity (IGE). However, estimating this parameter via the plug-in principle, e.g., using machine learning estimators for the conditional means, introduces model selection and regularization bias. To address this issue, we follow \citet{chernozhukov2022locally} and construct a debiased machine learning estimator for equation (\ref{eq:identification}). This estimator is based on an orthogonal moment function that corrects for the regularization bias in the estimation of $\beta_0$, which arises from the first-step estimation of the conditional expectations in our identification result.\par 
As illustrated in Appendix \ref{sec:illust}, to find the orthogonal moment function corresponding to equation (\ref{eq:ident}), it suffices to characterize the first step influence function of the identifying moment. To this end, we first define the following conditional expectations:\fontsize{10}{12} \selectfont
    \begin{align*} 
    \mu_{gt}\left(F_\tau\right)(z)&\coloneqq\mathbb{E}_\tau\left[Y_{gt}|Z_{t}=z\right], \quad \bm Z_{t}\coloneqq\left( \bm X_{gt}, D_{gt}\right), \quad g\in \{c,f\},\quad t=1, ..., T,\\
    \sigma_{tj}\left(F_\tau\right)(z)&\coloneqq\mathbb{E}_\tau\left[\left(Y_{ft}-\mu_f^P\right)\left(Y_{fj}-\mu_f^P\right)|Z_{tj}=z\right],\quad \bm Z_{tj}\coloneqq\left(\bm X_{ftj}, D_{ft}, D_{fj}\right),\quad t,j=1, ..., T,\\
    \mu_g^{1,T}\left(F_\tau\right)
    &\coloneqq \left(\mu_{1,t}\left(F_\tau\right)(z), ..., \mu_{gt}\left(F_\tau\right)(z)\right), \quad g\in \{c,f\},\\
    \sigma^{t,1,T}\left(F_\tau\right)
    &\coloneqq \left(\sigma_{t,1}\left(F_\tau\right)(z), ..., \sigma_{t,T}\left(F_\tau\right)(z)\right), \quad t=1,...,T\\\sigma^{1,T,1,T}\left(F_\tau\right)
    &\coloneqq \left(\sigma^{1,1,T}\left(F_\tau\right)(z), ..., \sigma^{T,1,T}\left(F_\tau\right)(z)\right),\\
      \gamma\left(F_\tau\right)
      &\coloneqq \left(\mu_g^{1,T}\left(F_\tau\right), \gamma^{f,1,T}\left(F_\tau\right),\gamma^{f,1,T,1,T}\left(F_\tau\right)\right),
\end{align*} \normalsize
where $\mathbb{E}_\tau$ denotes the expectation under $F_\tau=(1-\tau)F_0+\tau H$. Thus, equation~\eqref{eq:identification}, which identifies our parameter of interest $\beta_0$, can be rewritten as 
 \begin{align}\label{eq:ident_new}  \nonumber  \mathbb{E}\left[g_1\left(W,\gamma_0,\beta_0,\mu_c^P,\mu_f^P\right)\right]&=0, \\
\nonumber
g_1\left(W,\gamma_0,\beta_0,\mu_c^P,\mu_f^P\right)&=\beta_0\sum_{|t-j| \leq h} \sigma_{tj}\left(F_0\right)\left(\bm X_{ftj}, 1, 1\right)\\\nonumber
&+\beta_0\sum_{|t-j| > h}\left(\mu_{ft}\left(F_0\right)\left(\bm X_{ft}, 1\right)-\mu_f^P\right)\left(\mu_{fj}\left(F_0\right)\left(\bm X_{fj}, 1\right)-\mu_f^P\right)\\
&-\sum_{t=1}^T\left(\mu_{ct}\left(F_0\right)\left(\bm X_{ct}, 1\right)-\mu_c^P\right)\sum_{j=1}^T\left(\mu_{fj}\left(F_0\right)\left(\bm X_{fj}, 1\right)-\mu_f^P\right),
\end{align}   
where $\mathbb{E}\left[\cdot\right]$ is the expectation under the true distribution of $\bm W$ $\left(F_0\right)$ and $\gamma_0\coloneqq \gamma\left(F_0\right)$ is the probability limit  under $F_0$ of a first step estimator $\hat{\bm \gamma}$. Notice that $\beta_0$ also depends on the mean of children and parental income $\left(\mu_c^P,\mu_f^P\right)$. However, according to equation (\ref{eq:cond_means}) these two parameters  are identified by 
\begin{align*}
\mu_g^P&=\mathbb{E}\left[\sum_{t=1}^T\mu_{gt}\left(F_0\right)\left(\bm X_{gt}, 1\right)\right],  
\end{align*} so that the moment identifying $\mu_c^P$ can be expressed as
\begin{align}\label{eq:g2}\nonumber
\mathbb{E}\left[g_2\left(W,\gamma\left(F_0\right),\theta\right)\right]&=0,\\
g_2\left(W,\gamma\left(F_0\right),\theta\right)&=\sum_{t=1}^T\mu_{ct}\left(F_0\right)\left(\bm X_{ct}, 1\right)-\mu_c^P,  
\end{align} and analogously for $\mu_f^P$
\begin{align}\label{eq:g3}\nonumber
\mathbb{E}\left[g_3\left(W,\gamma\left(F_0\right),\theta\right)\right]&=0,\\
g_3\left(W,\gamma\left(F_0\right),\theta \right)&=\sum_{t=1}^T\mu_{ft}\left(F_0\right)\left(\bm X_{ft}, 1\right)-\mu_f^P.  
\end{align}
Accordingly, by defining the augmented parameter of interest $\theta_0\coloneqq \left(\beta_0,\mu_c^P,\mu_f^P\right)$, the identifying moment is given by 
\begin{align*}
 g\left(W,\gamma\left(F_\tau\right),\theta\right)= 
\begin{pmatrix}
g_1\left(W,\gamma\left(F_\tau\right),\theta\right) \\
g_2\left(W,\gamma\left(F_\tau\right),\theta\right) \\
g_3\left(W,\gamma\left(F_\tau\right),\theta\right)
\end{pmatrix}  .
\end{align*} Thus, to characterize the FSIF it suffices to find $\phi$ and $\alpha_0$ such that 
\begin{align}\label{eq:riesz_illustrate}
    \frac{d}{d\tau}\mathbb{E}\left[g\left(W,\gamma\left(F_\tau\right),\theta\right)\right]&=\int \phi\left(\omega,\gamma_0,\alpha_0,\theta\right)H(d\omega)
\end{align} holds. In other words, to derive the locally robust moment for the intergenerational elasticity, along with the means of permanent income, we must first characterize the influence function for each element in $\theta$, and then augment their corresponding identifying moments with it.\par We start by finding the FSIF for the nuisance parameter $\mu_{ct}\left(F_\tau\right)\left(\bm X_{ct}, 1\right)$ in the identifying equation $g_2\left(W,\gamma\left(F_\tau\right),\theta\right)$. The left-hand side of equation  (\ref{eq:riesz_illustrate}) is:
\begin{align}\label{eq:ref_muc}\nonumber
\frac{d}{d\tau}\mathbb{E}\left[g_2\left(W,\gamma\left(F_\tau\right),\theta\right)\right]&=\frac{d}{d\tau}\mathbb{E}\left[\sum_{t=1}^T\mu_{ct}\left(F_\tau\right)\left(\bm X_{ct}, 1\right)-\mu_c^P\right]\\
&=\sum_{t=1}^T\frac{d}{d\tau}\mathbb{E}\left[\mu_{ct}\left(F_\tau\right)\left(\bm X_{ct}, 1\right)\right],
\end{align} where the interchange of differentiation and expectation is justified by the dominated convergence theorem under standard regularity conditions. We now express the expectation as 
\begin{align}\label{eq:muc_if}\nonumber
\mathbb{E}\left[\mu_{ct}\left(F_\tau\right)\left(\bm X_{ct}, 1\right)\right]=&\mathbb{E}\left[\mathbb{E}_\tau\left[Y_{ct} \mid \bm X_{ct}, D_{ct}=1\right]\right]\\ \nonumber    
=&\mathbb{E}\left[\frac{D_{ct}}{p\left(D_{ct}=1|\bm X_{ct}\right)}\mathbb{E}_\tau\left[Y_{ct} \mid \bm X_{ct}, D_{ct}=1\right]\right]    \\   \nonumber  
=&\mathbb{E}\left[\frac{D_{ct}}{p\left(D_{ct}=1|\bm X_{ct}\right)}\mathbb{E}_\tau\left[Y_{ct} \mid \bm X_{ct}, D_{ct}\right]\right]\\  
 \coloneqq&\mathbb{E}\left[\alpha_{0c,t}\left(\bm X_{ct}, D_{ct}\right)\mu_{ct}\left(F_\tau\right)\left(\bm X_{ct}, D_{ct}\right)\right],
\end{align}
where the first equality follows by definition, the second by the law of iterated expectations, and the third one by the MAR Assumption \ref{as:unc_np}$.i$. Furthermore, the term $\alpha_{0c,t}\left(\bm X_{ct}, D_{ct}\right)$ is the Riesz representer of the functional $\mu_{ct}\left(F_\tau\right)\left(\bm X_{ct}, 1\right)$. \par
We now plug equation (\ref{eq:muc_if}) into (\ref{eq:ref_muc}) to characterize the FSIF for 
\begin{align}\label{eq:fsif_muc}\nonumber
 &\frac{d}{d\tau}\mathbb{E}\left[\mu_{ct}\left(F_\tau\right)\left(\bm X_{ct}, 1\right)\right]\\=& \frac{d}{d\tau}\mathbb{E}\left[\alpha_{0c,t}\left(\bm X_{ct}, D_{ct}\right)\mu_{ct}\left(F_\tau\right)\left(\bm X_{ct}, D_{ct}\right)\right]\nonumber\\ 
 =& -\frac{d}{d\tau}\mathbb{E}\left[\alpha_{0c,t}\left(\bm X_{ct}, D_{ct}\right)\left(Y_{ct}-\mu_{ct}\left(F_\tau\right)\left(\bm X_{ct}, D_{ct}\right)\right)\right]\nonumber\\ 
 =& \frac{d}{d\tau}\mathbb{E}_\tau\left[\alpha_{0c,t}\left(\bm X_{ct}, D_{ct}\right)\left(Y_{ct}-\mu_{ct}\left(F_\tau\right)\left(\bm X_{ct}, D_{ct}\right)\right)\right]\nonumber\\
 =& \int \alpha_{0c,t}\left(\bm X_{ct}, D_{ct}\right)\left(y_{ct} -\mu_{ct}\left(F_0\right)\left(\bm x_{ct}, d_{ct}\right)\right)H(d\omega)\nonumber\\
\coloneqq& \int \phi_{c,t}\left(\omega,\gamma,\alpha_{0c,t},\theta\right)H(d\omega),
\end{align} where the second equality follows by the fact that $Y_{ct}$ does not depend on $\tau$. The third equality exploits the fact that the prediction errors of children's annual income are uncorrelated to any function of $\left(\bm X_{ct}, D_{ct}\right)$, so that \begin{align*}
    \mathbb{E}_\tau\left[\alpha_{0c,t}\left(\bm X_{ct}, D_{ct}\right)\left(Y_{ct}-\mu_{ct}\left(F_\tau\right)\left(\bm X_{ct}, D_{ct}\right)\right)\right]=0,
\end{align*}
which in turn implies
\begin{align*}
    &\frac{d}{d\tau}\mathbb{E}_\tau\left[\alpha_{0c,t}\left(\bm X_{ct}, D_{ct}\right)\left(Y_{ct}-\mu_{ct}\left(F_\tau\right)\left(\bm X_{ct}, D_{ct}\right)\right)\right]=0\iff\\
&\frac{d}{d\tau}\mathbb{E}_\tau\left[\alpha_{0c,t}\left(\bm X_{ct}, D_{ct}\right)\left(Y_{ct}-\mu_{ct}\left(F_0\right)\left(\bm X_{ct}, D_{ct}\right)\right)\right]\\     &=-\frac{d}{d\tau}\mathbb{E}\left[\alpha_{0c,t}\left(\bm X_{ct}, D_{ct}\right)\left(Y_{ct}-\mu_{ct}\left(F_\tau\right)\left(\bm X_{ct}, D_{ct}\right)\right)\right].
\end{align*} Finally, the derivative of the expectation under the perturbed distribution $F_\tau$ corresponds to the integral with respect to the perturbation measure $H$ 
\begin{align*}
\frac{d}{d\tau}\mathbb{E}_\tau\left[\alpha_{0c,t}(Y_{ct}-\mu_{ct})\right] = \int \alpha_{0c,t}(y_{ct} - \mu_{ct}) H(d\omega),
\end{align*} because we consider the linear perturbation $F_\tau = F_0 + \tau(H-F_0)$. Thus, the derivative w.r.t $\tau $ isolates the perturbation $H-F_0$, the expectation under $H$ appears because we are evaluating the Gateaux derivative at $\tau=0$, and similarly the terms involving $F_0$ vanish, as they are constant w.r.t $\tau$.\par According to equations (\ref{eq:ref_muc}) and (\ref{eq:fsif_muc}), the FSIF of $\mu_c^P$ is thus given by 
\begin{align*}
    \sum_{t=1}^T\frac{D_{ct}}{p\left(D_{ct}=1 \mid \bm X_{ct}\right)}\left(Y_{ct}-\mu_{ct}\left(\bm X_{ct}, 1\right)\right), 
\end{align*} which corrects for the prediction errors in children's annual income, weighted by the propensity scores.\par Analogously, the the FSIF of $\mu_f^P$ is given by 
\begin{align*}
    \sum_{t=1}^T\frac{D_{ft}}{p\left(D_{ft}=1 \mid \bm X_{ft}\right)}\left(Y_{ft}-\mu_{ft}\left(\bm X_{ft}, 1\right)\right).
\end{align*}
\par We now turn attention to the moment identifying the intergenerational elasticity (equation (\ref{eq:ident_new})). In contrast to the two other identifying moments, this one involves the three nuisance parameters $\mu_{ct}, \mu_{ft},$ and $\sigma_{t,j}$. Thus, to find the FSIF, we start by decomposing the derivative on the left-hand side of \eqref{eq:riesz_illustrate} as follows: \fontsize{8.2}{8.2}\selectfont
\begin{align}\label{eq:riesz_lr}\nonumber
    \frac{d}{d\tau} \mathbb{E}\left[g\left(W,\gamma\left(F_\tau\right),\beta\right)\right]&=\beta \sum_{|t-j| \leq h} \underbrace{\frac{d}{d\tau}\mathbb{E}\left[\sigma_{tj}\left(F_\tau\right)\left(\bm X_{ftj}, 1, 1\right)\right]}_{(1)}\\\nonumber &+\beta\sum_{t=1}^T \sum_{j=t+1}^{\min(t+h, T)}\underbrace{\frac{d}{d\tau}\mathbb{E}\left[\left(\mu_{fj}\left(F_0\right)\left(\bm X_{fj}, 1\right)-\mu_f^P\right)\mu_{ft}\left(F_\tau\right)\left(\bm X_{ft}, 1\right)\right]}_{(2)}\\\nonumber &+\beta\sum_{t=1}^T \sum_{j=t+1}^{\min(t+h, T)}\underbrace{\frac{d}{d\tau}\mathbb{E}\left[\left(\mu_{ft}\left(F_0\right)\left(\bm X_{ft}, 1\right)-\mu_f^P\right)\mu_{fj}\left(F_\tau\right)\left(\bm X_{fj}, 1\right)\right]}_{(3)}\\\nonumber &-\sum_{t=1}^T \sum_{j=1}^{T}\underbrace{\frac{d}{d\tau}\mathbb{E}\left[\left(\mu_{ct}\left(F_0\right)\left(\bm X_{ct}, 1\right)-\mu_c^P\right)\mu_{fj}\left(F_\tau\right)\left(\bm X_{fj}, 1\right)\right]}_{(4)}\\ &-\sum_{t=1}^T \sum_{j=1}^T\underbrace{\frac{d}{d\tau}\mathbb{E}\left[\left(\mu_{fj}\left(F_0\right)\left(\bm X_{fj}, 1\right)-\mu_f^P\right)\mu_{ct}\left(F_\tau\right)\left(\bm X_{ct}, 1\right)\right]}_{(5)}.
\end{align}\normalsize 
The FSIF is obtained by observing that each term (1)--(5) in Equation (\ref{eq:riesz_lr}) can be expressed in the form of the left-hand side of Equation (\ref{eq:riesz_illustrate}). Accordingly, we proceed as follows: for each term, we (i) identify its Riesz representer $\alpha_0$, (ii) derive the corresponding FSIF $\phi$ by expressing the term as the right-hand side of Equation (\ref{eq:riesz_illustrate}), and (iii) substitute these results back into Equation (\ref{eq:riesz_lr}). This yields the required solution for $\phi$ and $\alpha_0$ in Equation (\ref{eq:riesz_illustrate}). Below, we implement this procedure.\par 
We start by analyzing the expectation in term (1) of equation (\ref{eq:riesz_lr}), which can be expressed as:
\begin{align}\label{eq:riesz1}\nonumber 
& \mathbb{E}\left[\sigma_{tj}\left(F_\tau\right)\left(\bm X_{ftj}, 1, 1\right)\right]\\\nonumber \coloneqq&\mathbb{E}\left[\mathbb{E}_\tau\left[\left(Y_{ft} - \mu_f^P\right)\left(Y_{fj} - \mu_f^P\right) \mid \bm{X}_{ftj}, D_{ft}=1, D_{fj}=1\right]\right]\\
\nonumber \coloneqq&\mathbb{E}\left[\mathbb{E}_\tau\left[U_{ft}U_{fj}
\mid \bm{X}_{ftj}, D_{ft}=1, D_{fj}=1\right]\right]\\
\nonumber =& \mathbb{E}\left[ \mathbb{E}_\tau\left[U_{ft}U_{fj} \mid \bm{X}_{ftj}, D_{ft}, D_{fj}\right]\right]\\\nonumber  =& \mathbb{E}\left[\frac{D_{ft}D_{fj}}{p\left(D_{ft}=1, D_{fj}=1 \mid \bm{X}_{ftj}\right)} \mathbb{E}_\tau\left[U_{ft}U_{fj}\mid \bm{X}_{ftj}, D_{ft}, D_{fj}\right]\right]\\ \coloneqq& \mathbb{E}\left[\alpha_{01,tj}\left(\bm X_{ftj}, D_{ft}, D_{fj}\right)\sigma_{tj}\left(F_\tau\right)\left(\bm X_{ftj}, D_{ft}, D_{fj}\right)\right],
\end{align} where in the second equality we have defined $U_{ft}\coloneqq Y_{fj} - \mu_f^P $,  the third equality follows by Assumption \ref{as:unc_np}$.iii$, and the fourth by LIE  and the fact that 
\begin{align*}
&D_{ft}D_{fj}\mathbb{E}_\tau\left[U_{ft}U_{fj} \mid \bm X_{ftj}, D_{ft}=1, D_{fj}=1\right]\\=& D_{ft}D_{fj}\mathbb{E}_\tau\left[U_{ft}U_{fj} \mid \bm X_{ftj}, D_{ft}, D_{fj}\right].
\end{align*}  
Having characterized the Riesz representer for $\mathbb{E}\left[\sigma_{tj}\left(F_\tau\right)\left(\bm X_{ftj}, 1, 1\right)\right]$, we now turn to derive its corresponding FSIF:\begin{align}\label{eq:fsif1}\nonumber
 &\frac{d}{d\tau}\mathbb{E}\left[\sigma_{tj}\left(F_\tau\right)\left(\bm X_{ftj}, 1, 1\right)\right]\\=& \frac{d}{d\tau}\mathbb{E}\left[\alpha_{01,tj}\left(\bm X_{ftj}, D_{ft}, D_{fj}\right)\sigma_{tj}\left(F_\tau\right)\left(\bm X_{ftj}, D_{ft}, D_{fj}\right)\right]\nonumber\\ 
=& -\frac{d}{d\tau}\mathbb{E}\left[\alpha_{01,tj}\left(\bm X_{ftj}, D_{ft}, D_{fj}\right)\left(U_{ft}U_{fj} -\sigma_{tj}\left(F_\tau\right)\left(\bm X_{ftj}, D_{ft}, D_{fj}\right)\right)\right]\nonumber\\ 
 =& \frac{d}{d\tau}\mathbb{E}_\tau\left[\alpha_{01,tj}\left(\bm X_{ftj}, D_{ft}, D_{fj}\right)\left(U_{ft}U_{fj} -\sigma_{tj}\left(F_\tau\right)\left(\bm X_{ftj}, D_{ft}, D_{fj}\right)\right)\right]\nonumber\\
 =&\int \alpha_{01,tj}\left(\bm X_{ftj}, d_{ft}, d_{fj}\right)\left(U_{ft}U_{fj} -\sigma_{tj}\left(F_0;\mu_f^P\right)\left(\bm x_{ftj}, d_{ft}, d_{fj}\right)\right)H(d\omega)\nonumber\\
\coloneqq &\int \phi_{1,tj}\left(\omega,\gamma,\alpha_{01,tj}\right)H(d\omega),\nonumber\\
\end{align} \normalsize
following the same arguments as equation (\ref{eq:fsif_muc}).
\par
We now derive the FSIF for the second term in equation (\ref{eq:riesz_lr})
 \fontsize{10}{12} \selectfont
\begin{align*}\nonumber
&\mathbb{E}\left[\left(\mu_{fj}\left(F_0\right)\left(\bm X_{fj}, 1\right)-\mu_f^P\right)\mu_{ft}\left(F_\tau\right)\left(\bm X_{ft}, 1\right)\right]\\\nonumber=&\mathbb{E}\left[\left(\mathbb{E}\left[Y_{fj}\mid \bm X_{fj}, D_{fj}=1\right]-\mu_f^P\right)\mathbb{E}_\tau\left[Y_{ft} \mid \bm X_{ft}, D_{ft}=1\right]\right]\\ \nonumber    
=&\mathbb{E}\left[\left(\mathbb{E}\left[Y_{fj}\mid \bm X_{fj}, D_{fj}=1\right]-\mu_f^P\right)\frac{D_{ft}}{p\left(D_{ft}=1|\bm X_{ft}\right)}\mathbb{E}_\tau\left[Y_{ft} \mid \bm X_{ft}, D_{ft}=1\right]\right]    \\   \nonumber  
=&\mathbb{E}\left[\left(\mathbb{E}\left[Y_{fj}\mid \bm X_{fj}, D_{fj}=1\right]-\mu_f^P\right)\frac{D_{ft}}{p\left(D_{ft}=1|\bm X_{ft}\right)}\mathbb{E}_\tau\left[Y_{ft} \mid \bm X_{ft}, D_{ft}\right]\right]       \\\nonumber
 =&\mathbb{E}\left[\left(\mathbb{E}\left[Y_{fj}\mid \bm X_{fj}\right]-\mu_f^P\right)\frac{D_{ft}}{p\left(D_{ft}=1|\bm X_{ft}\right)}\mathbb{E}_\tau\left[Y_{ft} \mid \bm X_{ft}, D_{ft}\right]\right]       \\  
 \coloneqq&\mathbb{E}\left[\alpha_{02,tj}\left(\bm X_{fj},  \bm X_{ft},D_{ft}\right)\mu_{ft}\left(F_\tau\right)\left(\bm X_{ft}, D_{ft}\right)\right],
\end{align*}\normalsize where we have followed the same arguments
as those of equation (\ref{eq:riesz1}). An analogous procedure to equation (\ref{eq:fsif1}), yields
\begin{align}\label{eq:fsif2new}\nonumber
 &\frac{d}{d\tau}\mathbb{E}\left[\left(\mu_{fj}\left(F_0\right)\left(\bm X_{fj}, 1\right)-\mu_f^P\right)\mu_{ft}\left(F_\tau\right)\left(\bm X_{ft}, 1\right)\right]\\=& \frac{d}{d\tau}\mathbb{E}\left[\alpha_{02,tj}\left(\bm X_{fj},  \bm X_{ft},D_{ft}\right)\mu_{ft}\left(F_\tau\right)\left(\bm X_{ft}, D_{ft}\right)\right]\nonumber\\ 
 =& -\frac{d}{d\tau}\mathbb{E}\left[\alpha_{02,tj}\left(\bm X_{fj},  \bm X_{ft},D_{ft}\right)\left(Y_{ft}-\mu_{ft}\left(F_\tau\right)\left(\bm X_{ft}, D_{ft}\right)\right)\right]\nonumber\\ 
 =& \frac{d}{d\tau}\mathbb{E}_\tau\left[\alpha_{02,tj}\left(\bm X_{fj},  \bm X_{ft},D_{ft}\right)\left(Y_{ft}-\mu_{ft}\left(F_\tau\right)\left(\bm X_{ft}, D_{ft}\right)\right)\right]\nonumber\\
 =& \int \alpha_{02,tj}\left(\bm x_{fj},  \bm x_{ft},d_{ft}\right)\left(y_{ft} -\mu_{ft}\left(F_0\right)\left(\bm x_{t}, d_{ft}\right)\right)H(d\omega)\nonumber\\
\coloneqq& \int \phi_{2,tj}\left(\omega,\gamma,\alpha_{02,tj}\right)H(d\omega).
\end{align} \par 
The key distinction between equation (\ref{eq:fsif1}) and equation (\ref{eq:fsif2new}) is that the latter requires
\begin{align*}
    \mathbb{E}\left[\alpha_{02,tj}\left(\bm{X}_{fj},\bm{X}_{ft}, D_{ft}\right)\left(Y_{ft} - \mathbb{E}\left[Y_{ft}\mid \bm X_{ft}\right]\right)\right] = 0.
\end{align*}
This orthogonality condition is satisfied by Assumption \ref{as:ortho_np}$.i$, since the conditional expectation $\mathbb{E}[Y_{ft} \mid \bm{X}_{ft}, D_{ft}, \bm{X}_{fj}]$ reduces to $\mathbb{E}_\tau[Y_{ft} \mid \bm{X}_{ft}, D_{ft}]$, rendering the prediction error $Y_{ft} - \mathbb{E}_\tau[Y_{ft} \mid \bm{X}_t, D_{ft}]$ orthogonal to any function of $\bm{X}_{fj}$, $\bm{X}_{ft}$, and $D_{ft}$. \par 
Observe that terms (2), (3), (4) and (5) in equation (\ref{eq:riesz_lr}) share an identical functional form, differing only in their superscripts (indicating generation $g \in \{c,f\}$) and time indices ($j$ or $t$). This structural similarity implies that the derivations for terms (3), (4), and (5) have the same structure as term (2). Consequently, the FSIFs and Riesz representers for these terms are given by 
\fontsize{8}{10} \selectfont \begin{align}\label{eq:phi3_new}
\phi_{3,tj}\left(\omega,\gamma,\alpha_{03,tj}\right) &= \alpha_{03,tj}\left(\bm{X}_{ft}, \bm{X}_{fj}, D_{fj}\right)\left(Y_{fj} - \mu_{fj}\left(\bm{X}_{fj}, D_{fj}\right)\right), &
\alpha_{03,tj} &= \left(\mu_{ft}\left(\bm{X}_{ft}, D_{ft}\right)-\mu_f^P\right)\frac{D_{fj}}{p\left(D_{fj}=1|\bm{X}_{fj}\right)} \\\label{eq:phi4_rev}
\phi_{4,tj}\left(\omega,\gamma,\alpha_{04,tj}\right) &= \alpha_{04,tj}\left(\bm{X}_{ct}, \bm{X}_{fj}, D_{fj}\right)\left(Y_{fj} - \mu_{fj}\left(\bm{X}_{fj}, D_{fj}\right)\right), &
\alpha_{04,tj} &= \left(\mu_{ct}\left(\bm{X}_{ct}, D_{ft}\right)-\mu_c^P\right)\frac{D_{fj}}{p\left(D_{fj}=1|\bm{X}_{fj}\right)} \\\label{eq:phi5_rev}
\phi_{5,tj}\left(\omega,\gamma,\alpha_{05,tj}\right) &= \alpha_{05,tj}\left( \bm{X}_{fj},\bm{X}_{ct}, D_{ct}\right)\left(Y_{ct} - \mu_{ct}\left(\bm{X}_{ct}, D_{ct}\right)\right), &
\alpha_{05,tj} &= \left(\mu_{fj}\left(\bm{X}_{fj}, D_{fj}\right)-\mu_f^P\right)\frac{D_{ct}}{p\left(D_{ct}=1|\bm{X}_{ct}\right)}.
\end{align}\normalsize
The orthogonality condition $\mathbb{E}[\phi_{k,tj}] = 0$ also holds for each $k \in \{3,4,5\}$ by Assumption \ref{as:ortho_np}$.i$.\par
Having characterized the first-step influence function for each term in equation (\ref{eq:riesz_lr}), we can plug equations (\ref{eq:fsif1})--(\ref{eq:phi5_rev})  into equation (\ref{eq:riesz_lr}): \fontsize{9}{11} 
\begin{align}\label{eq:char_fsif}\nonumber
\frac{d}{d\tau} \mathbb{E}\left[g\left(W,\gamma\left(F_\tau\right),\theta\right)\right]&=\beta \sum_{|t-j| \leq h}\int \phi_{1,tj}\left(\omega,\gamma,\alpha_{01,tj}\right)H(d\omega)+\beta \sum_{|t-j| > h}\int \phi_{2,tj}\left(\omega,\gamma,\alpha_{02,tj}\right)H(d\omega)\\&+\beta \sum_{|t-j| > h}\int \phi_{3,tj}\left(\omega,\gamma,\alpha_{03,tj}\right)H(d\omega)-\sum_{t=1}^T\sum_{j=1}^T\int \phi_{4,tj}\left(\omega,\gamma,\alpha_{04,tj}\right)H(d\omega)\nonumber\\
&\nonumber-\sum_{t=1}^T\sum_{j=1}^T\int \phi_{5,tj}\left(\omega,\gamma,\alpha_{05,tj}\right) H(d\omega)\\
&\coloneqq\int \phi_1\left(\omega,\gamma_0,\alpha_0,\theta\right)H(d\omega).
\end{align}\normalsize
 \par Equation (\ref{eq:char_fsif}) defines the first step influence function of estimating $\gamma$ on the moment identifying $\beta$, thereby allowing us to construct a locally robust moment to estimate the IGE. To illustrate this point, consider again equation (\ref{eq:lr_moment})
  \begin{align*}
\psi\left(W,\gamma,\alpha,\theta\right)=g\left(W,\gamma,\theta\right)+\phi\left(W,\gamma,\alpha,\theta\right).
 \end{align*} 
 Thus, we construct the locally robust moment by adding $\phi_1\left(W,\gamma,\alpha,\theta\right)$ from equation (\ref{eq:char_fsif}) to the identifying moment $g_1\left(W,\gamma,\theta\right)$ in equation (\ref{eq:ident_new})  
 \begin{align}\label{eq:important}\nonumber   \psi_1\left(W,\gamma,\alpha,\theta\right)&=\beta\sum_{|t-j| \leq h} \sigma_{tj}\left(\bm X_{ftj}, 1, 1\right)+\beta\sum_{|t-j| > h}\left(\mu_{ft}\left(\bm X_{ft}, 1\right)-\mu_f^P\right)\left(\mu_{fj}\left(\bm X_{fj}, 1\right)-\mu_f^P\right)\\
\nonumber &-\sum_{t=1}^T\left(\mu_{ct}\left(\bm X_{ct}, 1\right)-\mu_c^P\right)\sum_{j=1}^T\left(\mu_{fj}\left(\bm X_{fj}, 1\right)-\mu_f^P\right)\\\nonumber
 &+\beta \sum_{|t-j| \leq h}\frac{D_{ft}D_{fj}}{p\left(D_{ft}=1, D_{fj}=1 |\bm X_{ftj}\right)}\left(\left(Y_{ft}-\mu_f^P\right)\left(Y_{fj}-\mu_f^P\right) -\sigma_{tj}\left(\bm X_{ftj}, 1, 1\right)\right)\\ 
 \nonumber &+\beta \sum_{|t-j| > h}\left(\mu_{fj}\left(\bm X_{fj}, 1\right)-\mu_f^P\right)\frac{D_{ft}}{p\left(D_{ft}=1|\bm X_{ft}\right)}\left(Y_{ft}-\mu_{ft}\left(\bm X_{ft}, 1\right)\right)\\
  \nonumber &+\beta \sum_{|t-j| > h}\left(\mu_{ft}\left(\bm X_{ft}, 1\right)-\mu_f^P\right)\frac{D_{fj}}{p\left(D_{fj}=1|\bm X_{fj}\right)}\left(Y_{fj}-\mu_{fj}\left(\bm X_{fj}, 1\right)\right)\\
 \nonumber&-\sum_{t=1}^T\sum_{t=j}^T\left(\mu_{ct}\left(\bm X_{ct}, 1\right)-\mu_c^P\right)\frac{D_{fj}}{p\left(D_{fj}=1|\bm X_{fj}\right)}\left(Y_{fj}-\mu_{fj}\left(\bm X_{fj}, 1\right)\right)\\
 &-\sum_{t=1}^T\sum_{t=j}^T\left(\mu_{fj}\left(\bm X_{fj}, 1\right)-\mu_f^P\right)\frac{D_{ct}}{p\left(D_{ct}=1|\bm X_{ct}\right)}\left(Y_{ct}-\mu_{ct}\left(\bm X_{ct}, 1\right)\right).
  \end{align} Similarly for $\mu_f^P$ and $\mu_c^P$, we have 
  \begin{align}\label{eqs:12}\nonumber
   \psi_2\left(W,\gamma,\alpha,\theta\right)&=\sum_{t=1}^T\mu_{ct}\left(\bm X_{ct}, 1\right)-\mu_c^P+\sum_{t=1}^T\frac{D_{ct}}{p\left(D_{ct}=1|\bm X_{ct}\right)}\left(Y_{ct}-\mu_{ct}\left(\bm X_{ct}, 1\right)\right) \\
   \psi_3\left(W,\gamma,\alpha,\theta\right)&=\sum_{t=1}^T\mu_{ft}\left(\bm X_{ft}, 1\right)-\mu_f^P+\sum_{t=1}^T\frac{D_{ct}}{p\left(D_{ft}=1|\bm X_{ft}\right)}\left(Y_{ft}-\mu_{ft}\left(\bm X_{ft}, 1\right)\right).
  \end{align}
Thus, our locally robust moment for the parameter $\theta$ is given by
  \begin{align}\label{eq:phi_ref}
       \psi\left(W,\gamma,\alpha,\theta\right)=\left( \psi_1\left(W,\gamma,\alpha,\theta\right), \psi_2\left(W,\gamma,\alpha,\theta\right), \psi_3\left(W,\gamma,\alpha,\theta\right)\right),
  \end{align} which yields a locally robust moment for the IGE, incorporating that it depends on $\mu_f^P$ and $\mu_c^P$.\par 
A debiased GMM estimator for \(\theta\) is thus given by
\[
\hat{\theta} = \arg\min_{\theta \in \Theta} \hat{\psi}(\theta)^\prime \hat{\Upsilon} \, \hat{\psi}(\theta),
\]
where \(\hat{\Upsilon}\) is a positive semi-definite weighting matrix and \(\Theta\) denotes the parameter space.
\subsection{Asymptotic Properties of the Locally Robust Estimator}\label{sec:as_p}
To establish consistency for the locally robust estimator, we will impose the following assumption
\begin{assumption1}{C}{NP}(Boundedness and Regularity Conditions for  Consistency)
\label{ass:clr}
\begin{enumerate}[(i)]
    \item \textbf{Identification:} \( \mathbb{E}[\psi\left(W, \gamma_0, \alpha_0, \theta\right)] = 0 \) if and only if \( \theta = \theta_0 \);
    \item \textbf{Compactness:} The parameter space \( \Theta\subset \mathbb{R}^3 \) is compact;
    \item \textbf{Regularity of the Identifying Moment:} \( \mathbb{E}[\left\|g\left(W, \gamma_0, \theta\right)\right\|] < \infty \) and \\\( \displaystyle \int \left\|g\left(w, \hat{\gamma}^{(\ell)}, \theta\right) - g\left(W, \gamma_0, \theta\right)\right\| dF_0(w) \overset{p}{\to} 0 \)  for all \( \theta \in \Theta \), 
where \( F_0 \) denotes the unknown cumulative distribution function of the data \( W \).
    \item \textbf{Local Stability in \( \theta\):} There exist a constant \( C > 0 \) and a function \( d(W, \gamma) \) such that for \( \| \gamma - \gamma_0 \| \) sufficiently small and \( \hat{\theta}^{LR}_n, \theta \in \Theta \),
    \[
    \left\|g\left(W, \gamma, \hat{\theta}^{LR}_n\right) - g\left(W, \gamma, \theta\right)\right\| \leq d(W, \gamma) \left\|\hat{\theta}^{LR}_n - \theta\right\|^{1/C}, \quad \text{with } \mathbb{E}[d(W, \gamma)] < C;
    \]
    \item \textbf{Regularity of the Orthogonal Moment:}
    \begin{itemize}
        \item[(a)] \( \mathbb{E}[\left\|\psi\left(W, \gamma_0, \alpha_0, \theta_0\right)\right\|] < \infty \);
        \item[(b)] The following hold:
        \begin{align*}
        &\displaystyle \int \left\|\phi\left(w, \hat{\gamma}^{(\ell)}, \alpha_0, \theta_0\right) - \phi\left(w, \gamma_0, \alpha_0, \theta_0\right)\right\|^2 dF_0(w) \overset{p}{\to} 0, \\
        &\displaystyle \int \left\|\phi\left(w, \gamma_0, \hat{\alpha}^{(\ell)}, \theta_0\right) - \phi\left(w, \gamma_0, \alpha_0, \theta_0\right)\right\|^2 dF_0(w) \overset{p}{\to} 0, \\
        &\displaystyle \int \left\|\hat{\Delta}_\ell(w)\right\| dF_0(w) \overset{p}{\to} 0,
        \end{align*}
        where
        $
        \hat{\Delta}_\ell(w) := \phi(w, \hat{\gamma}^{(\ell)}, \hat{\alpha}^{(\ell)}, \hat{\theta}^{LR}_n)
        - \phi(w, \gamma_0, \hat{\alpha}^{(\ell)}, \hat{\theta}^{LR}_n)
        - \phi(w, \hat{\gamma}^{(\ell)}, \alpha_0, \theta_0)$ \\
        $+ \phi(w, \gamma_0, \alpha_0, \theta_0)$.
    \end{itemize} 
\end{enumerate}
\end{assumption1}
Identification ensures we are solving a well-posed moment problem.  The compactness assumption is economically meaningful as the intergenerational income elasticity is theoretically bounded on the interval $\left(0, 1\right)$, reflecting imperfect but positive persistence of income across generations. The assumed bounds align with cross-country evidence, where estimates range from approximately 0.14 (Denmark) to 0.58 (Brazil), with most developed economies exhibiting elasticities between 0.2 and 0.5 \citep{stuhler2018review}. Moreover, permanent income cannot exceed the highest observed income in the data, nor can it be negative for individuals with any labor market participation. \par 
The regularity of the identifying moment assumes integrability and $L^1$ continuity. The former guarantees the moment function remains well-defined in expectation across the entire parameter space, while the latter ensures that the difference between the moment function evaluated at the estimated nuisance parameter and its true value becomes negligible. Local Stability controls how the moment function varies with $\theta$, preventing extreme sensitivity to parameter changes when nuisance estimates are near their true values. The integrability condition for the orthogonal moment matches that of the identifying moment, while the stronger $L^2$ convergence (compared to $L^1$) ensures the difference between the orthogonal moment function evaluated at the estimated nuisance parameters and their true value becomes negligible faster than for the identifying moment. \par 
As regards the conditions for assymptotic normality, we start by imposing regularity conditions that translate into concrete requirements within our intergenerational mobility framework. Specifically, we require the orthogonal moment function in equation (\ref{eq:phi_ref}) to be square-integrable at the true parameter values. This condition implies two key substantive requirements: first, the inverse propensity weights must be bounded, ensured by Assumption~\ref{as:unc_np}, which restricts $p(D_{gt}=1|\bm X_{gt})$ and $p(D_{ft}=1, D_{fj}|\bm X_{fjt})$ away from zero and one; second, the income process must exhibit sufficient regularity, captured by weak temporal dependence (via mixing conditions) and finite higher-order moments, particularly for the cross-product terms $Y_{ft}Y_{fj}$ that enter the moment function.\par 
In addition, we assume that the machine learning estimators for the nuisance parameters, such as the conditional expectations $\mathbb{E}[Y_{ct}|\bm X_{ct}, D_{ct}=1]$ and the propensity scores, converge at suitable rates in mean-square error. These conditions are typically satisfied in longitudinal data settings where income dynamics are moderately dependent over time and income observation probability varies smoothly with covariates.\par 
We now turn to establishing the assumptions required for asymptotic normality.
\begin{assumption1}{3}{LR}(Boundedness and Regularity Conditions for Consistency)
\label{ass:4lr}
(i) The  orthogonal moment function is square-integrable:
\[
\mathbb{E}[\left\|\psi\left(W, \beta_0, \gamma_0, \alpha_0\right)\right\|^2] < \infty.
\]
(ii) The nuisance estimators are consistent in mean-square error:
\[
\int \left\|g\left(w, \hat{\gamma}^{(\ell)}, \theta_0\right) - g\left(w, \gamma_0, \theta_0\right)\right\|^2 dF_0(w) \xrightarrow{p} 0,
\]
\[
\int \left\|\phi\left(w, \hat{\gamma}^{(\ell)}, \alpha_0, \theta_0\right) - \phi\left(w, \gamma_0, \alpha_0, \theta_0\right)\right\|^2 dF_0(w) \xrightarrow{p} 0,
\]
\[
\int \left\|\phi\left(w, \gamma_0, \hat{\alpha}^{(\ell)}, \hat{\theta}^{LR}_n\right) - \phi\left(w, \gamma_0, \alpha_0, \theta_0\right)\right\|^2 dF_0(w) \xrightarrow{p} 0,
\] 
where \( F_0 \) denotes the unknown cumulative distribution function of the data \( W \).
\end{assumption1}
We further impose a regularity condition that controls the remainder term arising from the interaction of first-step estimation errors. Specifically, Assumption~\ref{ass:5lr} requires the correction term \( \hat\Delta_\ell(w) \), which captures the deviation from exact orthogonality due to estimation of the nuisance parameters, to vanish sufficiently quickly in sample averages. This condition imposes a rate requirement on the interaction remainder \( \hat\Delta_\ell(w) \), namely that its average must converge to zero faster than \( 1/\sqrt{n} \). This ensures that the remainder is asymptotically negligible and does not affect the limiting distribution of the estimator.
\begin{assumption1}{4}{LR}(First-Step Remainder Control)
\label{ass:5lr}
For each fold \( \ell = 1, \dots, L \), define the correction term
\[
\hat\Delta_\ell(w) := \phi(w, \hat{\gamma}^{(\ell)}, \hat{\alpha}^{(\ell)}, \hat{\theta}^{LR}_n )
- \phi(w, \gamma_0, \hat{\alpha}^{(\ell)}, \hat{\theta}^{LR}_n )
- \phi(w, \hat{\gamma}^{(\ell)}, \alpha_0, \theta_0)
+ \phi(w, \gamma_0, \alpha_0, \theta_0).
\]
We assume that this term satisfies at least one of the following conditions:

\begin{enumerate}
    \item \( \sqrt{n} \displaystyle\int \hat\Delta_\ell(w) \, dF_0(w) \xrightarrow{p} 0 \) and \( \displaystyle\int \left\|\hat\Delta_\ell(w)\right\|^2 \, dF_0(w) \xrightarrow{p} 0 \)
    \item \( \frac{1}{\sqrt{n}} \sum_{i \in I_\ell} \left\|\hat\Delta_\ell(W_i)\right\| \xrightarrow{p} 0 \)
    \item \( \frac{1}{\sqrt{n}} \sum_{i \in I_\ell} \hat\Delta_\ell(W_i) \xrightarrow{p} 0 \).
\end{enumerate}
\end{assumption1}
To establish valid inference after machine learning estimation of nuisance parameters, we require a Neyman orthogonality condition that ensures our moment function remains insensitive to small estimation errors. This assumption requires the moment condition to hold at estimated nuisance parameters $\hat{\alpha}^{(\ell)}$, and (2) specifying alternative bias control conditions that adapt to different estimation scenarios. The first condition (affine linearity) covers classical doubly robust estimators, while the second (quadratic bound with $n^{-1/4}$ convergence) handles many semiparametric cases. The third condition provides a weaker alternative when the bias vanishes asymptotically. In our framework, these conditions will be satisfied through either the double robustness properties of our moment function or the convergence rates of our machine learning estimators, similar to standard results in the semiparametric literature.
\begin{assumption1}{5}{LR}(Neyman Orthogonality and Bias Control)
\label{ass:lr_new}
For each fold \( \ell = 1, \dots, L \), we require the orthogonality condition
\[
\int \phi(w, \gamma_0, \hat{\alpha}^{(\ell)}, \beta) \, dF_0(w) = 0
\]
to hold with probability approaching one. In addition, one of the following conditions must be satisfied:
\begin{enumerate}
    \item \( \bar\psi(\gamma, \alpha, \beta) \coloneqq \mathbb{E}\left[\psi\left(W, \gamma, \alpha, \beta\right)\right] \) is affine in \( \gamma \)
    \item \( |\bar\psi(\gamma, \alpha_0, \theta_0)| \leq C \|\gamma - \gamma_0\|^2 \) for all \( \gamma \) such that \( \|\gamma - \gamma_0\| \) is sufficiently small, and \( \|\hat{\gamma}^{(\ell)} - \gamma_0\| = o_p(n^{-1/4}) \)
    \item \( \sqrt{n} \cdot \bar\psi(\hat{\gamma}^{(\ell)}, \alpha_0, \theta_0) \xrightarrow{p} 0 \).
\end{enumerate}
\end{assumption1}
The following assumption ensures the consistency of the auxiliary components required for valid variance estimation. It serves two key purposes: first, it guarantees that the estimation error in \( \hat\theta^{LR}_n \) becomes asymptotically negligible when substituted into the moment function; second, it requires the first-step remainder term \( \hat{\Delta}_\ell(w) \) to vanish fast enough. Together, these conditions ensure that the variability introduced by cross-fitting and parameter estimation does not distort the asymptotic variance calculations.\par 
\begin{assumption1}{6}{LR}(Square-integrability)
\label{ass:lr_other}
For each fold \( \ell = 1, \dots, L \)

$\displaystyle\int \left\| g\left(w, \hat{\gamma}^{(\ell)}, \hat\beta^{LR}_n\right) - g\left(w, \hat{\gamma}^{(\ell)}, \beta_0\right) \right\|^2 \, dF_0(w) \xrightarrow{p} 0$, and $
\int \left\| \hat{\Delta}_\ell(w) \right\|^2 \, dF_0(w) \xrightarrow{p} 0$.
\end{assumption1}
Finally, we impose an assumption to guarantee the stability of the Jacobian matrix \( G(\beta) \), ensuring the asymptotic normality of our estimator. Specifically, it requires: (1) convergence of the nuisance parameter estimates \(\hat{\gamma}^{(\ell)}\) to their true values \(\gamma_0\), (2) differentiability of the moment function \(\psi(W,  \gamma,\theta)\) in a neighborhood of \(\theta_0\), and (3) uniform convergence of the Jacobian over cross-fitting folds. These conditions ensure that the first-stage estimation of \(\gamma\) does not distort the asymptotic behavior of the estimator, even when machine learning methods are employed. Moreover, by controlling the sensitivity of the moment function to perturbations in both \(\theta\) and \(\gamma\), this assumption underpins the validity of inference in our cross-fitted setting.
\begin{assumption1}{7}{LR}(Jacobian Stability)
\label{ass:lr8}
The Jacobian matrix \( G(\beta) := \mathbb{E}\left[\partial_\theta \psi(W, \gamma_0, \theta)\right] \) exists, and there is a neighborhood \( \mathcal{N} \) around \( \theta_0 \) and a norm \( \|\cdot\| \) such that the following conditions hold:
\begin{enumerate}
    \item For each fold \( \ell \), the nuisance parameter estimate satisfies \( \|\hat{\gamma}^{(\ell)} - \gamma_0\| \xrightarrow{p} 0 \);
    \item For all \( \|\gamma - \gamma_0\| \) sufficiently small, the function \( \psi(W, \gamma,\theta) \) is differentiable with respect to \( \theta \) in \( \mathcal{N} \) with probability approaching one. Moreover, there exists a constant \( C > 0 \) and a function \( d(W, \gamma) \) such that for all \( \theta \in \mathcal{N} \) and \( \|\gamma - \gamma_0\| \) sufficiently small, 
    \[
    \left\| \frac{\partial \psi(W, \gamma,\theta)}{\partial \beta} - \frac{\partial \psi(W, \gamma,\theta_0)}{\partial \beta} \right\| \leq d(W, \gamma) |\theta - \theta_0|^{1/C}, \quad \text{with } \mathbb{E}[d(W, \gamma)] < C;
    \]
    \item For each fold \( \ell = 1, \dots, L \),
    \[
    \int \left\| \frac{\partial \psi(w, \hat{\gamma}^{(\ell)}, \theta_0)}{\partial \beta} - \frac{\partial \psi(w, \gamma_0, \theta_0)}{\partial \beta} \right\| dF_0(w) \xrightarrow{p} 0.
    \]
\end{enumerate}
\end{assumption1}\par Having established the assumptions for asymptotic normality, we now turn to derive a closed form solution for the asymptotic variance of \(\sqrt{n}(\hat{\theta} - \theta_0)\), which according to Lemma \ref{lemma:asymptotic_normality} is given by
\[
V = \left(G' \Upsilon G\right)^{-1},
\] where \[
G = \mathbb{E} \left[
  \frac{\partial g(W, \gamma, \theta)}{\partial \theta}
\right] = \mathbb{E} \left[
  \begin{pmatrix}
    \frac{\partial g_1(W,\gamma, \theta)}{\partial \beta} &
    \frac{\partial g_1(W,\gamma, \theta)}{\partial \mu_c^P} &
    \frac{\partial g_1(W,\gamma, \theta)}{\partial \mu_f^P} \\[6pt]
    \frac{\partial g_2(W,\gamma, \theta)}{\partial \beta} &
    \frac{\partial g_2(W,\gamma, \theta)}{\partial \mu_c^P} &
    \frac{\partial g_2(W,\gamma, \theta)}{\partial \mu_f^P} \\[6pt]
    \frac{\partial g_3(W,\gamma, \theta)}{\partial \beta} &
    \frac{\partial g_3(W,\gamma, \theta)}{\partial \mu_c^P} &
    \frac{\partial g_3(W,\gamma, \theta)}{\partial \mu_f^P}
  \end{pmatrix}
\right]
\]  is the Jacobian matrix, and
 $\Upsilon$ is the efficient weighting matrix defined as $\Upsilon = \Psi^{-1}$, $$\Psi = \frac{1}{n} \sum_{\ell=1}^L \sum_{i \in \mathcal{I}_\ell} \sum_{(tj) \in \mathcal{J}_i} \psi_{i,tj}^{(\ell)}\psi_{i,tj}^{(\ell)'}.$$
For the identifying moments in equations (\ref{eq:ident_new}), (\ref{eq:g2}), and (\ref{eq:g3}), we have
\begin{align}\label{eq:jaco}\nonumber
\frac{\partial g_1(W,\gamma, \theta)}{\partial \beta} &= 
\sum_{|t-j| \leq h} \sigma_{tj}\left(\bm X_{ftj}, 1, 1\right)
+ \sum_{|t-j| > h} \left(\mu_{ft}(\bm X_{ft}, 1) - \mu_f^P\right)\left(\mu_{fj}(\bm X_{fj}, 1) - \mu_f^P\right), \\\nonumber
\frac{\partial g_1(W,\gamma, \theta)}{\partial \mu_c^P} &= 
-\sum_{j=1}^T \left(\mu_{fj}(\bm X_{fj}, 1) - \mu_f^P\right), \\\nonumber\frac{\partial g_1(W,\gamma, \theta)}{\partial \mu_f^P} &= 
-\beta\sum_{t=1}^T\sum_{j=1}^T\left(\mu_{ft}(\bm X_{fj}, 1)+\mu_{fj}(\bm X_{fj}, 1) - 2\mu_f^P\right) 
+ \sum_{t=1}^T \left(\mu_{ct}(\bm X_{ct}, 1) - \mu_c^P\right), \\\nonumber
\frac{\partial g_2(W,\gamma, \theta)}{\partial \beta} &= 0, \quad
\frac{\partial g_2(W,\gamma, \theta)}{\partial \mu_c^P}= -1, \quad
\frac{\partial g_2(W,\gamma, \theta)}{\partial \mu_f^P} = 0, \quad
\frac{\partial g_3(W,\gamma, \theta)}{\partial \beta} = 0, \\\nonumber
\frac{\partial g_3(W,\gamma, \theta)}{\partial \mu_c^P} &= 0, \quad
\frac{\partial g_3(W,\gamma, \theta)}{\partial \mu_f^P} = -1,\\
\end{align} where we have used 
\fontsize{10}{12} \selectfont
\begin{align*}
\sigma_{tj}(X_{tj}, D_{ft}, D_{fj})
&= \mathbb{E} \left[ Y_{ft} Y_{fj} - \mu_f^P(Y_{ft} + Y_{fj}) + \left(\mu_f^P\right)^2 \mid \bm{X}_{ftj}, D_{ft}, D_{fj} \right]\\&=
\mathbb{E} \left[ Y_{ft} Y_{fj}\mid \bm{X}_{ftj}, D_{ft}, D_{fj} \right]  + \left(\mu_f^P\right)^2\\&- \mu_f^P \left(\mathbb{E} \left[Y_{ft}\mid \bm{X}_{ftj}, D_{ft}, D_{fj} \right] + \mathbb{E} \left[Y_{fj}\mid \bm{X}_{ftj}, D_{ft}, D_{fj} \right]\right)\\&=
\mathbb{E} \left[ Y_{ft} Y_{fj}\mid \bm{X}_{ftj}, D_{ft}, D_{fj} \right]  + \left(\mu_f^P\right)^2\\&- \mu_f^P \left(\mathbb{E} \left[Y_{ft}\mid \bm{X}_{ftj} \right] + \mathbb{E} \left[Y_{fj}\mid \bm{X}_{ftj} \right]\right)\\&=
\mathbb{E} \left[ Y_{ft} Y_{fj}\mid \bm{X}_{ftj}, D_{ft}, D_{fj} \right]  + \left(\mu_f^P\right)^2\\&- \mu_f^P \left(\mathbb{E} \left[Y_{ft}\mid \bm{X}_{ft}\right] + \mathbb{E} \left[Y_{fj}\mid \bm{X}_{fj} \right]\right)\\&=
\mathbb{E} \left[ Y_{ft} Y_{fj}\mid \bm{X}_{ftj}, D_{ft}, D_{fj} \right]  + \left(\mu_f^P\right)^2\\&- \mu_f^P \left(\mathbb{E} \left[Y_{ft}\mid \bm{X}_{ft},D_{ft}=1\right] + \mathbb{E} \left[Y_{fj}\mid \bm{X}_{fj},D_{fj}=1 \right]\right),
\end{align*} \normalsize to find $\frac{\partial g_1(W,\gamma, \theta)}{\partial \mu_f^P}$. In the last expression, the third and fourth equality follow by the MAR Assumption \ref{as:unc_np}$.iii$, and Assumption \ref{as:ortho_np}$.i$, which ensures $\mathbb{E}[Y_{ft} \mid \bm{X}_{ftj}] = \mathbb{E}[Y_{ft} \mid \bm{X}_t]$. Finally, the last equation also uses the MAR Assumption \ref{as:unc_np}$.iii$.\par
Combining the results above, we obtain the closed-form solution for Jacobian, and thus for the asymptotic variance  in Lemma \ref{lemma:asymptotic_normality}. Accordingly, the asymptotic variance can be estimated as
\begin{align*}
 \hat{V}= \left(\hat{G}' \hat{\Upsilon} \hat{G}\right)^{-1},   
\end{align*} where $\hat{G}$ is a consistent estimator of the Jacobian characterized by equation (\ref{eq:jaco}), and $\hat{\Upsilon}=\hat{\Psi}^{-1},$ where $$\hat{\Psi} = \frac{1}{n} \sum_{\ell=1}^L \sum_{i \in \mathcal{I}_\ell} \sum_{(tj) \in \mathcal{J}_i} \hat{\psi}_{i,tj}^{(\ell)}\hat{\psi}_{i,tj}^{(\ell)'}.$$
\subsection{Derivation of the $t-$ test for Assumptions \ref{as:ortho_np}$.ii$}\label{sec:as_test}
We begin by constructing a test for the orthogonality condition between children's income prediction errors and parental permanent income. The formal hypothesis is specified as:  
\begin{align}\label{eq:test1_app} H_0: \frac{1}{T}\sum_{t=1}^T\mathbb{E}\left[\epsilon_{ct}Y_{f}^P\right]=0, \quad  \quad \text{vs}\quad  H_1: \frac{1}{T}\sum_{t=1}^T\mathbb{E}\left[\epsilon_{ct}Y_{f}^P\right] \neq 0,   
\end{align}
where $\epsilon_{ct}\coloneqq Y_{ct}-\mathbb{E}\left[Y_{ct}\mid \bm X_{ct}\right]$ denotes the children's income prediction errors at time $t$ and $Y_f^P$ represents parental permanent income. The main challenge in testing this hypothesis is that both random variables are unobserved, and their machine learning estimation introduces regularization and model selection bias when testing $H_0$. To address these issues, we propose a three stages procedure. First, we establish identification of the object of interest \(\theta_{cf} \coloneqq \frac{1}{T}\sum_{t=1}^T\mathbb{E}\left[\epsilon_{ct}Y_{f}^P\right]\). Second, we construct a locally robust estimator \(\hat{\theta}_{cf}\). Finally, we provide a $t-$test based on $\hat{\theta}_{cf,n}$.\par 
The assumptions required for identifying $\theta_{cft}$ differ from those needed for identifying the IGE. Accordingly, we now present variants of Assumptions \ref{as:ortho_np} and \ref{as:unc_np}.
\begin{assumption1}{1}{NP'}(Conditional Mean Independence)\label{as:ortho_np_n}     The observable characteristics satisfy: $$     \mathbb{E}\left[Y_{ft} \mid \bm{X}_{ft},   \bm{X}_{cj}\right] = \mathbb{E}\left[Y_{ft} \mid \bm{X}_{ft}\right] \quad \text{ for } ,\quad  t,j=1,...T.$$ \end{assumption1} 
\begin{assumption1}{2}{NP'}(Missing At Random)\label{as:unc_np_new} 
\begin{enumerate}[i.]    \item The missingness of children's and parents annual income $Y_{gt}$ is  as good as random once we control for $\bm X_{gt}$\begin{align*}       Y_{gt}&\perp D_{gt}\mid \bm X_{gt}, \quad  t=1,..., T.    \end{align*}     \item  Given family characteristics, there is both missing and non-missing children and fathers' incomes for every age   \begin{align*}     0<&p\left(D_{ct}=1\mid \bm X_{ct} \right)<1 \quad a.s, \quad t=1,..., T.    \end{align*}     \item The missingness of child–parent income pairs is as good as random once we control for covariates:    \begin{align*}         \left(Y_{ct}, Y_{fj}\right) \perp \left(D_{ct}, D_{fj}\right) \mid \left(\bm X_{ct}, \bm X_{fj}\right), \quad t,j=1,\dots,T.    \end{align*}    \item Given covariates, there is both missing and non-missing child–parent income pairs:
    \begin{align*}        0 < \Pr\left(D_{ct}=1, D_{fj}=1 \mid \bm X_{ct}, \bm X_{fj}\right) < 1 \quad \text{a.s.}, \quad t,j=1,\dots,T.    \end{align*}\end{enumerate}\end{assumption1}
With this variants of the assumptions in place, we now show identification of $\theta_{cft}$: \begin{align}\label{eq:id_theta} \theta_{cf}&= \frac{1}{T}\sum_{t=1}^T\mathbb{E}\left[\epsilon_{ct}Y_{f}^P\right]\nonumber \\ &= \frac{1}{T}\sum_{t=1}^T\mathbb{E}\left[\epsilon_{ct} \frac{1}{T}\sum_{j=1}^TY_{fj}\right]\nonumber \\ &=\frac{1}{T^2}\sum_{t=1}^T \sum_{j=1}^T\mathbb{E}\left[\left(Y_{ct}-\mathbb{E}\left[Y_{ct}\mid \bm X_{ct}\right]\right)Y_{fj}\right] \nonumber \\ &= \frac{1}{T^2}\sum_{t=1}^T\sum_{j=1}^T\mathbb{E}\left[Y_{ct}Y_{fj}\right]-\mathbb{E}\left[\mathbb{E}\left[Y_{ct}\mid \bm X_{ct}\right]Y_{fj}\right] \nonumber \\ &=\frac{1}{T^2}\sum_{t=1}^T \sum_{j=1}^T\mathbb{E}\left[\mathbb{E}\left[Y_{ct}Y_{fj}\mid \bm X_{ct}, \bm X_{fj}\right]\right]-\mathbb{E}\left[\mathbb{E}\left[Y_{ct}\mid \bm X_{ct}\right]\mathbb{E}\left[ Y_{fj}\mid \bm X_{ct}, \bm X_{fj}\right]\right]\nonumber \\ &=\frac{1}{T^2}\sum_{t=1}^T\sum_{j=1}^T\mathbb{E}\left[\mathbb{E}\left[Y_{ct}Y_{fj}\mid \bm X_{ct}, \bm X_{fj},D_{ct}=1,D_{fj}=1\right]\right]\nonumber \\ &-\frac{1}{T^2}\mathbb{E}\left[\mathbb{E}\left[Y_{ct}\mid \bm X_{ct}, D_{ct}=1\right]\mathbb{E}\left[ Y_{fj}\mid \bm X_{fj}, D_{fj}=1\right]\right]\nonumber \\ &\coloneqq \mathbb{E}\left[\frac{1}{T^2}\sum_{t=1}^T\sum_{j=1}^T\left(\mu_{cftj}\left(\bm X_{ct},\bm X_{fj},1, 1\right)-\mu_{ct}\left(\bm X_{ct}, 1\right)\mu_{fj}\left(\bm X_{fj}, 1\right)\right)\right],\end{align} 
 where the second equality follows by the definition of permanent income. The fifth equality follows by LIE, while the sixth one follows by Assumption \ref{as:unc_np_new}$.iv.$\par  Having established identification, we now construct a moment for $\theta_{cf}$ that is locally robust to the nuisance parameters $\gamma_{cf} \coloneqq \left(\mu_f^{1,T},\mu_c^{1,T},\mu_{cf}^{t,1,T}\right)$, where 
 \begin{align*}
     \mu_g^{1,T}&\coloneqq \left(\mu_{g1},...,\mu_{gT}\right), \quad g\in \{c,f\},
     \\\mu_{cf}^{t,1,T}
    &\coloneqq \left(\mu_{cft}, ..., \mu_{cft}\right), \quad t=1,...,T\\\sigma^{1,T,1,T}&\coloneqq \left(\mu_{cf}^{1,1,T}, ..., \mu_{cf}^{T,1,T}\right).
 \end{align*}
 Building on the Riesz representer characterization for $\mu_{ct}$ in equation (\ref{eq:muc_if}) and following the arguments from Appendix \ref{sec:lrm}, the first-step influence function for $\mu_{ct}$ in the the identifying moment in equation (\ref{eq:id_theta}) is \begin{align*} -\frac{1}{T^2}\sum_{t=1}^T\sum_{j=1}^T\mu_{fj}\left(\bm X_{fj}, 1\right)\frac{D_{ct}}{p\left(D_{ct}=1|\bm X_{ct}\right)}\left(Y_{ct}-\mu_{ct}\left(\bm X_{ct}, 1\right)\right).\end{align*}
Similarly, for $\mu_{ft}$, we have \begin{align*} -\frac{1}{T^2}\sum_{t=1}^T\mu_{ct}\left(\bm X_{ct}, 1\right)\sum_{j=1}^T\frac{D_{fj}}{p\left(D_{fj}=1|\bm X_{fj}\right)}\left(Y_{fj}-\mu_{fj}\left(\bm X_{fj}, 1\right)\right). \end{align*} Following the same argument in equation (\ref{eq:riesz1}), the FSIF for $\mu_{cftj}$ is given by
\begin{align*}    \frac{1}{T^2}\sum_{t=1}^T\sum_{j=1}^T\frac{D_{ct}D_{fj}}{p\left(D_{ct}=1, D_{fj}=1 |\bm X_{ct}, \bm X_{fj}\right)}\left(Y_{ct}Y_{fj} -\mu_{cftj}\left(\bm X_{ct},\bm X_{fj},1, 1\right)\right)\end{align*}
Accordingly, the locally robust moment for $\theta_{cf}$ is given by  \fontsize{10}{12}\begin{align}\label{eq:cf}    \psi_{cf}\left(W, \gamma_{cf}, \theta_{cf}\right)&=g_{cf}\left(W,\gamma_{cf},\theta_{cf}\right)+\phi_{cf}\left(W, \gamma_{cf}, \alpha_{cf}, \theta_{cf}\right)\nonumber \\\nonumber    &=\frac{1}{T^2}\sum_{t=1}^T\sum_{j=1}^T\left(\mu_{cftj}\left(\bm X_{ct},\bm X_{fj},1, 1\right)-\mu_{ct}\left(\bm X_{ct}, 1\right)\mu_{fj}\left(\bm X_{fj}, 1\right)\right) -\theta_{cf}\\    &+\frac{1}{T^2}\sum_{t=1}^T\sum_{j=1}^T\frac{D_{ct}D_{fj}}{p\left(D_{ct}=1, D_{fj}=1 |\bm X_{ct}, \bm X_{fj}\right)}\left(Y_{ct}Y_{fj} -\mu_{cftj}\left(\bm X_{ct},\bm X_{fj},1, 1\right)\right) \nonumber\\\nonumber    &-\frac{1}{T^2}\sum_{t=1}^T\sum_{j=1}^T\mu_{ct}\left(\bm X_{ct}, 1\right)\frac{D_{fj}}{p\left(D_{fj}=1|\bm X_{fj}\right)}\left(Y_{fj}-\mu_{fj}\left(\bm X_{fj}, 1\right)\right)\\    &-\frac{1}{T^2}\sum_{t=1}^T\sum_{j=1}^T\mu_{fj}\left(\bm X_{fj}, 1\right)\frac{D_{ct}}{p\left(D_{ct}=1|\bm X_{ct}\right)}\left(Y_{ct}-\mu_{ct}\left(\bm X_{ct}, 1\right)\right).
\end{align} \normalsize  and the debiased moment function is then computed as
\begin{gather*} \hat{\psi}_{cf}\left(\theta_{cf}\right) = \frac{1}{n} \sum_{\ell=1}^L \sum_{f \in \mathcal{F}_\ell} \sum_{i \in \mathcal{P}_f} \sum_{(t,j) \in \mathcal{J}_i} \hat{\psi}_{cf,i,t,j}^{(\ell)},\quad \hat{\psi}_{i,tj}^{(\ell)} \coloneqq g_{cf}\big(W_{i,tj}, \hat{\gamma}_{cf}^{(\ell)}, \theta_{cf}\big) + \phi_{cf}\big(W_{i,tj}, \hat{\gamma}_{cf}^{(\ell)}, \hat{\alpha}_{cf}^{(\ell)}, \theta_{cf}\big), \end{gather*} where $\mathcal{J}_i$ denotes the set of all tuples $(t, j)$ observed for child--father pair $i$. Since the system is exactly identified, there is no need to compute fold-specific $\hat{\theta}_{cf}^{(\ell)}$. Accordingly, the locally robust estimator $\hat{\theta}_{cf,n}$ is the solution to the sample moment condition $ \hat{\psi}_{cf}\left(\theta_{cf}\right)=0.$\par  To test the  null hypothesis $H_0: \theta_{cf} = 0$ , we implement a $t-$test based on the estimator $\hat{\theta}_{cf,n}$. Accordingly, the $t$ statistic is  given by \[
t_{cf,n} = \frac{\hat{\theta}_{cf,n}}{\sqrt{\hat{V}_{cf,n}/n}},
\] where $\hat{V}_{cf,n}$ is a consistent estimator of the asymptotic variance of $\hat{\theta}_{cf,n}$.\par  Consistency of $\hat{\theta}_{cf,n}$ follows by Lemma \ref{lem:consistency}. In particular, under Assumptions \ref{as:ortho_np_n}, \ref{as:unc_np_new} and \ref{ass:clr}, we have \[ \hat{\theta}_{cf,n} \overset{p}{\to} \theta_{cf0}. \]
 Similarly, under Assumptions \ref{as:ortho_np_n}, \ref{as:unc_np_new}, \ref{ass:4lr}-\ref{ass:lr_new} and \ref{ass:lr8}, $\hat{\theta}^{LR}_n\xrightarrow{p}\theta_0$, the asymptotic normality of $\hat{\theta}_{cf,n}$  directly follows from Theorem 9 of \cite{chernozhukov2022locally}. Specifically, we have:
\[ \sqrt{n}\left(\hat{\theta}_{cf,n} - \theta_{cf0}\right) \xrightarrow{d} \mathcal{N}(0, V_{cf}), \] where \( V_{cf} =\mathbb{E}\left[\psi_{cf}^2\left(W, \gamma_{cf}, \theta_{cf}\right)\right] \). In addition, if Assumption \ref{ass:lr_other} holds, then  \( \hat{V}_{cf,n} \xrightarrow{p} V_{cf} \).\par  Under $H_0$ in equation (\ref{eq:test1_app}) and Assumptions \ref{as:ortho_np_n}, \ref{as:unc_np_new}, \ref{ass:4lr}-\ref{ass:lr_other}, we have
\begin{align*}\sqrt{n}\left(\hat{\theta}_{cf,n} - \theta_{cf0}\right) \xrightarrow{d} \mathcal{N}(0, V_{cf}) \Rightarrow \frac{\hat{\theta}_{cf,n}}{\sqrt{\hat{V}_{cf,n}/n}} \xrightarrow{d} \mathcal{N}(0, 1),
\end{align*}
where $\hat{V}_{cf,n}$ is a consistent estimator of $V_{cf}$ that accounts for dependence within families.\par 
Having established the asymptotic distribution of the test, we now turn to show that under Assumptions \ref{as:ortho_np_n}, \ref{as:unc_np_new}, \ref{ass:4lr}-\ref{ass:lr8} and \ref{ass:clr}, the test that rejects $H_0$ when $|t_{cf,n}| > z_{1-\alpha/2}$ is consistent.\par 
\begin{align*}t_{cf,n} &= \frac{\hat{\theta}_{cf,n}}{\sqrt{\hat{V}_{cf,n}/n}} \\&= \frac{\theta_{cf0} + O_p(n^{-1/2})}{\sqrt{(V_{cf}+o_p(1))/n}} \\&= \frac{\sqrt{n}\theta_{cf0} + O_p(1)}{V_{cf}^{1/2} + o_p(1)} \\&= \sqrt{n}\, \theta_{cf0}\, V_{cf}^{-1/2} + O_p(1).
\end{align*}
Thus, under $H_1$ where $\theta_{cf0} \neq 0$:
\begin{align*}\Pr\left(|t_{cf,n}| > z_{1-\alpha/2} \mid \theta_{cf0}\right) &= \Pr\left(\sqrt{n}\, \theta_{cf0}\, V_{cf}^{-1/2} + O_p(1) > z_{1-\alpha/2}\right) \\&\to 1,\end{align*}
since $\sqrt{n}\, \theta_{cf0}\, V_{cf}^{-1/2}$ diverges as $n \to \infty$ and dominates the $O_p(1)$ term.\par Having established the consistency of the test, we now analyze its behavior under local alternatives of the form
\[H_{1n}:\theta_{cf0} = \frac{\delta}{\sqrt{n}}, \quad \delta \in \mathbb{R} \ \text{fixed}.\]
Under Assumptions \ref{as:ortho_np_n}, \ref{as:unc_np_new}, \ref{ass:4lr}-\ref{ass:lr8} and \ref{ass:clr}, we have \[
\sqrt{n}\left(\hat{\theta}_{cf,n} - \theta_{cf0}\right) \xrightarrow{d} \mathcal{N}\left(0, V_{cf}\right), \quad \hat{V}_{cf,n} \xrightarrow{p} V_{cf}.\]
Therefore, the $t$-statistic satisfies
\begin{align*} t_{cf,n} &= \frac{\hat{\theta}_{cf,n}}{\sqrt{\hat{V}_{cf,n}/n}} \\ &= \frac{\theta_{cf0} + (\hat{\theta}_{cf,n} - \theta_{cf0})}{\sqrt{\hat{V}_{cf,n}/n}} \\ &= \frac{\delta/\sqrt{n} + (\hat{\theta}_{cf,n} - \theta_{cf0})}{\sqrt{V_{cf}/n} + o_p(n^{-1/2})} \\ &= \frac{\delta + \sqrt{n}(\hat{\theta}_{cf,n} - \theta_{cf0})}{V_{cf}^{1/2} + o_p(1)} \\ &= \frac{\delta}{V_{cf}^{1/2}} + \frac{\sqrt{n}(\hat{\theta}_{cf,n} - \theta_{cf0})}{V_{cf}^{1/2}} + o_p(1) \\ &= \frac{\delta}{V_{cf}^{1/2}} + O_p(1).
\end{align*} By Slutsky's theorem, under $H_{1n}$ we then have
\[t_{cf,n} \xrightarrow{d} \mathcal{N}\left( \frac{\delta}{V_{cf}^{1/2}}, 1 \right),\]
so that the $t$-test has asymptotic power \[ \lim_{n\to\infty}\Pr\left(|t_{cf,n}| > z_{1-\alpha/2} \mid \theta_{cf0} = \frac{\delta}{\sqrt{n}}\right) = 2\left[1 - \Phi\left(z_{1-\alpha/2} - \frac{|\delta|}{V_{cf}^{1/2}}\right)\right] > \alpha \quad \text{whenever } \delta \neq 0,\]
where $\Phi$ denotes the CDF of the standard normal distribution. \par 
Thus, the $t$-test has asymptotic power strictly greater than its size against any local alternative with $\delta \neq 0$, and the power approaches one under fixed alternatives.\par  \section{Tables} 
\begin{table}[ht]
\centering
\caption{Sample Sizes by Birth Cohort Window} 
\label{tab:sample_sizes}
\begin{tabular}{lcccc}
  \toprule
Cohort Window & Child-Father Pairs & Families & Child Obs. & Father Obs. \\ 
  \midrule
1954-1963 & 1,099 & 574 & 19,694 & 16,802 \\ 
1955-1964 & 1,089 & 578 & 19,053 & 17,561 \\ 
1956-1965 & 1,083 & 587 & 18,390 & 18,316 \\ 
1957-1966 & 1,071 & 588 & 17,601 & 19,117 \\ 
1958-1967 & 1,097 & 618 & 17,292 & 20,656 \\ 
1959-1968 & 1,085 & 636 & 16,729 & 21,073 \\ 
1960-1969 & 1,062 & 647 & 15,836 & 21,562 \\ 
1961-1970 & 1,042 & 650 & 15,065 & 21,812 \\ 
1962-1971 & 1,045 & 659 & 14,459 & 22,590 \\ 
1963-1972 & 1,038 & 676 & 13,717 & 22,986 \\ 
1964-1973 & 1,060 & 701 & 13,312 & 23,909 \\ 
1965-1974 & 1,067 & 718 & 12,766 & 24,259 \\ 
1966-1975 & 1,074 & 727 & 12,284 & 24,641 \\ 
1967-1976 & 1,088 & 742 & 11,912 & 25,092 \\ 
1968-1977 & 1,103 & 757 & 11,747 & 25,400 \\ 
   \midrule
Range & 1038-1103 & 574-757 & 11,747-19,694 & 16,802-25,400 \\ 
   \bottomrule
\end{tabular}
\end{table}
\begin{table}[ht]
\centering
\caption{Results of Locally Robust Test for Covariance of Children Prediction Errors and Parental Permanent Income Across Cohorts $\left(H_0: \frac{1}{T}\sum_{t=1}^T\mathbb{E}\left[\epsilon_{ct}Y_{f}^P\right]=0\right)$.} 
\label{tab:cohort_estimates}
\begin{tabular}{lccccc}
  \hline
Cohort & Number of Parents & Estimate & Standard Error & $t-$ statistics & $p$-value \\ 
  \hline
1954-1963 & 1099 & 0.322 & 0.234 & 1.374 & 0.170 \\ 
  1955-1964 & 1089 & 0.305 & 0.142 & 2.150 & 0.032 \\ 
  1956-1965 & 1083 & -0.008 & 0.123 & -0.067 & 0.946 \\ 
  1957-1966 & 1071 & 0.192 & 0.119 & 1.610 & 0.107 \\ 
  1958-1967 & 1097 & 0.166 & 0.175 & 0.951 & 0.341 \\ 
  1959-1968 & 1085 & 0.180 & 0.135 & 1.336 & 0.182 \\ 
  1960-1969 & 1062 & 0.171 & 0.105 & 1.625 & 0.104 \\ 
  1961-1970 & 1042 & 0.087 & 0.070 & 1.237 & 0.216 \\ 
  1962-1971 & 1045 & 0.013 & 0.068 & 0.186 & 0.852 \\ 
  1963-1972 & 1038 & 0.086 & 0.069 & 1.235 & 0.217 \\ 
  1964-1973 & 1060 & 0.056 & 0.064 & 0.878 & 0.380 \\ 
  1965-1974 & 1067 & 0.009 & 0.072 & 0.124 & 0.901 \\ 
  1966-1975 & 1074 & 0.074 & 0.065 & 1.150 & 0.250 \\ 
  1967-1976 & 1088 & 0.085 & 0.071 & 1.199 & 0.231 \\ 
  1968-1977 & 1103 & 0.003 & 0.071 & 0.042 & 0.967 \\ 
   \hline
\end{tabular}
\end{table}
 \end{document}